\documentclass[a4paper,fleqn,usenatbib]{mnras}
\usepackage{graphics}
\usepackage{graphicx}
\usepackage{soul}
\usepackage{multicol} 
\usepackage{amsmath}
\usepackage{amssymb}
\usepackage{url}
\usepackage{ulem}
\usepackage[T1]{fontenc}
\usepackage{ae,aecompl}
\usepackage{xcolor}
\usepackage{widetext}

\newcommand\chisq{\ifmmode{\chi\sp{2}}\else\math{\chi\sp{2}}\fi}
\newcommand\redchisq{\ifmmode{ \chi\sp{2}\sb{\rm red}}
    \else\math{\chi\sp{2}\sb{\rm red}}\fi}
\newcommand\Teq{\ifmmode{T\sb{\rm eq}}\else$T$\sb{eq}\fi}
\newcommand\mjup{\ifmmode{M\sb{\rm Jup}}\else$M$\sb{Jup}\fi}
\newcommand\rjup{\ifmmode{R\sb{\rm Jup}}\else$R$\sb{Jup}\fi}
\newcommand\msun{\ifmmode{M\sb{\odot}}\else$M\sb{\odot}$\fi}
\newcommand\rsun{\ifmmode{R\sb{\odot}}\else$R\sb{\odot}$\fi}
\newcommand\mearth{\ifmmode{M\sb{\oplus}}\else$M\sb{\oplus}$\fi}
\newcommand\rearth{\ifmmode{R\sb{\oplus}}\else$R\sb{\oplus}$\fi}

\newcommand\degree{\degr}
\newcommand\degrees{\degree}

\title[CSs-induced CMB in light of Weighted Morphology] {Cosmic Strings-induced CMB anisotropies in light of Weighted Morphology}
\author[Adeela Afzal, M. Alakhras, M. H. Jalali Kanafi and S. M. S. Movahed]
{Adeela Afzal$^{1,2}$, M. Alakhras$^{3,4}$, M. H. Jalali Kanafi$^{3,5}$ \& S. M. S. Movahed$^{3,6,7}$\thanks{E-mail: m.s.movahed@ipm.ir}
\\
$^{1}$Department of Physics, Quaid-i-Azam University, Islamabad, 45320, Pakistan\\
$^{2}$National Centre for Physics (NCP), Quaid-i-Azam University Campus, Shahdra Valley Road, Islamabad 44000, Pakistan\\
$^{3}$Department of Physics, Shahid Beheshti University,  1983969411, Tehran, Iran\\
$^{4}$Department of Physics, Faculty of Sciences, Damascus University, Damascus, Syria\\
$^{5}$ School of Physics, Institute for Research in Fundamental Sciences (IPM), P. O. Box 19395-5531, Tehran, Iran\\
$^{6}$ School of Astronomy, Institute for Research in Fundamental Sciences (IPM), P. O. Box 19395-5531, Tehran, Iran\\
$^{7}$Department of Mathematics and Statistics, The University of Lahore,
1-KM Defence Road Lahore-54000, Pakistan
}

\begin{document}
    \maketitle

    \begin{abstract}

  Motivated by the morphological measures in assessing the geometrical and topological properties of a generic cosmological stochastic field, we propose an extension of the weighted morphological measures, specifically the $n$th conditional moments of derivative (cmd-$n$). This criterion assigns a distinct weight to each excursion set point based on the associated field. We apply the cmd-$n$ on the Cosmic Microwave Background (CMB) to identify the cosmic string networks (CSs) through their unique Gott-Kaiser-Stebbins effect on the temperature anisotropies. We also formulate the perturbative expansion of cmd-$n$ for the weak non-Gaussian regime up to $\mathcal{O}(\sigma_0^3)$. We propose a comprehensive pipeline designed to analyze the morphological properties of string-induced CMB maps within the flat sky approximation. To evaluate the robustness of our proposed criteria, we employ string-induced high-resolution flat-sky CMB simulated patches of $7.2$ deg$^2$ size with a resolution of $0.42$ arcminutes. Our results demonstrate that the minimum detectable value of cosmic string tension is $G\mu\gtrsim 1.9\times 10^{-7}$ when a noise-free map is analyzed with normalized cmd-$n$. Whereas for the ACT, CMB-S4, and Planck-like experiments at 95.45\% confidence level, the normalized cmd-$n$ can distinguish the CSs network for  $G\mu\gtrsim2.9 \times 10^{-7}$, $G\mu\gtrsim 2.4\times 10^{-7}$ and $G\mu\gtrsim 5.8\times 10^{-7}$, respectively. The normalized cmd-$n$ exhibits a significantly enhanced capability in detecting CSs relative to the Minkowski Functionals.

    \end{abstract}




\section{Introduction}
\label{introduction}

The emergence of distinct phases resulting from the interaction among various components within a system is ubiquitous in nature. A spectacular consequence of the transition between multiple phases is defect formation. The topology of the underlying field's potential plays a crucial role in determining the stochastic distribution of stable topological defects, which include domain walls, monopoles, and CSs, as a consequence of spontaneous symmetry breaking  \citep{Vilenkin:2000jqa,2012PhRvX...2d1022G,2007spf..book.....K,2018lopt.book.....G}.
Achieving a comprehensive understanding of the occurrence of a phase transition in the early universe and exploiting associated consequences across a wide range of temporal and spatial scales requires model building, based on either fundamental principles or phenomenological descriptions and comparing with appropriate observational data sets. Moreover, it is critical to use simulations, generate synthetic data, and utilize techniques based on simulation-based inference (SBI) \citep[and references therein]{tejero2020sbi,2016arXiv160506376P,2019MNRAS.488.4440A,2016arXiv160506376P,2019MNRAS.488.4440A,2020PNAS..11730055C,2022mla..confE..24H}.

The extensive collection of rich, multi-dimensional, and high-precision data sets has become available through a multitude of observations and surveys, particularly those focused on the study of our cosmos. This includes a collection of surveys that are currently operational, ongoing and planned for the future, such as:
DESI\footnote{\texttt{http://www.desi.lbl.gov}} \citep{Dey_2019}, PFS\footnote{\texttt{http://pfs.ipmu.jp}} \citep{2016SPIE.9908E..1MT}, the Roman Space Telescope\footnote{\texttt{http://wfirst.gsfc.nasa.gov}} \citep{10.1093/mnras/stab1762}, Euclid\footnote{\texttt{http://sci.esa.int/euclid}} \citep{2020A&A...642A.191E}, and CSST\footnote{\texttt{http://nao.cas.cn/csst}} \citep{2011SSPMA,2019ApJ883}, Planck\footnote{\texttt{https://www.esa.int}} \citep{aghanim2020planck}, ACT\footnote{\texttt{https://act.princeton.edu/}} \citep{2020JCAP...12..047A}, SPT\footnote{\texttt{https://pole.uchicago.edu/}} \citep{2019JCAP...02..056A}, BICEP/Keck Array\footnote{\texttt{https://www.cfa.harvard.edu/CMB/bicepkeck/}} \citep{2019JCAP...02..056A}, Simons Observatory (SO)\footnote{\texttt{https://simonsobservatory.org/}} \citep{2019JCAP...02..056A},  CMB stage IV\footnote{\texttt{https://cmb-s4.org/}}  \citep{2019arXiv190704473A,2017arXiv170602464A}, CMB-HD\footnote{\texttt{https://cmb-hd.org/}} \citep{2022arXiv220305728T}, LiteBIRD\footnote{\texttt{https://www.isas.jaxa.jp/}} \citep{2020JLTP..199.1107S}, the Probe of Inflation and Cosmic Origins (PICO) \citep{2019arXiv190210541H}, AliCPT \citep{2017arXiv170909053L,2017arXiv171003047L,10.1093/nsr/nwy019,2024ApJS..274...26Z}, CCAT-prime\footnote{\texttt{https://www.ccatobservatory.org/}} \citep{2024SPIE13102E..22H}, SPIDER\footnote{\texttt{https://spider.princeton.edu/}} \citep{2022ApJ...927..174A}, CLASS (Cosmology Large Angular Scale Surveyor) \citep{10.1117/12.2056701}, Einstein Telescope (ET)\footnote{\texttt{https://www.et-gw.eu/}} \citep{Punturo_2010}, LISA (Laser Interferometer Space Antenna)\footnote{\texttt{https://www.lisamission.org/}} \citep{2017arXiv170200786A}, LIGO-Virgo-KAGRA\footnote{\texttt{https://ligo.org/}} \citep{2021PhRvD.104j2001A}, SKA (Square Kilometre Array)\footnote{\texttt{https://www.skao.int/en}} \citep{5136190}, LSST (Legacy Survey of Space and Time) \citep{2017arXiv170804058L}.

Alongside observational data, the advent of numerous simulation suites has enabled the effective use of big data to inform field-level inferences. This development includes forward modeling and the introduction of sophisticated measures that focus on the size, shape, connectivity, and boundaries represented by the un-weighted and weighted morphologies along with Topological Based Data analysis (TDA) \citep[and references therein]{2024ApJ...963...31K,2024MNRAS.535..657J, 2024JCAP...09..034Y}.
A Typical cosmological field (a short list of cosmological fields whose properties are elucidated by employing the statistical notion of random fields \citep{matsubara03} includes the Cosmic Microwave Background (CMB) \citep{bond1987statistics,hu2002cosmic,dodelson2020modern,durrer2020cosmic, Lesgourgues:2013qba,akrami2020planck}; the large-scale structure (LSS) of the universe \citep{Bernardeau:2001qr,Cooray:2002dia,peebles2020large}; stochastic gravitational wave background \citep{2018JCAP...11..038K,2019JCAP...11..017C,2023ApJ...951L...8A,2020ApJ...905L..34A}) acquires stochastic characteristics as a result of either the initial conditions, the evolutionary process, or both. The mathematical description of a generic stochastic field is given by: $\mathfrak{F}=\left\{\mathcal{F}_j\;|\; \mathcal{F}_j:\Pi_j\to\mathbb{R}, \;\Pi_j\subset\mathbb{R}^{D} \right\}_{j=1}^{d}$. Where $\mathcal{F}^{(d+D)}$, is a measurable mapping from probability space into a $\sigma$-algebra of $\mathbb{R}^{d}$-valued function on $\mathbb{R}^{D}$-Euclidean space \citep{adler81,adler2011topological,adler2010persistent}. Decomposition of $\mathcal{F}^{(d+D)}$ into orthogonal and complete sets can lead to ambiguities in cosmological inferences. Particularly, the clarity of the localized characteristics of the cosmological stochastic field is often reduced through the use of Fourier or Spherical Harmonics transformations. In addition, the one-point probability distributions are inherently limited in extracting the higher-order statistics.

A common extension of the aforementioned measure is provided by the weighted Two-Point Correlation Function (TPCF) \citep{rice1954selected,szalay88,desjacques2018large}. The incorporation of excess probability in the context of identifying feature pairs is represented through the weighted two-point correlation function (TPCF), which can be expressed in terms of the unweighted TPCF framework \citep{peeb80,kaiser1984spatial,Bardeen:1985tr} (see also \citep[and references therein]{2021MNRAS.503..815V}). Bispectrum and trispectrum are some famous criteria beyond two-point statistics. From a geometrical and topological perspective, one can define the excursion sets associated with $\mathcal{F}^{(d+D)}$ as:  $\mathcal{E}_{\mathcal{F}}(\vartheta^{(d)}):\{\boldsymbol{r}^{D}\in \mathcal{E}\; |\; \mathcal{F}^{(d)}(\boldsymbol{r}^{D})\ge \vartheta^{(d)} \}$  \citep{matsubara03,Pogosyan:2008jb,gay2012non,codis2013non,matsubara2020statistics,2021MNRAS.503..815V,2024MNRAS.528.1604S}. Imposing additional constraints on the common excursion sets results in e.g. critical sets, crossing statistics \citep{rice44a,rice44b,Bardeen:1985tr,bond1987statistics,ryden1988area,ryd89,matsubara03}. It is worth mentioning that under the assumption of the central limit theorem and the statistical isotropy, we can adopt a perturbative approach to describe these stochastic fields \citep{matsubara03,matsubara2020statistics,codis2013non}.

To quantify the morphology of $\mathcal{F}^{(d+D)}$, we should measure size, shape, connectedness, and boundaries performed by so-called un-weighted morphology \citep{mecke1993robust,schmalzing1997beyond,2001PhyA..293..592B,Beisbart2002}. Imposing the principles of motion invariance, additivity, and conditional continuity and according to the {\it Hadwiger's theorem}, the $(1+D)$ Minkowski Functionals (MFs) describe the morphology of field \citep{mecke1993robust,schmalzing1997beyond,2001PhyA..293..592B,Beisbart2002}. More recently, the generalized formalism to predict statistics of cosmological tensor fields has been introduced in \citep{2024PhRvD.110f3543M,2024PhRvD.110f3544M,2024PhRvD.110f3545M,2024PhRvD.110f3546M}.
Relaxing motion invariance, the Minkowski Tensors (MTs), have been introduced with diverse applications \citep[and references
therein]{matsubara1996genus,codis2013non,appleby2018minkowski,appleby2019ensemble,Appleby2023}. The application of scalar and tensor types of Minkowski Functionals, the un-weighted morphology, have a long history for CMB and LSS analysis, see e.g. \citep{1998MNRAS.297..355S,matsubara03,hikage2006primordial,matsubara2010analytic, gay2012non,codis2013non,planck2013results, Ade:2013xla,2017JCAP...06..023G,matsubara2021weakly,matsubara2022minkowski,appleby2018minkowski,pranav2019topology,2024MNRAS.527..756C,2024JCAP...01..039C}. Recently,  a novel approach known as the weighted morphological measures, {\it {\underline{c}}onditional {\underline{m}}oments of {\underline{d}}erivative} (cmd),  has been developed and implemented for redshift space distortions (RSD) \citep{2024ApJ...963...31K}.  This method allocates the specific weight through the amount of the field associated with each excursion set point. Subsequently, from this point of view, a new window has been unveiled that provides insight into the cosmological field, particularly when we are looking for exotic features such as CSs with subdominant contributions.

It has been confirmed by various observations \citep{2020A&A...641A..10P} that the dominant part of initial density perturbations is sourced by quantum fluctuations of the inflaton field generated during the so-called inflationary era and shortly after the Big Bang \citep{Guth:1980zm,Liddle:1993fq,Steinhardt:1995uf,Liddle:1999mq}. The inflationary model suggests that quantum fluctuations of a scalar field became "frozen in" during the rapid expansion and became stretched beyond the observable horizon, eventually seeding the density variations observed today.  Nevertheless, due to the expansion and cooling of the early universe, we expect some phase transitions between different states to take place and consequently, depending on the topology of the underlying field's potential, a series of stable topological defects such as domain walls, monopoles and CSs can be generated \citep{Kibble:1976sj,Kibble:1980mv22,Hindmarsh:1994re,Vilenkin:2000jqa,Copeland:2009ga,Polchinski:2004hb}. The line-like defects (CSs) is a prediction of some particular model in context of inflation such as  hybrid inflation, brane-word and super-string theories  \citep{Kibble:1976sj,Zeldovich:1980gh,Vilenkin:1981iu,    Vachaspati:1984dz,Vilenkin:1984ib,Shellard:1987bv,Hindmarsh:1994re,Vilenkin:2000jqa,Sakellariadou:2006qs,Bevis:2007gh,Depies:2009im,Bevis:2010gj,Copeland:1994vg,Sakellariadou:1997zt,Sarangi:2002yt,Copeland:2003bj,Pogosian:2003mz,Majumdar:2002hy,Dvali:2003zj,Kibble:2004hq,HenryTye:2006uv}.

The remnants of cosmic phase transitions, when the universe expanded and cooled down, contribute to the density fluctuations \citep{Kibble:1976sj,Kibble:1980mv22}.
The CS network is composed of infinite strings, loops, and junctions that can generate GWs as it evolves over cosmic time. Generally, observational signatures of CSs depend primarily on two factors: (1) the probability of inter-commutation events, where strings interact and exchange segments,  and (2) the dimensionless string tension, represented as: $G\mu/c^2=\mathcal{O}\left(\varpi^2/M_{\rm Planck}^2\right)$, in which,  $M_{\rm Planck}\equiv\sqrt{\hbar c/G}$, $\varpi$ and $c$ are respectively, the Planck's mass, the energy of symmetry breaking scale and light speed. Also, $\mu$ is the mass per unit length of the CS. In this paper, we choose to work in natural units with $\hbar=c=1$.
Constraining the CSs parameters, particularly determining the $G\mu$, is crucial, as it provides essential limits on the fundamental parameters governing CSs formation theories. Exploring the imprint of the CSs network takes different theoretical, statistical, and observational routes, thanks to their diverse imprints on cosmological data sets \citep[for a review]{2014NuPhS.246...45B}. A comprehensive review of the existing literature indicates that CMB, LSS and galaxy formation, 21cm intensity map, Pulsar Timing Arrays (PTAs), gravitational waves (GWs) (stochastic and resolved), gravitational lensing, Gamma-ray burst and cosmic rays, and cross-correlation between different tracers are almost widely-used approaches to put upper and lower bounds on the various types of CSs (see e.g. \cite[and references therein]{Ade:2013xla,2013JCAP...02..045M,Blanco-Pillado:2017rnf,2021JCAP...10..046T,LIGOScientific:2021nrg,2021PhRvL.126d1304E,2023ApJ...951L..11A,2024PhRvD.109l3524J,2024PhRvD.110b3534R}).

The stochastic gravitational wave background (SGWB) is generated by the superposition of gravitational waves radiated by CS networks through various mechanisms. Gravitational wave bursts are generated by cusps that form on closed loops in the stable-c string model \citep{1985PhRvD..31.3052V}. Stable-k model generates the GW bursts from the kinks on the closed loops of CS networks \citep{2001PhRvD..64f4008D}. Furthermore, stable-m, which emits monochromatic GWs from closed loops, and stable-n \citep{2011PhRvD..83h3514B,2015PhRvD..92f3528B} can also contribute to the SGWB. The details of stable-string models and their comparison with NANOGrav $12.5$-year results are found in \cite[and references therein]{2021PhRvD.103j3512B}. PTAs place bounds on the four mentioned stable-string models in the absence of supermassive black hole binaries and the maximum values of corresponding posterior distributions for $G\mu$ lie within the range of  $G\mu\sim 10^{-10.5}...10^{-10.0}$ \citep{2023ApJ...951L..11A}. Notably, data from advanced GW observatories, such as the third LIGO-VIRGO run, yield an upper limit of $G\mu \leq 1.5\times10^{-7}$, reflecting improvements over prior constraints \citep{LIGOScientific:2021nrg}. Emitting GWs by Nambu-Goto CSs loop caused to $10^{-14} \le G\mu \le 1.5 \times 10^{-10}$ ~\citep{Ringeval:2017eww,Blanco-Pillado:2017oxo,Blanco-Pillado:2017rnf}. The new interval for CSs tension,  $10^{-15}<G\mu<10^{-8}$,  has been reported by exploring PTAs \citep{Jenet:2006sv,Pshirkov:2009vb,Tuntsov:2010fu, Damour:2004kw,Battye:2010xz,Oknyanskij:2005pd, Kuroyanagi:2012jf}. The NANOGrav 15-year data revealed a stringent upper limit on the $G\mu$. Nevertheless, by considering the fraction of CS loops, following the Nambu-Goto dynamics ($f_{\rm NG}$) and exclusively emitting gravitational waves GWs introduces a degeneracy in the $(G\mu, f_{\rm NG})$ parameter plane. This degeneracy results in an extension of the posterior distribution indicating $G\mu\gtrsim 10^{-7}$. This demonstrates the necessity of further information about the fraction of long-lived loops in a CS \citep{2024JCAP...12..001K}. Also, the Multi-messenger constraints on the Abelian-Higgs CSs demonstrated
$G\mu f_{\rm NG}^{2.6}\gtrsim 3.2 \times 10^{-13}$ at
95\% confidence \citep{2023JCAP...04..045H}.

Based on gravitational lensing of CSs, there was no evidence for CSs with tension $G\mu<3.0\times 10^{-7}$ out to redshifts $z>0.6$ from the COSMOS survey imaged 1.64 square degrees \citep{Christiansen:2010zi}.
Studying the consequences of CSs on the early structure and galaxy formation has a substantial historical background \citep{1984PhRvL..53.1700S} and as illustrations through the ionization history of the universe, JWST observation and performing N-body simulation have been examined in \cite[and references therein]{1991PhRvL..67.1057V,2006PhRvD..73d3515M,Shlaer:2012rj,2023PhRvD.108d3510J,2024PhRvD.109l3524J}.
The imprint of CS wakes on the 21cm intensity map has also been extensively examined by  \cite{Brandenberger:2010hn, Hernandez:2011ym,Hernandez:2012qs,Pagano:2012cx,2013JCAP...02..045M,Hernandez:2014jda,2019JCAP...09..009B,2013JCAP...02..045M,2021JCAP...10..046T,2021MNRAS.508..408H}. Also cross-correlation of the 21cm redshift with CMB polarization to search the contribution of CSs has been performed in \citep{2022JCAP...11..012B}. Searching the CSs from the black hole has been done in \citep{2022MNRAS.517.2221C,2024arXiv240704743A}.

The CMB field is a premier laboratory to explore the early and late eras of the universe, and it is affected by diverse phenomena that encompass both high-energy, primordial events to processes at low-energy scales \citep{bond1987statistics,hu2002cosmic,dodelson2020modern,durrer2020cosmic,aghanim2020planckvi}. The randomness in CMB, sourced by quantum fluctuations and inflationary dynamics, offers insights into the early universe, energy density fluctuations, and subsequent structure formation. Both the inflationary model and the topological defects scenario predict similar features in the CMB power spectrum on large angular scales, where the initial conditions play a crucial role. However, significant differences emerge on intermediate and small scales, as the super-horizon scale behavior of perturbations differs substantially between the two theories. The inflationary models often predict a smooth spectrum with characteristic acoustic peaks from sound waves in the primordial plasma; topological defect models yield distinct patterns, influenced by the non-Gaussian signatures and discontinuities produced by the CSs and other defect types. These differing predictions offer unique observational pathways, potentially detectable with high-resolution CMB data to distinguish between the inflationary and defect-based origins of cosmic structure \citep[and references therein]{1984Natur.310..391K,Bouchet:1988hh,Bennett:1990,Ringeval:2005kr,Fraisse:2007nu,Ade:2013xla,vafaei2017multiscale,sadr2018cosmic,2021MNRAS.503..815V}.

A straightforward approach to assess the contribution of CSs is concentrating on the various orders of the CMB spectrum (power spectrum, bispectrum, trispectrum, and so on), by taking the fraction of the power at $\ell=10$ due to CSs contribution \citep{Pen:1997ae,Bevis:2006m,Hindmarsh:2009qk,Hindmarsh:2009es,Bevis:2010gj,Lazanu:2014eya,Regan:2015cfa,Lizarraga:2016onn}. A part of main results for CSs tension according to the extensive evaluation done by \cite{Ade:2013xla} are as follows: a conservative bound  $G\mu < 9.0 \times 10^{-7}$ at (95\% confidence) was derived for \texttt{SMICA} component separation algorithm based on the bispectrum; the wavelet decomposition revealed $G\mu< 4 \times 10^{-7}$;  according to the power spectrum and for the Abelian-Higgs CSs with $f_{10}<0.024$, the upper bound is $G\mu<3.0\times10^{-7}$, while for the $f_{10}<0.010$ a little tight constraint, $G\mu<1.3\times10^{-7}$, has been obtained. Joint analysis of Planck and WMAP polarization power spectra approved the $G\mu<1.49\times 10^{-7}$ at 95\% confidence interval for Numbo-Goto CSs \citep{Lazanu:2014eya}. Planck 2015 temperature and polarization data gave rise to $G\mu < 1.1\times 10^{-7}$ ~\citep{Charnock:2016nzm}. The non-Gaussian effect of the CS-induced CMB map has been evaluated in ~\citep{Hobson:1998av,Ringeval:2010ca, Ducout:2012it}. With modal bispectrum estimation of Planck CMB map, the upper bound has been achieved as $G\mu < 9.0 \times 10^{-7}$ at $95\%$ level of confidence \citep{Ade:2013xla}.  More recently, the distinguishability between the current-carrying CSs from their uncharged (Nambu-Goto) counterparts through the CMB has been done in \citep{2024PhRvD.110b3534R}. Current-carrying CSs in the PTA band and LIGO O3 run have been done by \cite{2024PhLB..85038516A}.

The distinctive geometry and topology inherent in networks of CSs result in notable discontinuities on the CMB map at the field level. These attributes also contribute to the integrated Sachs-Wolfe (ISW) effect, which is known as the Gott-Kaiser-Stebbins effect~\citep{1984Natur.310..391K,Gott:1985, Stebbins:1988, Bouchet:1988hh, Allen:1997ag, Pen:1997ae}. Taking into account the phenomena outlined earlier, the statistical and morphological features of excursion and critical sets in the CMB map have been introduced as alternative strategies for analyzing the footprint of the CSs network. The identification of line-like patterns associated with straight CSs and super-strings in the temperature fluctuations of the CMB can be achieved through the use of edge-detection algorithms. These algorithms revealed a detection threshold of \(G\mu \gtrsim 5.5\times 10^{-8}\) in the observations made by the South Pole Telescope (SPT)~\citep{Amsel:2007ki,Stewart:2008zq,2010IJMPD..19..183D,2010JCAP...02..033D,2010IJMPD..19..183D}. A sensitivity of $G\mu\gtrsim 1.4 \times 10^{-7}$ for the SPT-3G (third generation) has been found by using wavelet and curvelet methods~\cite[]{Hergt:2016xup}. For the Nambu-Goto string simulations, the contribution of CSs is sensitive to $G\mu\gtrsim5\times 10^{-7}$ via wavelet-Bayesian inference~\citep{2017MNRAS.472.4081M}. Also \cite{Hammond:2008fg} used wavelet-domain Bayesian
de-noising and they finally identified the CSs to $G\mu \gtrsim 6.3\times 10^{-10}$ for noise-free and ideal case for forthcoming arcminute-resolution experiments. Including secondary anisotropies increased the lower limit as $G\mu \gtrsim 1.0\times 10^{-7}$ and $G\mu \gtrsim 2.5\times 10^{-7}$ for accounting the thermal Sunyaev-Zel'dovich and Rayleigh-Jeans, respectively. Concrete results for applying wavelets in the real space for constraining the CSs tension can be found in~\citep{Ade:2013xla}.

A neural network-based approach has been applied by
\citep{Ciuca:2017wrk} on the noiseless arcminute-resolution random-kick maps to reach a detection level
of $G\mu \gtrsim 2.3 \times 10^{-9}$.  According to a convolutional neural network, the lower detectable tension is $G\mu \gtrsim 5 \times 10^{-9}$ \citep{Ciuca:2017gca}. The crossing statistics of random kicks of CSs network model on the flat sky CMB synthetic data demonstrated $G\mu\gtrsim 4.0\times 10^{-9}$ \citep{movahed2011level}. Taking into account the clustering of the local maxima of the CMB map induced by CSs in the real space, indicated $G\mu\gtrsim 1.2\times 10^{-8}$ for the noise-free maps with one arc-minute resolution \citep{Movahed:2012zt}.

\begin{figure*}
    \centering
    \includegraphics[width=0.95\textwidth,height=4cm]{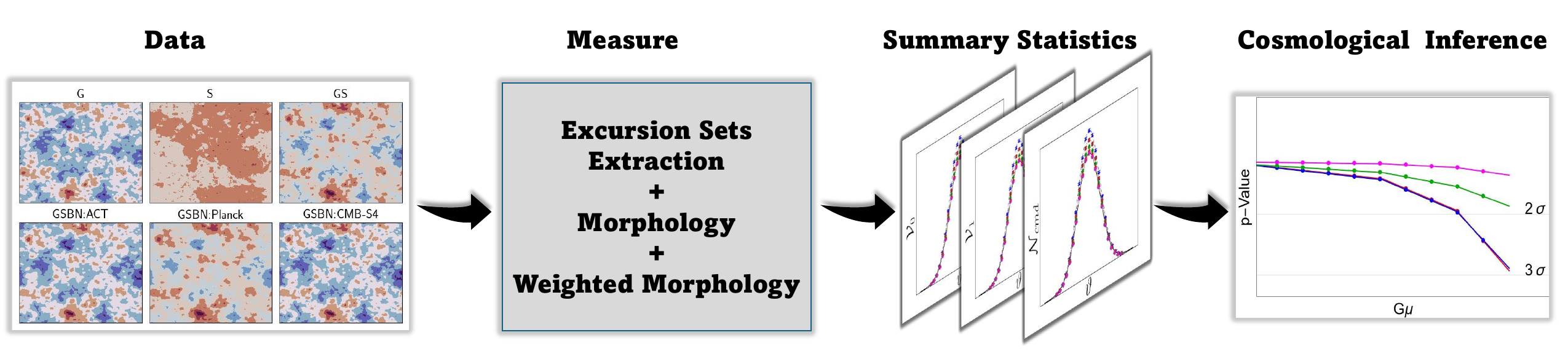}
    \caption{The workflow of the paper from left to right. We first consider the simulated CMB maps with the effect of CSs and other foreground effects like beam and noise. We then analyze the weighted morphological measures known as the cmd statistics to quantify the effect of CSs in the CMB. Finally, we present cosmological inferences like p-values to constrain the lower limit of CS tension in contrast with the unweighted morphology.}
    \label{fig:fig_Work_Flow}%
\end{figure*}

Implementing the multiscale edge-detection algorithm, accompanied by statistical measures  on the CMB map modified by a series of Nambu-Goto string networks using the Bennett-Bouchet-Ringeval
code~\citep{Bennett:1990, Ringeval:2005kr} confirmed that $G\mu\gtrsim 4.3\times 10^{-10}$ for the ideal case, while for peak-peak correlation function on the Planck-like simulation, the minimum detectable value of CSs tension is $G\mu\gtrsim8.9\times 10^{-7}$ \citep{vafaei2017multiscale}. The LightGBM and CNN machine learning algorithms with proper feature vector revealed $G\mu\gtrsim 1.9 \times 10^{-7}$ at  $3\sigma$ for CMB-S4 simulation. While for the observed
Planck \texttt{SMICA} data when LightGBM and CNN are used the $G\mu\lesssim 8.6 \times 10^{-7}$\; ($3\sigma$) has been achieved \citep{2022MNRAS.509.2169T}. On the synthetic string-induced CMB-S4 map, the $G\mu\gtrsim 1.2 \times 10^{-7}$ and $G\mu\gtrsim 3.0 \times 10^{-8}$ at ($3\sigma$) have been reported for gradient boosting (GB) and random forest (RF) machine learning algorithms, respectively. The clustering of local maxima of Planck data set provided the upper bound on different component separation algorithms, namely for  \texttt{NILC}  $G\mu\lesssim 8.38 \times 10^{-7}$, $G\mu\lesssim 6.71 \times 10^{-7}$, $G\mu\lesssim5.59 \times 10^{-7}$ and $G\mu\lesssim7.17 \times 10^{-7}$ at $2\sigma$ confidence interval for \texttt{NILC}, \texttt{SEVEM}, \texttt{SMICA} and \texttt{CR}, respectively \citep{2021MNRAS.503..815V}. The MFs in the context of unweighted morphology are one of the real space tests for CSs, which has been carried out by \cite{Ade:2013xla}. Marginalizing over relevant parameters of MFs gave  $G\mu < 7.8\times 10^{-7}$ at 95\% confidence level for \texttt{SMICA} map.

In principle, to ensure the reliability of measures for assessing CSs, it is important to test their performance on the synthetic data sets before applying them to the realistic observational data. Consequently, the primary objective of this research is to analyze the capability of the weighted morphological measure termed by ``cmd-$n$'' which is a specific extension of Minkowski Valuations~\citep{2024ApJ...963...31K}, in distinguishing the network of CSs from the temperature anisotropies of the CMB represented by the total intensity of radiation \footnote{Through the mechanism of cosmic birefringence, CSs networks can modify the polarization of CMB, thus providing another potential indicator of such networks~\citep{2007PhRvD..76d3005B, 2022JCAP...06..033Y}.}. More specifically, the lower detectable bound on the $G\mu$ will be inferred, as the conclusion of the investigation.  A comprehensive and well-defined approach to this matter should incorporate at least the following parts:\\
1) Generating synthetic CMB maps with and without CSs imprint with more realistic beam effect, systematic noise, and foreground;\\
2) Adopting robust algorithms for data reduction and data analysis;\\
3) Constructing the specific summary statistics\footnote{Among the approaches to put pristine and stringent constraints on the desired cosmological parameter, and going beyond statistical summary estimators, a Bayesian map-based or field-level inference is considered. This method uses a forward modeling \citep{fieldlevelLSS,2023ApJ...952..145J,2024PhRvL.133v1006N}. As an illustration, to evaluate the cosmic string tension, the role of forward modeling is to make a connection between early universe fluctuations, the presence of phase transition and intervening foreground phenomena at the field level, such that all pixels are also considered as free parameters. This idea remains open for further investigation.};\\
4) Determining the summary statistics theoretically and numerically;\\
5) Computing the significance of detecting the CSs intensity, or in other words, the minimum value of $G\mu$ that is recognized from other portions for mock data, thereby demonstrating the capability of detection;\\
6) And finally, implementation of the proposed pipeline on the observed data to put the upper bound on $G\mu$\footnote{When dealing with observational data sets, we are interested in achieving the upper bound on the contribution of the desired mechanism. In contrast, the lower bound is reasonable when synthetic maps are utilized to elucidate the detection performance of introduced summary statistics, like those done in this research.}.

In light of the mentioned parts, we classify the main novelties and advantages of our research as follows:\\
I) The primary version of the {\it {\underline{c}}onditional {\underline{m}}oments of {\underline{d}}erivative} (cmd) was proposed to encapsulate the anisotropic and non-Gaussian aspects of redshift space distortions (RSD) through the relaxation of certain properties of the {\it Hadwiger's theorem} \citep{2024ApJ...963...31K}. Here, we extend this definition by adding a specific condition that is proper to magnify the sharp edges in a generic cosmological field, such as a CMB map. We also use the perturbative formalism to derive a theoretical prediction for the generalized cmd-$n$ up to $\mathcal{O}(\sigma_0^3)$. \\
II) Given the fact that the effect of CSs on the CMB map is dominated at small scales, we carefully, generate the synthetic Gaussian CMB maps in the flat sky approximation and by utilizing the reliable CSs simulations, combine the various parts of CMB components including temperature fluctuation sourced by inflationary initial conditions as the Gaussian part and CSs part, taking into account the beam effect and noise level of three surveys such as CMB-S4, ACT, and Planck.\\
III) A comprehensive pipeline is proposed to ensure the input map is adequately prepared for the implementation of our exclusive summary statistics, the weighted morphology measure, and the cmd-$n$ analysis (Fig.~\ref{fig:fig_Work_Flow}).\\
IV) Also, the significance of detection with the cmd-$n$ criterion by determining the p-value would be performed to obtain the detectability of CSs network by lens of a weighted morphology from the mock CMB observation. A conservative statistical error in the measurement of $G\mu$ from the CMB map, corresponding to a defined relative uncertainty of the cmd-$n$, is also exhibited.\\
V) In addition to presenting a comprehensive survey in the literature and for the sake of clarity, we will also compare the capability of the summary statistics, including the MFs and crossing statistics, with the weighted morphology measures (cmd-$n$).\\
We argue that CSs would be detected at high significance through the cmd-$n$ measures if $G\mu\gtrsim 1.9\times 10^{-7}$ for a noise-free map. While for ACT, CMB-S4, and Planck experiments, the minimum detectable value of cosmic string tension is $G\mu\gtrsim2.9 \times 10^{-7}$, $G\mu\gtrsim 2.4\times 10^{-7}$ and $G\mu\gtrsim 5.8\times 10^{-7}$, respectively up to 2$\sigma$ confidence interval. We also advocate using the cmd-$n$ instead of other complicated emulators because the contribution of interesting cosmological parameters is captured by spectral moments, making it a new and valuable element of the summary statistics utilized in the SBI approach.

The rest of the paper is organized as follows: in Sec.~\ref{sec:Weighted_Morphology}, we will revisit the weighted morphology and introduce the cmd-$n$ measures for the CMB field. The theoretical form of the cmd-$n$ for weakly non-Gaussian fields through the perturbative framework to the order of $\mathcal{O}(\sigma_0^3)$ will be given in the mentioned section. Sec.~\ref{sec:CMB_data_Analysis} is devoted to data description and details of generating mock string-induced CMB maps, including instrumental beam and noise. The summary statistics and cosmological inference related to the capability of the cmd-$n$ are given in Sec. \ref{Sec:Cosmological_Inference}. We also clarify a comparison between the cmd-$n$ and the MFs along with other approaches documented in the literature that pursue analogous aims in Sec. \ref{Sec:Cosmological_Inference}. We will give our summary and conclusions in Sec.~\ref{Sec_Conclusions}.

\section{Theoretical Foundation of the cmd-{\it n} Weighted Morphology for CMB map}\label{sec:Weighted_Morphology}
\begin{figure}
    \centering
    \includegraphics[width=1\columnwidth]{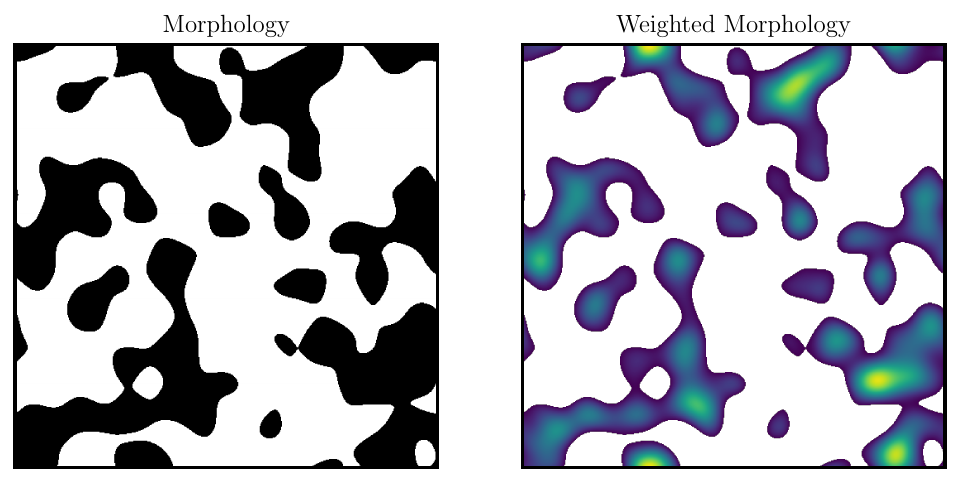}
    \caption{The binary map on the left panel illustrates a typical representation of the excursion sets, which serves as the basis for computing morphological properties. For weighted morphology, however, the analysis extends beyond the geometric and topological structure of the excursion sets. It incorporates the explicit values of the field and/or its derivatives evaluated on the excursion sets, which are treated as weighting factors to enhance the morphological characterization (right panel).  For morphological measures, we took the binary color, whereas for the weighted morphology, we adopted the color spectrum whose brightness represents the value of the field (and/or its derivatives) inside and on the boundary of the excursion sets. }
    \label{fig:wm}%
\end{figure}

The CMB field is expressed by $\mathbf{T}\in{\rm L}^2(\mathbb{R}^2)$ and $\mathbf{T}$ generally exhibits the $2\times2$ tensor. It is convenient to write CMB field using the Stokes parameters by $\mathbf{T}:\{\delta_T,Q+iU,Q-iU\}$ on the $\hat{e}_{\theta}\otimes \hat{e}_{\phi}$ subspace. The $\delta_T\equiv(T-\langle T\rangle)/\langle T\rangle$ represents the temperature fluctuations summed over polarization states (total intensity), $Q$ and $U$ denote to plus and cross polarization, respectively. As a result, $\mathbf{T}^{(3+2)}$ is known as $(3+2)$-dimensional stochastic field. The examining (weighted) morphology of temperature fluctuation is indeed the objective of this research, therefore, hereafter we ignore the polarization parts such that $\mathbf{T}^{(1+2)}\equiv\delta_T(\theta,\phi)$. We also smooth  $\mathbf{T}^{(1+2)}$ by convolving it with a typical smoothing window function as:
\begin{equation}\label{eq:smoothed}
\delta_{T}^{\rm smoothed}(\theta,\phi)=\int d\Omega'\; \mathcal{W}(\Omega,\Omega';\Delta\Omega)\;\delta_T(\theta',\phi')
\end{equation}
where $\Delta \Omega\equiv \cos^{-1}|\Omega.\Omega'|$ is smoothing scale in polar coordinate. To construct the mock map, we use the proper smoothing kernel ($\mathcal{W}$) associated with each survey. For the convenience, hereafter, we omit the superscript ``smoothed''. The excursion set associated with $\delta_T(\theta,\phi)\ge \vartheta \sigma_0$ reads as:
\begin{equation}\label{eq:excursion1}
    \mathcal{E}_T(\vartheta)=\left\{(\theta,\phi) \in {\rm L}^2(\mathbb{R}^2)\;|\; \delta_T(\theta,\phi)\ge \vartheta \sigma_0 \right\}
\end{equation}
The boundary of excursion sets, $\partial\mathcal{E}_T$, reveals the iso-height temperature fluctuations contours. The critical sets and crossing statistics are well-known geometrical sets derived by imposing additional constraints on the general definition of excursion sets expressed by Eq. (\ref{eq:excursion1})~ \citep{rice44a,rice44b,Bardeen:1985tr,bond1987statistics,ryden1988area,ryd89,matsubara03}.
According to the {\it Hadwigers's theorem}, 
a convex ring embedded in $D$-dimension can be described by $D+1$ functionals with well-knows geometrical and topological interpretations \citep{schmalzing1995minkowski,mecke1993robust,schmalzing1997beyond,2001PhyA..293..592B,Beisbart2002}. Tensions and anomalies recently reported in the field of cosmology (see e.g \citep{2022NewAR..9501659P,2023CQGra..40i4001A}) have prompted significant extension for scalar MFs to encompass vector and tensor forms \citep{Beisbart2002,2024ApJ...963...31K}. For the sake of completeness, here we briefly describe the mathematical backbone of weighted morphology specifically adapted for the CMB field. We begin with a broad definition of morphology as:
\begin{eqnarray}\label{eq:MVs1}
    \Xi &\equiv&\frac{1}{A}\int_{A}dA\;\mathcal{G}(s_{\nu};{\boldsymbol{r}},\delta_T({\boldsymbol{r}}),\boldsymbol{\nabla}\delta_T({\boldsymbol{r}}),\cdots)
\end{eqnarray}
where $A$ is the area of 2-dimensional map, $s_{\nu}$ depends on the local curvature at each ${\boldsymbol{r}}$ on the CMB map and $\nu=0,1,2$. The $\mathcal{G}$ is a general functional form of $(s_{\nu};{\boldsymbol{r}},\delta_T({\boldsymbol{r}}),\boldsymbol{\nabla}\delta_T({\boldsymbol{r}}),\cdots)$ \citep{2024ApJ...963...31K}. A specific form of $\Xi$ which is known as an extension of scalar MFs is given by the Minkowski Valuations (MVs) for $\nu>0$ as \citep{McMullen1997,Alesker1999,Hug2007TheSO,2017JCAP...06..023G}:
\begin{widetext}
    \begin{eqnarray}\label{eq:MVscom1}
        \mathcal{W}_{\nu}^{(p,q)} \equiv\frac{1}{A}\int_{\partial {\mathcal{E}}_T(\vartheta)}\;dl\; s_{\nu} \overbrace{{\boldsymbol{r}}\otimes{\boldsymbol{r}}\otimes...\otimes {\boldsymbol{r}}}^{p-times}
        \otimes\; \underbrace{\frac{\boldsymbol{\nabla}\delta_T({\boldsymbol{r}})}{|\boldsymbol{\nabla}\delta_T({\boldsymbol{r}})|}\otimes\frac{\boldsymbol{\nabla}\delta_T({\boldsymbol{r}})}{|\boldsymbol{\nabla}\delta_T({\boldsymbol{r}})|}\otimes ...\otimes \frac{\boldsymbol{\nabla}\delta_T({\boldsymbol{r}})}{|\boldsymbol{\nabla}\delta_T({\boldsymbol{r}})|}}_{q-times}
    \end{eqnarray}
\end{widetext}
and for $\nu=0$, Eq. (\ref{eq:MVscom1})  becomes\footnote{It is noteworthy that due to the presence of the Dirac delta function, $\delta_D\left(\delta_T-\vartheta\sigma_0\right)$, or the step function, $\Theta\left(\delta_T-\vartheta\sigma_0\right)$, in the corresponding definition of $\mathcal{G}$ for the MVs, the integral over the entire surface has transformed into an integral over the excursion sets ($\mathcal{E}_T$) or its boundary ($\partial\mathcal{E}_T$).}:
\begin{eqnarray}\label{eq:MVscom2}
    \begin{split}
        \mathcal{W}_{0}^{(p,0)} \equiv&\frac{1}{A}\int_{ {\mathcal{E}}_T(\vartheta)}\;dA\; \overbrace{{\boldsymbol{r}}\otimes{\boldsymbol{r}}\otimes...\otimes {\boldsymbol{r}}}^{p-times}
    \end{split}
\end{eqnarray}
Here $\otimes$ shows the tensor product. Also ${\boldsymbol{r}}$ is the position vector on the ${\partial {\mathcal{E}}_T(\vartheta)}$ and $dl$ is the line element along ${\partial {\mathcal{E}}_T(\vartheta)}$. The MFs which are generic scalar functionals to quantify CMB morphology, are defined as  \citep{1998MNRAS.297..355S,hikage2006primordial,matsubara2010analytic}:
\begin{align}\label{eq:Minkowski1}
    V_0(\vartheta) &= \int_{ {\mathcal{E}}_T(\vartheta)} dA \\\notag
    V_1(\vartheta) &= \dfrac{1}{4} \int_{{ {\partial\mathcal{E}}_T(\vartheta)}} dl  \\\notag
    V_2(\vartheta) &= \dfrac{1}{2\pi}\int_{{ {\partial\mathcal{E}}_T(\vartheta)}} \kappa dl
\end{align}
where $\kappa$ is geodesic curvature on the ${ {\partial\mathcal{E}}_T(\vartheta)}$. Looking at Eqs. (\ref{eq:MVscom1}), (\ref{eq:MVscom2}) and (\ref{eq:Minkowski1}) and based on the nature of morphology, the formation of excursion sets distinctly reveals the morphological properties of the CMB field, making it unnecessary to determine the exact values of the field at all points within the area and along the excursion set boundaries \citep{1998MNRAS.297..355S,hikage2006primordial,matsubara2010analytic}. As discussed in the introduction, inspired by the level crossing statistic introduced in \citep{rice44a,rice44b,Bardeen:1985tr,bond1987statistics,ryd89,ryden1988area,matsubara03}, the original version of the cmd was introduced to quantify the anisotropy and non-Gaussianity nature of RSD  \citep{2024ApJ...963...31K}. These measures allocate the specific weight through either the amount of the field or its derivative associated with each excursion set point. In contrast to the morphological measures, the cmd measure represents weighted statistics. Generally, the cmd incorporates the amount of the field's first derivative at excursion sets.  Fig. \ref{fig:wm} illustrates the schematic difference between common morphological measures and the cmd as a weighted version of morphology. In the following subsection, we will clarify the specific functional form of the extended cmd measures.
\subsection{The cmd-{\it n} measures for  CMB}
\begin{figure}
    \centering
    \includegraphics[width=1\columnwidth]{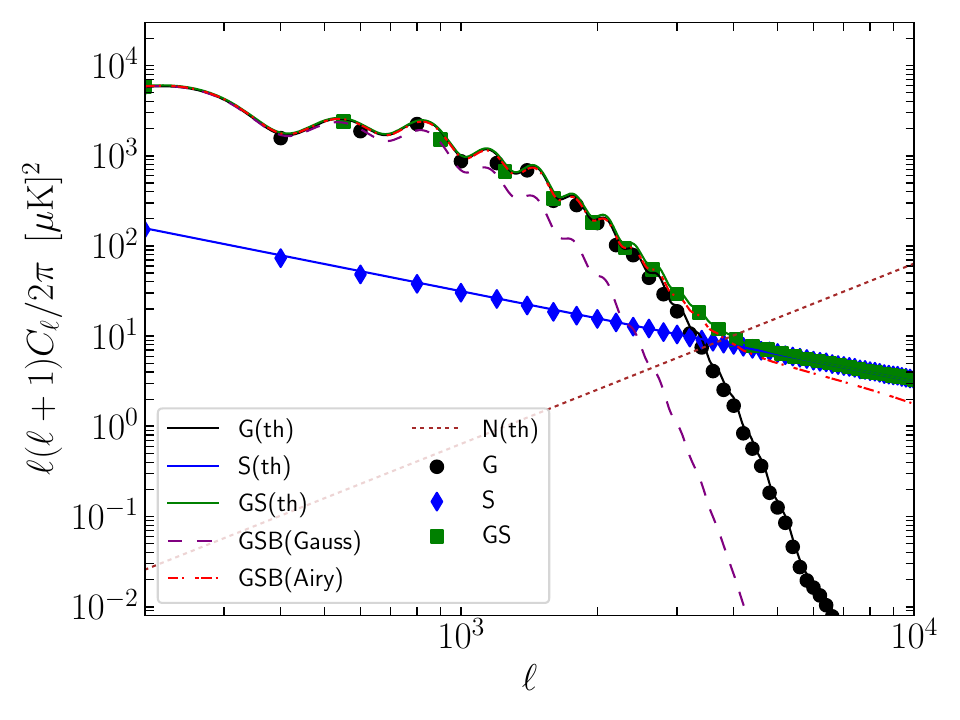}
    \caption{Angular power spectrum of temperature fluctuations along with the effect of cosmic strings. The black solid line corresponds to the Gaussian fluctuations sourced by the inflation scenario with the Planck fiducial $\Lambda$CDM model (from CAMB). The filled black circle symbols illustrate numerical calculation of $\ell(\ell+1)C^{TT}_{\ell}/2\pi$ from a synthetic G map with resolution $R=0.42'$. The blue solid line shows the contribution from cosmic strings having $G\mu=8\times10^{-8}$ predicted by theoretical analysis with power-law behavior. The green solid line and green filled rectangle symbols are associated with the theoretical and numerical calculation of the power spectrum for the GS map, respectively.  The purple long-dashed and red dot-dashed lines indicate the GS map affected by the Gaussian and the Airy beams (GSB), respectively.  The dotted line depicts the noise power spectrum.  }
    \label{fig_PSCMBCS}%
\end{figure}

In order to establish the theoretical framework for the cmd weighted morphology of the CMB map, we adopt a probabilistic framework. We define the feature vector for CMB field by incorporating both the field itself and its first derivatives as: $\mathcal{A}\equiv\{\delta_T(\theta,\phi),\eta_{\theta},\eta_{\phi}\}$, where $\eta_{\theta}\equiv\partial \delta_T/\partial \theta$ and $\eta_{\phi}\equiv\partial \delta_T/\sin(\theta)\partial \phi$. For the CSs-induced map, the small scales possess a drastic potential to investigate the contribution of CSs network compared to large scales (small $\ell$) (see Fig. \ref{fig_PSCMBCS}).  Thereby, we convert the polar coordinate on the surface of the sphere to the Cartesian coordinates, therefore, the feature vector becomes $\mathcal{A} = \{\delta_T(x,y),\eta_{x},\eta_{y}\}$. For a square map with $N_{\rm pix}$ and resolution equates to $R$, we have, $x=R\times n_x$ and $y=R\times n_y$ with $(n_x, n_y)=1,...,N_{\rm pix}$.

The joint probability density function (JPDF) of the set $\mathcal{A}$ expanded around the Gaussian multivariate distribution is given by \citep{matsubara03}:
\begin{align}
    \mathcal{P}(\mathcal{A})= \exp\left( \sum_{q=3}^{\infty}\dfrac{(-1)^q}{q!} \sum_{\mu_1\cdots \mu_q=1}^{3} \mathcal{K}^{(q)}_{\mu_1\cdots\mu_q}\right.\left.\dfrac{\partial^q}{\partial\mathcal{A}_{\mu_1}\cdots\partial\mathcal{A}_{\mu_q}} \right)\mathcal{P}^{\mathbb{G}}(\mathcal{A})
\end{align}
here, $\mathcal{K}^{(q)}_{\mu_1\cdots\mu_q}\equiv\langle\mathcal{A}_{\mu_1\cdots\mu_q}\rangle_c$ is the cumulant with $\langle\rangle_c$ denoting the connected moment and the multivariate Gaussian JPDF, $\mathcal{P}^{\mathbb{G}}(\mathcal{A})$ is defined in terms of covariance matrix, $\mathcal{K}^{(2)}\equiv \langle\mathcal{A}\otimes\mathcal{A}\rangle_c$ as:
\begin{align}
    \mathcal{P}^{\mathbb{G}}(\mathcal{A})=\dfrac{1}{(2\pi)^{3/2}\sqrt{\text{det}\,\mathcal{K}^{(2)}}}\exp\left(-\frac{\mathcal{A}^T\,.\,[\mathcal{K}^{(2)}]^{-1}\,.\,\mathcal{A}}{2}\right)
\end{align}
The matrix form of $\mathcal{K}^{(2)}$ in $(1+2)-$dimensions is:
\begin{align}
    \label{cummat}
    \mathcal{K}^{(2)}=\begin{bmatrix}
        \langle\delta_T\delta_T\rangle_{c}       & \langle\delta_T\eta_{x}\rangle_{c} & \langle\delta_T\eta_{y}\rangle_{c} \\
        \langle\eta_{x}\delta_T\rangle_{c}      & \langle\eta_{x}\eta_{x}\rangle_{c} & \langle\eta_{x}\eta_{y}\rangle_{c}\\
        \langle\eta_{y}\delta_T\rangle_{c}       & \langle\eta_{y}\eta_{x}\rangle_{c} & \langle\eta_{y}\eta_{y}\rangle_{c}
    \end{bmatrix}
\end{align}
With regard to the statistical isotropy, the distinct components of the covariance matrix (Eq. (\ref{cummat})) become:
\begin{align}
    \langle\delta_T\delta_T\rangle_c
    &=\sigma_0^2 \\\notag
    \langle\delta_T\eta_{i}\rangle_c
    &=0 \\\notag
    \langle\eta_{i}\eta_{j}\rangle_c
    &=\delta_{ij}\,\sigma_{1i}^2
\end{align}
Here, $i,j\in\{x,y\}$ and $\delta_{ij}$ is the Kronecker delta function.
The first order of spectral moment is $\sigma_{1i}^2\equiv\langle\eta_{i}^2\rangle_c$. This allows us to write $\sigma_1^2=\langle\eta_{x}^2+\eta_{y}^2\rangle_c=\sigma_{1x}^2+\sigma_{1y}^2$. For the isotropic field, we can write
$\sigma_{1x}=\sigma_{1y}=\sigma_1/\sqrt{2}$. The spectral moments is defined as:
\begin{align}\label{eq:spectra1}
    \sigma_m^2=\dfrac{1}{(2\pi)^2}\int d^2\mathbf{k}\, k^{2m}P(k)
\end{align}
where $P(k)$ is the power spectrum of the flat sky map as $\langle \delta_T(\mathbf{k})\delta_T(\mathbf{k}')\rangle=(2\pi)^2\delta_d(\mathbf{k}+\mathbf{k}') P(k)$ and it is related to the CMB full sky power spectrum through $\ell(\ell+1)C^{TT}_{\ell}\sim k^2P(k)$. The statistical ensemble average of a certain quantity function of the smoothed CMB field $\mathcal{F(\mathcal{A})}$ is expressed by an average as:
\begin{eqnarray}
    \label{averagegauss}
    \langle\mathcal{F}(\mathcal{A})\rangle=\left\langle{\hat{\mathcal{L}}} \mathcal{F}(\mathcal{A})\right\rangle^{\mathbb{G}}
\end{eqnarray}
where
\begin{eqnarray}
    \label{averagegauss1}
    {\hat{\mathcal{L}}}\equiv\exp\left(\sum_{q=3}^{\infty}\dfrac{(-1)^q}{q!}\sum_{\mu_1\cdots \mu_q=1}^{3}\mathcal{K}^{(q)}_{\mu_1\cdots\mu_q}
    \frac{\partial^q}{\partial\mathcal{A}_{\mu_1}\cdots\partial\mathcal{A}_{\mu_q}}\right)\nonumber\\
\end{eqnarray}
and  $\big\langle X(\mathcal{A})\big\rangle^{\mathbb{G}} \equiv \int d\mathcal{A}\; X(\mathcal{A})\; \mathcal{P}^{\mathbb{G}}(\mathcal{A})$. Subsequently, for the weak non-Gaussian field, the expectation value of  $\mathcal{F}(A)$ can be expressed in terms of Gaussian integrations based on the perturbative formalism. To calculate the expectation value of the cmd number density in its original form, the functional configuration of $\mathcal{G}$ in Eq. (\ref{eq:MVs1}) is\footnote{About selecting a typical integrand among various options see \cite{2024ApJ...963...31K}}:
\begin{equation}
    \mathcal{G}\equiv\delta_D\left(\delta_T-\vartheta\sigma_0\right)\left(\eta_{i}\right)^n
\end{equation}
where $\vartheta\equiv\delta_T/\sigma_0$ is a dimensionless threshold and $\sigma_0^2=\langle \delta_T^2\rangle$. This definition conveys that the cmd measure is recognized as the $n^{\rm th}$-moment of the first derivative, which is determined at the border of the excursion sets, a condition guaranteed by the delta Dirac function. For $n > 1$, the cmd measure is no longer merely morphological and it also provides weight to the excursion sets. For the Gaussian statistics, this definition only provides a non-zero average for the even moments of the derivative (see Eq.19 in \cite{2024ApJ...963...31K}). To overcome this issue, we impose further conditions to account for only positive values of the derivative in an isotropic field\footnote{Only for an isotropic field, the $2\Theta (\eta_i)\eta_i^n$ is equivalent to $|\eta_i|^n$}:
\begin{align}\label{eq:cmd1}
    \mathcal{G}\equiv2\delta_D\left(\delta_T-\vartheta\sigma_0\right)\Theta\left(\eta_{i}\right)\left(\eta_{i}\right)^n,
\end{align}
where $\Theta(.)$ is step-function. Considering the $\mathcal{F}(\mathcal{A})\equiv\mathcal{G}$ and by utilizing  Eq.~(\ref{averagegauss}), we calculate the expectation value of the cmd-$n$, $\mathcal{N}_{\text{cmd},n}$, for the CMB map in the Gaussian and isotropic regime:
\begin{eqnarray}
    \label{NcmdGauss}
    \langle \mathcal{N}_{\text{cmd},n}\rangle^{\mathbb{G}}&=&\frac{1}{2}\sum_{i=1}^2\langle 2\delta_D\left(\delta_T-\vartheta\sigma_0\right)\Theta\left(\eta_{i}\right)\left(\eta_{i}\right)^n\rangle^{\mathbb{G}}\nonumber\\
    &=&\dfrac{1}{\sqrt{2}\,\pi}\,\dfrac{\sigma_{1}^n}{\sigma_0}\,\Gamma\left(\dfrac{n+1}{2}\right)\,\exp(-\vartheta^2/2),
\end{eqnarray}
where $i=1,2$ indicates the $x$ and $y$ directions. Also, $\Gamma$ is the Gamma function. It is evident that the $\mathcal{N}_{{\rm cmd},n=1}$ corresponds to the crossing statistics, and, with the exception of a coefficient, it is equivalent to the first MFs, denoted as $V_1$. A similar approach to define MFs and crossing statistics for the CMB map has been given in \citep{matsubara2010analytic,vafaei2017multiscale,2021MNRAS.503..815V}. We will employ these to compare cmd's performance in the recognition of CSs. For mildly non-Gaussian distribution at isotropic regime and keeping additional terms which quantify the non-Gaussianity, we achieve\footnote{in the appendix, we derive the same expansion for a 3-dimensional field.}:
\begin{eqnarray}\label{nongaussskewness}
    &&\langle\mathcal{N}_{\text{cmd},n}\rangle^{\mathbb{NG}}=\nonumber\\
    &&\langle\mathcal{N}_{\text{cmd},n}\rangle^{\mathbb{G}}\times
    [1 + \left( \frac{1}{6} S^{(0)} H_3 (\vartheta) +  \frac{n}{3} S^{(1)} H_1 (\vartheta) \right) \sigma_0 \nonumber\\
    &&+( \frac{1}{72} (S^{(0)})^2 H_6 (\vartheta) + \left( \frac{1}{24} K^{(0)} + \frac{n}{18} S^{(0)} S^{(1)} \right)  H_4 (\vartheta)\nonumber\\
    && + \left( \frac{n}{8} K^{(1)} + \frac{n(n-2)}{18} (S^{(1)})^2 \right) H_2 (\vartheta)\nonumber\\
    && -\frac{5n(n-2)}{96} K^{(2)} H_0 (\vartheta)) \sigma _0^2 + \mathcal{O} (\sigma_0^3)]
\end{eqnarray}
where $H_n(\vartheta)$ represent the probabilistic Hermite polynomials. The $S^{(0)}$ and $K^{(0)}$ are the skewness, kurtosis, respectively, and $S^{(i)}$ and $K^{(i)}$ denote their higher-order derivatives, with
$i$ indicates the order of the derivative. Mathematically, they are given by:
\begin{align}\label{eq:skewness1}
    S^{(0)}&\equiv \dfrac{\langle \delta_T^3\rangle_c}{\sigma_0^4},\,\,\,\,\,\,\,\,\,\,S^{(1)}\equiv\frac{3}{2}\dfrac{\langle \delta_T^2|\boldsymbol{\nabla} \delta_T|^2\rangle_c}{\sigma_0^2\sigma_1^2},\\\notag
    K^{(0)}&\equiv \dfrac{\langle \delta_T^4\rangle_c}{\sigma_0^{6}},\,\,\,\,\,\,\,\,\,\,K^{(1)}\equiv\dfrac{2\langle \delta_T^2|\boldsymbol{\nabla} \delta_T|^2\rangle}{\sigma_0^4\sigma_1^2},\,\,\,\,\,\,\,\,\,\, K^{(2)}\equiv-\frac{6}{5}\dfrac{\langle|\boldsymbol{\nabla} \delta_T|^4\rangle_c}{\sigma_0^2\sigma_1^4}.
\end{align}
Having obtained these results, we can proceed to compute the $\langle\mathcal{N}_{\text{cmd},n}\rangle^{\mathbb{NG}}$ directly from the data. Moreover, by determining the spectral moments (Eq. (\ref{eq:spectra1})), we can clarify the theoretical expectations concerning the number density of the cmd-$n$ through Eqs. (\ref{nongaussskewness}) and (\ref{eq:skewness1}). It is worth mentioning that the cmd-$n$ measure given by Eq. (\ref{nongaussskewness}) is inherently a dimensionful quantity. To make it dimensionless, we normalize it by dividing the maximum value of the CMB Gaussian measure as:
$\overline{\mathcal{N}}_{{\rm cmd},n}^{\mathbb{NG}}(\vartheta,G\mu)\equiv\langle\mathcal{N}_{\text{cmd},n}(\vartheta,G\mu)\rangle^{\mathbb{NG}}/\langle\mathcal{N}_{\text{cmd},n}(\vartheta=0,G\mu=0)\rangle^{\mathbb{G}}$  and referring to it as the normalized cmd-$n$ measure.

\begin{figure*}
    \centering
    \includegraphics[width=2\columnwidth]{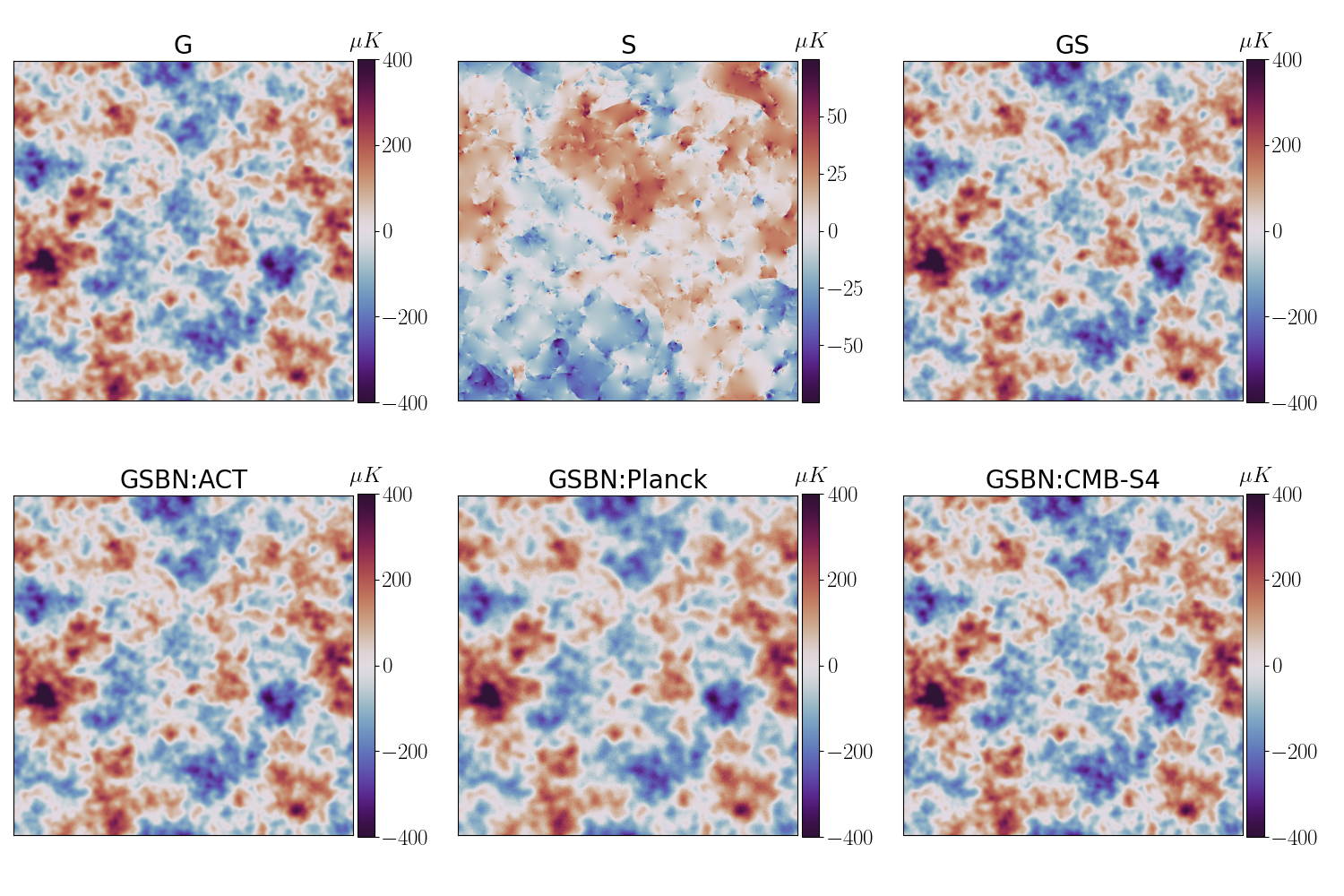}
    \caption{Components of simulated CMB maps. Each map has a size of $7.2\degree \times 7.2\degree$ with a resolution of $0.42$ arcminutes. The upper row shows: (left) the Gaussian random field for the $\Lambda$CDM model, (middle) the pure temperature fluctuations induced by cosmic strings with $G\mu = 5 \times 10^{-7}$, and (right) the superposition of the pure string map with the Gaussian map (GS). The bottom row shows the beam and noise effects on the combined map (GS) for the ACT, Planck, and CMB-S4 experiments (left to right). The beam parameters and noise levels for these experiments are detailed in Table~\ref{beamparameters}.}
    \label{fig_Maps}%
\end{figure*}
\section{Synthetic String-induced CMB maps}
\label{sec:CMB_data_Analysis}

Although the different observational data sets (e.g.
\cite{Ade:2013xla}) reveal that the CSs network is a subdominant
source of the large angle anisotropy, depending on the fraction of
the CMB power spectrum by CSs, $f_{\ell=10}$, the intercommuting
probability, the number of distinct strings at each Hubble volume as
well as underlying theory for topological defects production, we
anticipate to have observable CSs contribution on the small angular
scales. Therefore, we focus on the fluctuations generated by
discontinuities of the CS networks on the CMB temperature map at
small angular scales. We pursue the same recipe for Nambu-Goto
string networks introduced and performed by
\cite{Bouchet:1988hh,Bennett:1990,
Ringeval:2005kr,Fraisse:2007nu,Ringeval:2012tk} and also carried out
by \cite{vafaei2017multiscale,sadr2018cosmic} for making
high-resolution flat-sky CMB patches modified by the CS network.

To achieve accurate cosmic string-induced CMB maps, we superimpose simulated maps from various redshifts. Furthermore, we employ the small-angle approximation, an appropriate method for modeling small-scale features. In other words, the imprint of CSs on the Gaussian CMB temperature fluctuations is accomplished by computing the CSs' Integrated Sachs-Wolfe effect generated by each string along the line of sight \citep{Stebbins:1988,Hindmarsh:1993pu,Stebbins:1995}. The contribution of CSs in CMB map for a given $G\mu$ is denoted by ``S'', ($\delta_T^{{\rm S}}(G\mu)$). This component would be constructed by multiplying the synthetic CS network template with the given value of $G\mu$.  We assume that the dominant parts of the CMB temperature map are sourced by Gaussian initial conditions supported by adiabatic scale-invariant slow-roll-inflationary fluctuations and are labeled by ``G'', ($\delta_T^{{\rm G}}$) (e.g. \cite{Ade:2013xla}). To simulate the Gaussian part of the CMB map, we compute the temperature power spectrum ($C_{\ell}$) for the parameter set of the  $\Lambda$CDM model consistent with {\it Planck} 18 \citep{aghanim2020planckvi}  through the CAMB software\footnote{\texttt{http://camb.info}} \citep{Lewis:1999bs}. Under the flat-sky approximation, the flat power spectrum, $P_{TT}(k)$, is related to the full sky power spectrum via  $\ell(\ell+1)C_{\ell}^{TT}\sim k^2P_{TT}( k)$ \citep{White:1997wq,Hindmarsh:2009qk,Fraisse:2007nu}. A Gaussian realization, $\delta_T^{{\rm G}}(x,y)$, is generated through the inverse Fourier transform of $\delta_T^{{\rm G}}(\mathbf k) = \sqrt{\frac{P_{TT}(k)}{2}}(\mathcal{R}_1+i\mathcal{R}_2)$. Here, the $\mathcal{R}_1$ and $\mathcal{R}_2$ are independent random variables with zero mean and unit variance, and $k=|\mathbf{k}|$. Finally, for the string-induced CMB map without the noise and instrumental beam effects, we use the superposition as $\delta_T^{{\rm GS}}\equiv\delta_T^{{\rm G}}+\delta_T^{{\rm S}}(G\mu)$.  The associated power spectrum of the GS map reads as: $C_\ell^{{\rm GS}}=C_\ell^{\text{G}}+C_\ell^{\text{S}}$.  We have constructed $100$ realizations of CMB maps that contain the imprint of CSs. The map size is $7.2\degrees\times 7.2\degrees$ at resolution $R = 0.42$ arc-minute.


Fig. \ref{fig_PSCMBCS} illustrates the angular power spectrum of different components of the synthetic map. The blue solid line is for the theoretical prediction of pure CSs power spectrum such that $\ell(\ell+1)C_{\ell}^{\rm S}\sim \ell^{-\epsilon}G\mu^2$ for $\epsilon=1$ with $G\mu = 5\times 10^{-7}$~\citep{2010PhRvD..82f3518Y}. However the precise value of exponent has been reported as $\epsilon=0.889^{+0.001}_{-0.090}$  for $\ell\gg 1$ and at $1\sigma$ confidence level~\citep{Fraisse:2007nu,2010PhRvD..82f3527R,Bevis:2010gj,2010JCAP...10..003L}. The black solid line is the theoretical (th) Gaussian CMB power spectrum for the $\Lambda$CDM model (from CAMB). The green solid line is the net power spectrum after including the CS effects. The red dot-dashed line includes the Airy beam, and the purple dashed line includes the Gaussian beam effect for the Planck experiment.
The different symbols indicate the numerical computation of power spectra of the simulated maps for 100 realizations, following the same color coding as mentioned for the theoretical predictions.


\begin{table}
    \centering
    \begin{tabular}{|l| c| c| c|}
        \hline
        Parameters & Planck & CMB-S4 & ACT \\
        \hline
        f (GHz) &   270 & 150 & $277$ \\
        d (m) & --- & --- & $6$      \\
        $\theta$ (degree) & --- & --- & $70$     \\
        FWHM (arcminutes) & $5$ & $1$ & ---      \\
        $\sigma_{\rm noise}$ ($\mu K$-arcminutes) & $46.8$ & $3.07$ & $8$ \\
        \hline
    \end{tabular}
    \caption{Relevant parameters for the two types of beam effects defined in section~\ref{sec:beam_noise} and corresponding noise level.
    }
    \label{beamparameters}
\end{table}


\subsection{Instrumental Beam and Noise effects}
\label{sec:beam_noise}
The observed CMB temperature fluctuations are influenced by the convolution of the instrumental beam with the underlying sky temperature distribution because of the constrained resolution of the telescopes. The beam effect in the Fourier space will be simplified to normal multiplication with the beam window function $W^{\rm B}_\ell$ \citep{bond1987statistics}:
\begin{align}
    \label{Beam_definition}C_\ell^{\text{GSB}}=\left(C_\ell^{\text{G}}+C_\ell^{\text{S}}\right)(W^{\rm B}_\ell)^2
\end{align}
We consider two types of beam effects, namely the Gaussian and the Airy beams. The Gaussian beam function is given by:
\begin{align}
    \label{GaussBeam}W^{\rm B}_\ell= e^{-\frac{\ell(\ell+1)\Delta^2}{2}}
\end{align}
with $\Delta$ being the width of the beam in radians, which is related to the full-width half maximum (FWHM) of the beam as $\Delta=\text{FWHM}/\sqrt{8\ln 2}$ \citep{bond1987statistics}. This type of beam effect is usually considered for the Planck \citep{aghanim2020planck} and CMB-S4 \citep{2016arXiv161002743A,2019arXiv190704473A} experiments. We also consider another type of beam effect known as the Airy pattern. Following the work of \citep{Fraisse:2007nu}, we define the normalized primary beam window function as:
\begin{align}
    W^{\rm B}_\ell\equiv\dfrac{\mathcal{A}{\rm i}(\ell)}{\mathcal{A}{\rm i}(0)}
\end{align}
with $\mathcal{A}{\rm i}$ being an Airy beam pattern defined as:
\begin{align}
    \mathcal{A}{\rm i}(\ell)=\dfrac{2}{\pi^4\,d^2}\left(\arccos{\dfrac{\ell}{\ell_c}}-\dfrac{\ell}{\ell_c}\sqrt{1-\left(\dfrac{\ell}{\ell_c}\right)^2}\right)
\end{align}
where, $\ell_c=2\,\pi\,d/(\lambda\,\theta)$. Here, $d$ is the diameter, $\lambda$ is the beam wavelength, and $\theta$ is the maximum opening angle of the telescope. This type of beam effect is relevant for the ACT experiment \citep{2006NewAR..50..969K}.

\begin{figure}
    \centering
    \includegraphics[width=1\columnwidth]{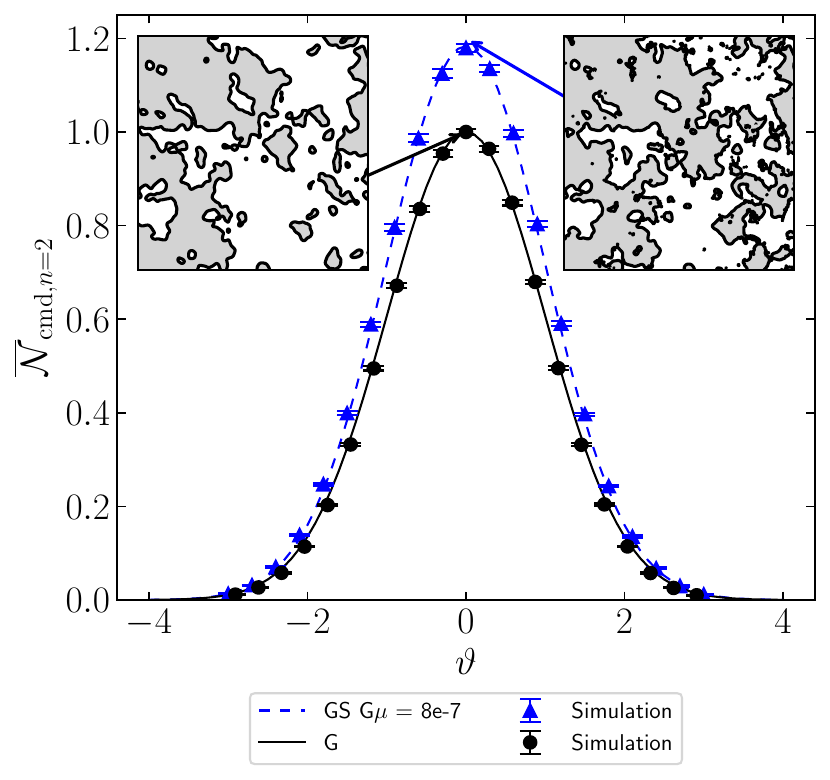}
    \caption{The normalized cmd-$n$ statistics as a function of $\vartheta\equiv\delta_T/\sigma_0$ for G (filled circle symbol and black solid line) and GS (filled triangle symbols and dashed line) for $G\mu=8\times 10^{-7}$ with the inset maps correspond to the areas bounded by zero threshold, presented for the same seed number and window size. The lines illustrate the theoretical prediction by computing the spectral moments associated with the simulated maps through Eq. (\ref{eq:spectra1}) and inserting in Eq. (\ref{NcmdGauss}) according to the normalized definition.}
    \label{fig_New_stat_Gauss_BN}%
\end{figure}
To check the further robustness of our summary statistics,  we additionally consider the effects of instrumental noise through the following analysis:
\begin{align}
    \label{experiment_power_spectrum}
    C_\ell^\text{GSBN}=\left(C_\ell^{\text{G}}+C_\ell^{\text{S}}\right)(W^{\rm B}_\ell)^2 + \sigma_{\rm noise}^2
\end{align}
The choice of parameters for the Planck, CMB-S4, and ACT experiments is given in Table~\ref{beamparameters}.
The different elements of a typical synthetic map realization are presented in Fig.~\ref{fig_Maps} and the corresponding power spectra are given in Fig.~\ref{fig_PSCMBCS}, both for theory (lines) and simulation (symbols). To better comparison between observed maps by various experiments, we took the same seed for map generation illustrated in Fig.~\ref{fig_Maps}.  The beam size of the Planck experiment is greater than that of the other surveys included in this analysis. As a result, the lower middle panel in Fig.~\ref{fig_Maps} reveals a more pronounced smearing effect when compared to ACT and CMB-S4. the higher smearing compared to ACT and CMB-S4, thereby making the detection of the cosmic structures' imprint more challenging.

\begin{figure*}
    \centering
    \includegraphics[width=1\columnwidth]{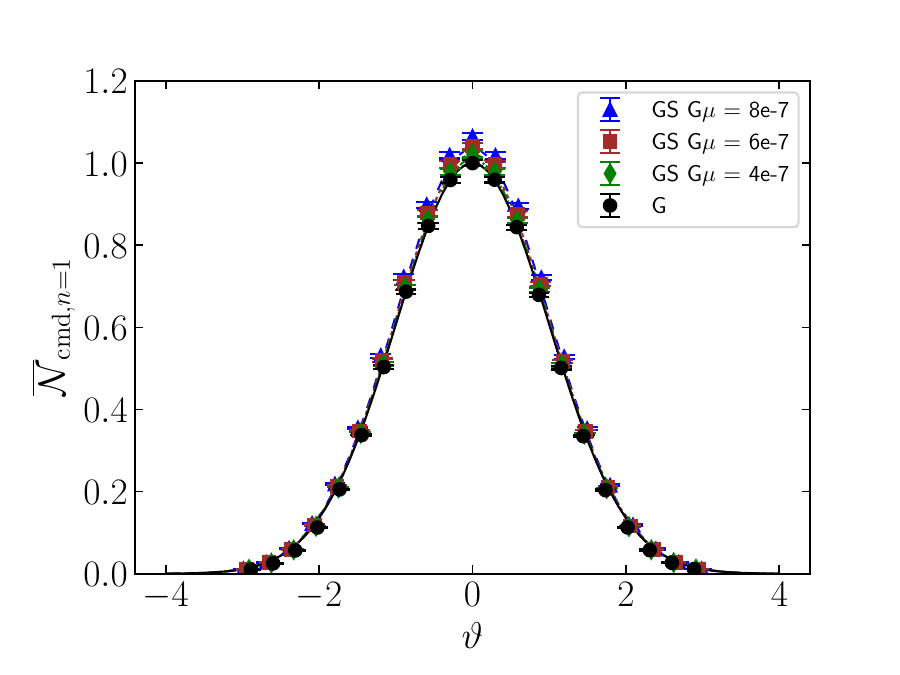}
    \includegraphics[width=1\columnwidth]{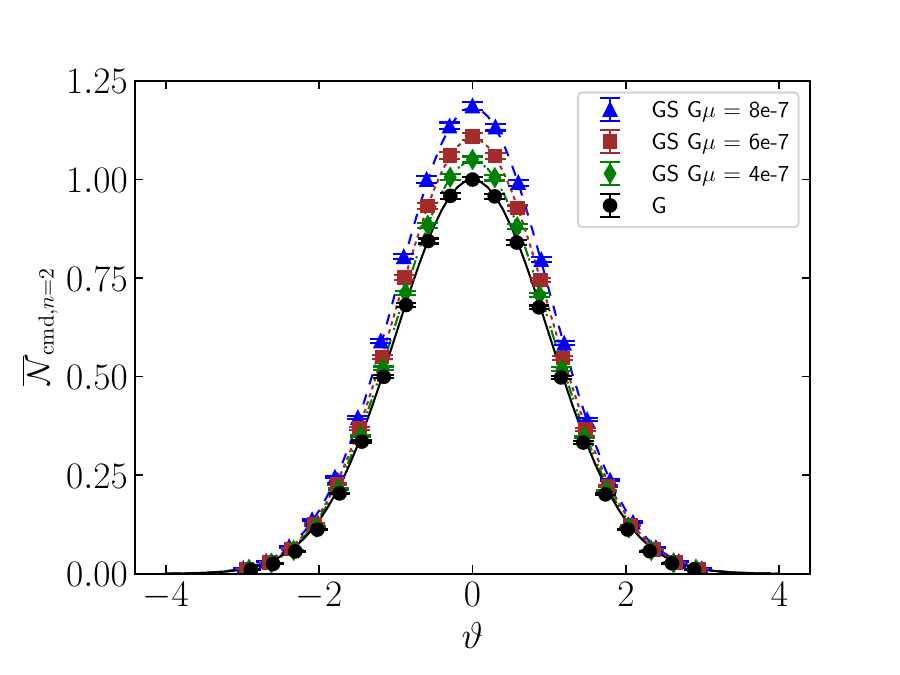}
    \includegraphics[width=1\columnwidth]{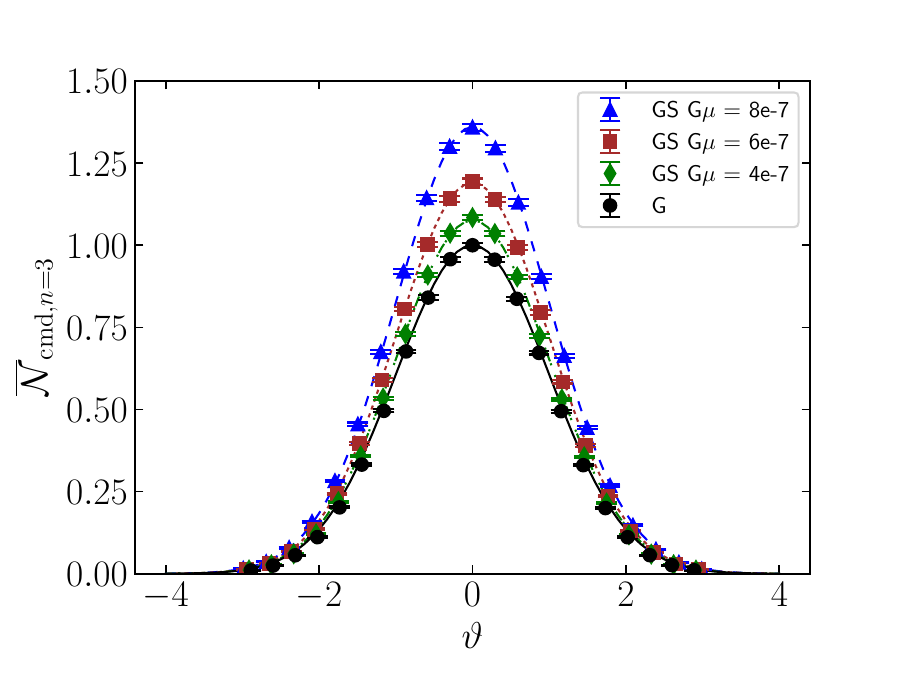}
    \includegraphics[width=1\columnwidth]{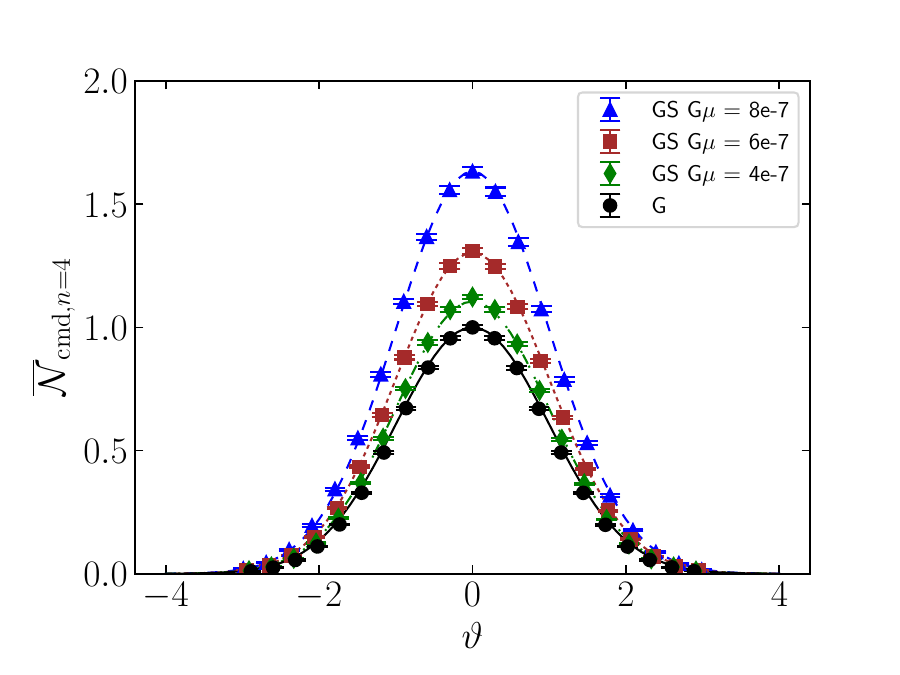}
    \includegraphics[width=1\columnwidth]{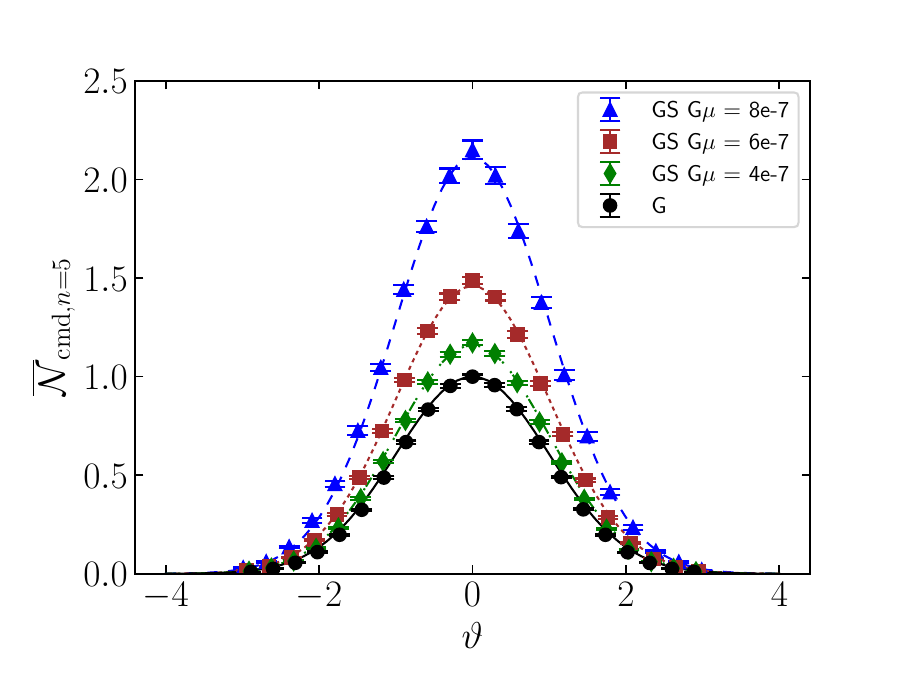}
    \includegraphics[width=1\columnwidth]{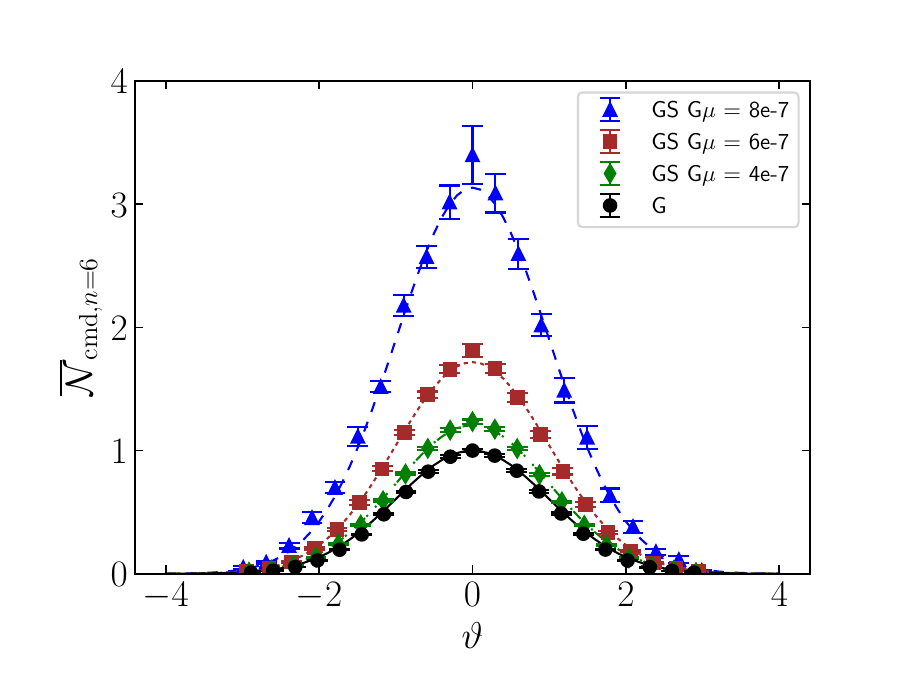}
    \caption{The theoretical normalized cmd-$n$ statistics are presented as a function of the threshold for various values of $n$ and cosmic string tension. In each panel, the filled circle, diamond, square, and triangle symbols are associated with G map and GS with $G\mu=4\times 10^{-7}, \;6\times 10^{-7}, \;8\times 10^{-7}$, respectively. The numerical results have been determined by doing an ensemble average across 100 realizations with the $2\sigma$ confidence interval. The lines represent the Gaussian theoretical prediction (Eq. (\ref{NcmdGauss})).}
    \label{fig_Ncmd_Stats_Sim}
\end{figure*}
\begin{figure*}
    \centering
    \includegraphics[width=1\columnwidth]{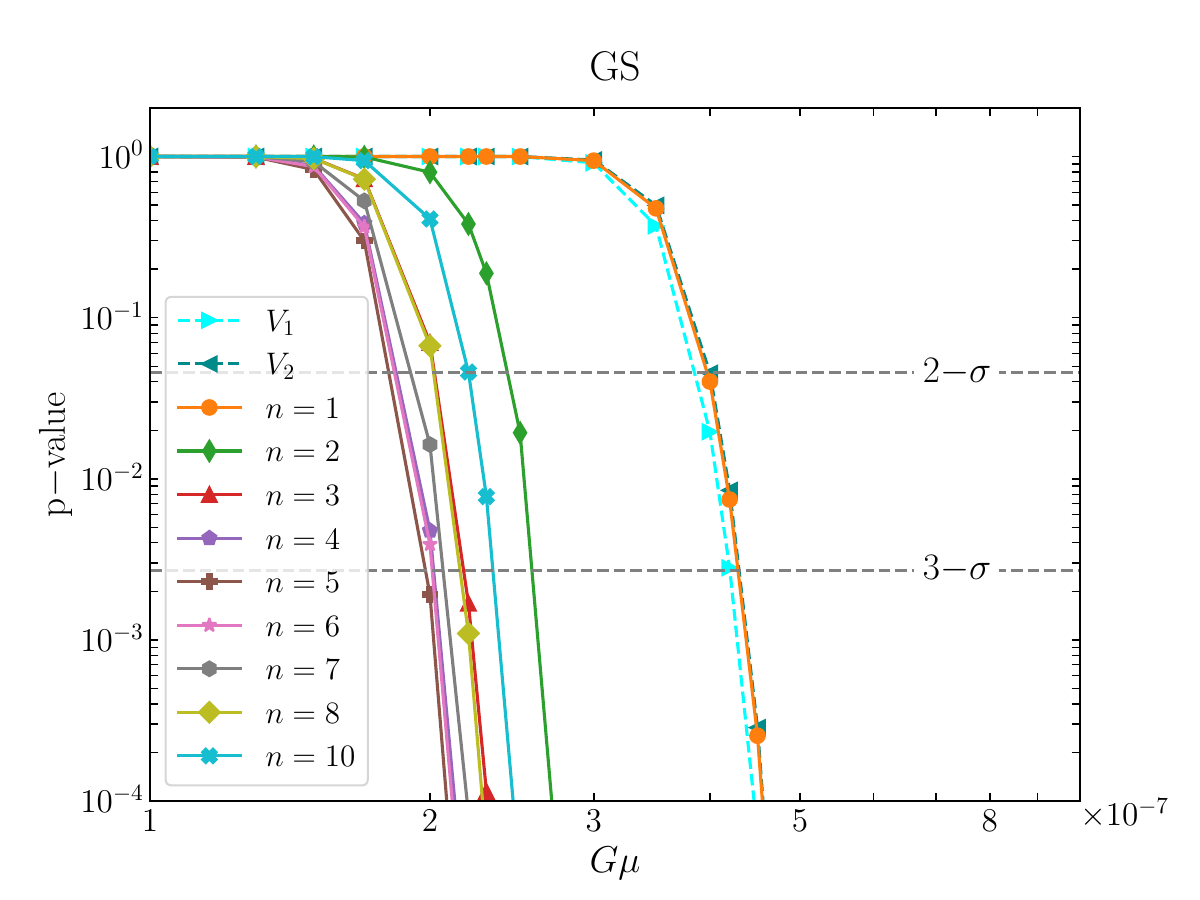}
    \includegraphics[width=1\columnwidth]{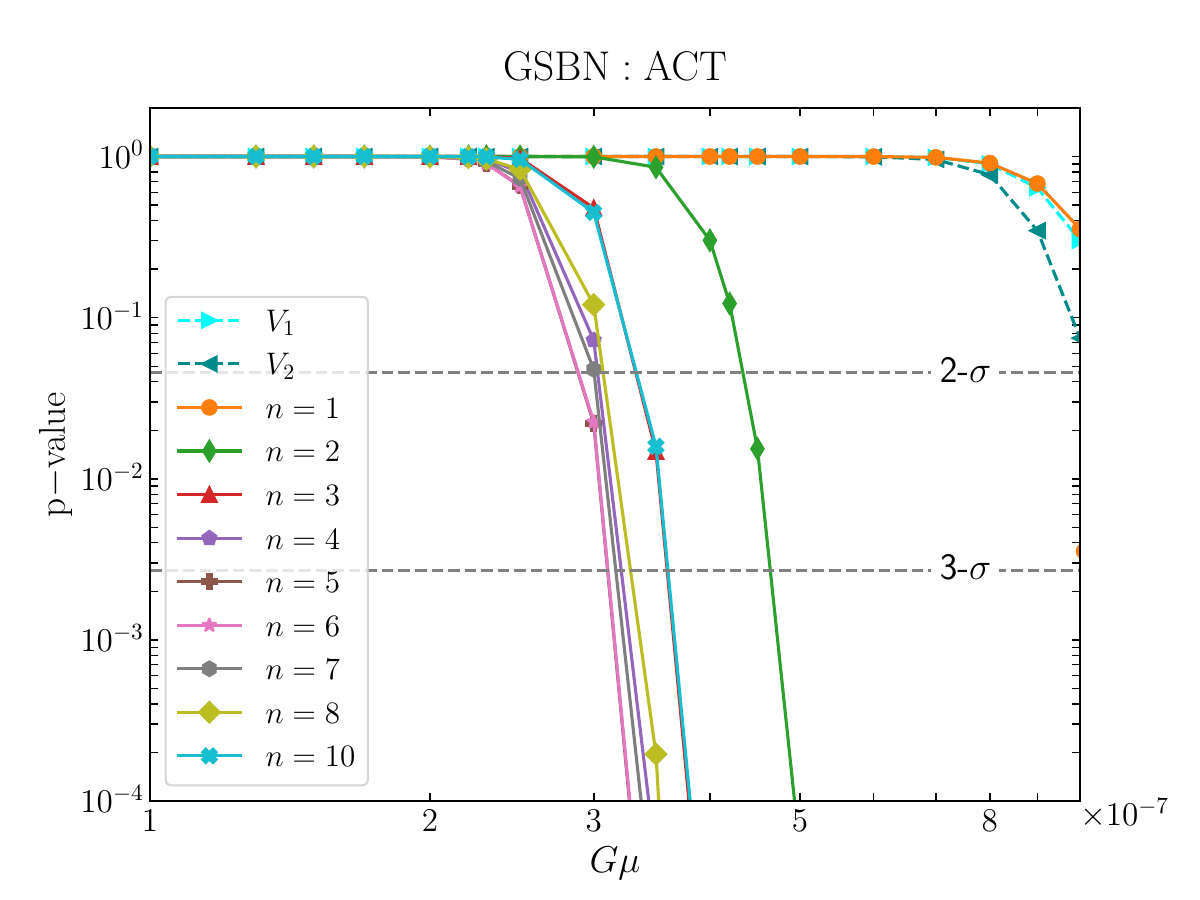}
    \includegraphics[width=1\columnwidth]{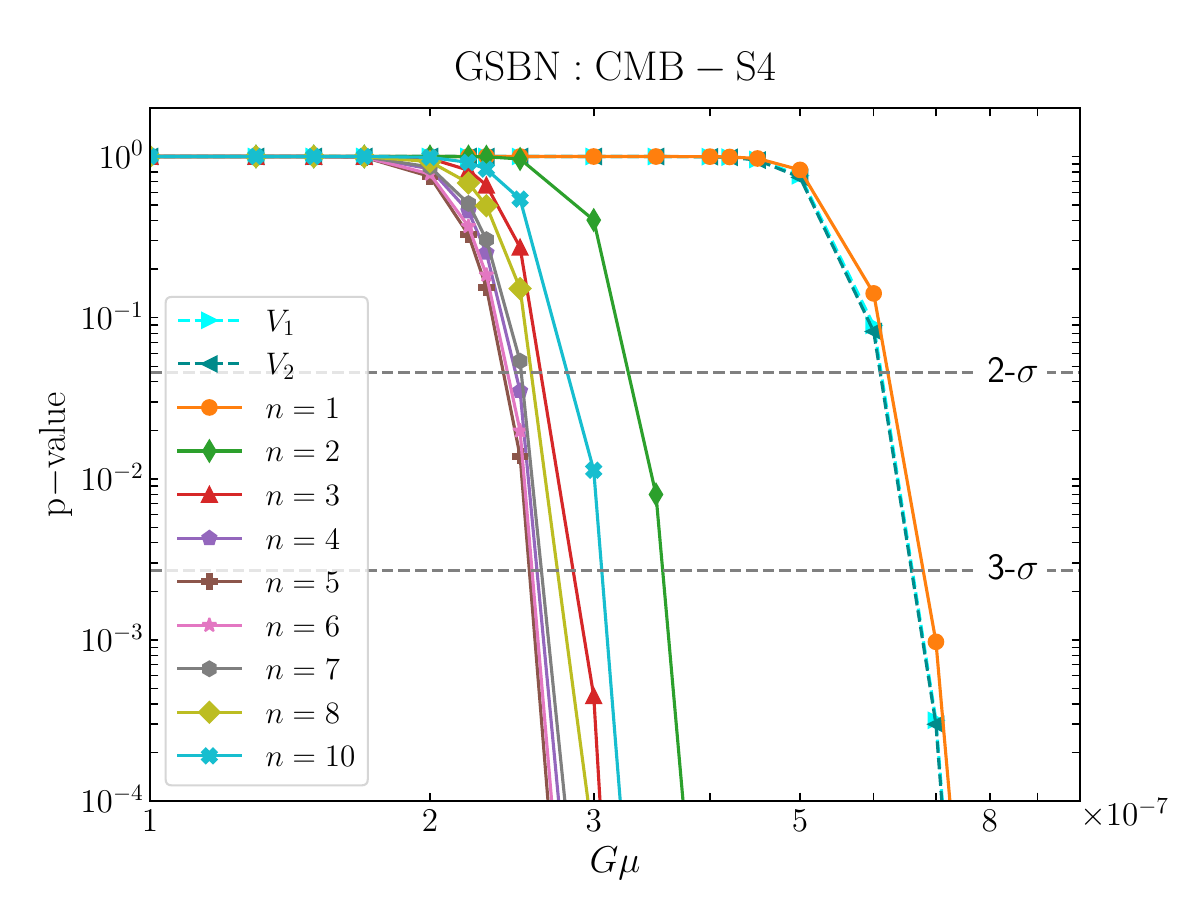}
    \includegraphics[width=1\columnwidth]{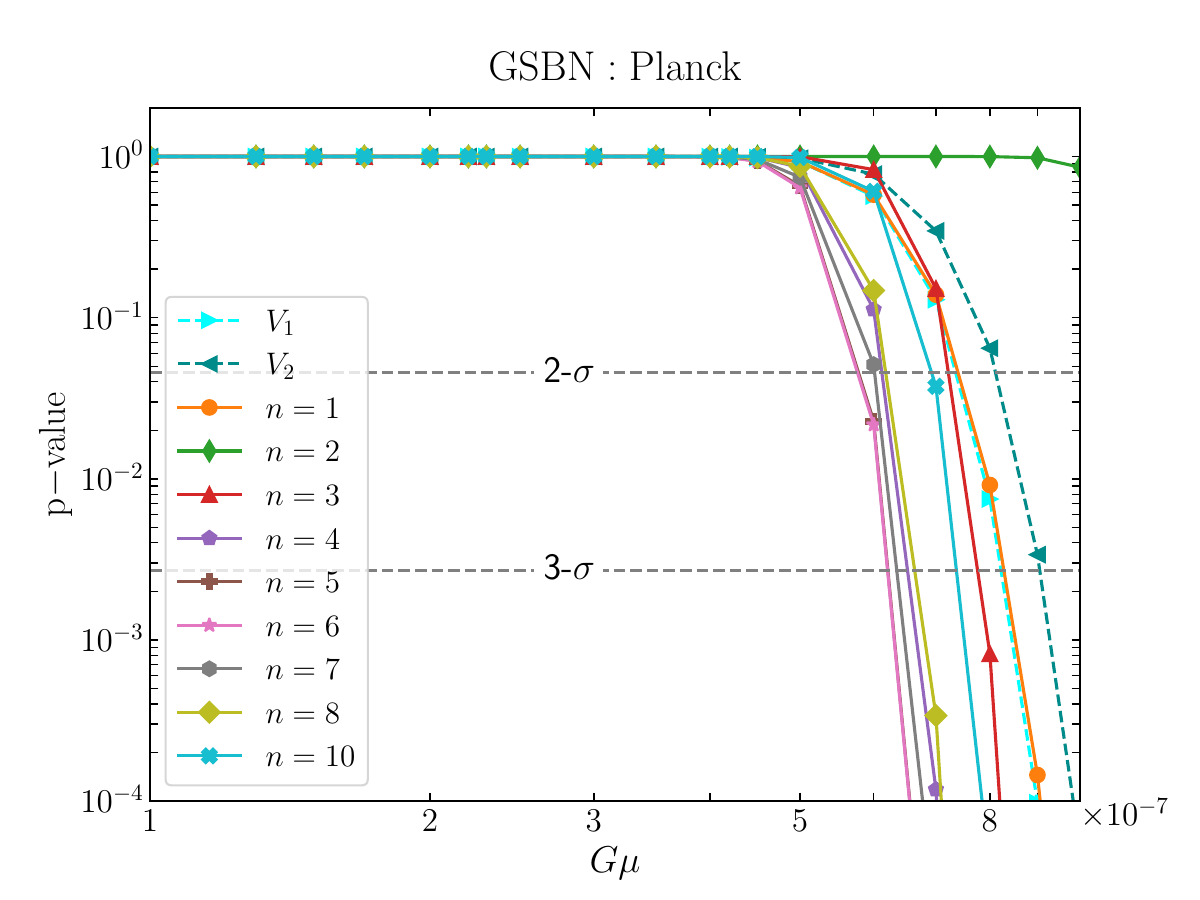}
    \caption{The p-value as function of $G\mu$ for our summary statistics as discussed in Sec.\ref{sec:Weighted_Morphology}.  The upper left panel is for the GS simulation. The level of detectability for ACT, CMB-S4, and Planck-like experiments is depicted in the upper right, lower left, and lower right panels, respectively. We have also included the first ($V_1$) and the second ($V_2$) MFs for comparison.  }
    \label{fig_pvalues}
\end{figure*}
\section{Summary Statistics and Cosmological Inference}
\label{Sec:Cosmological_Inference}

As mentioned in the introduction, the specific goal of this study is to elaborate on the performance of the extension form of weighted morphological measures (cmd-$n$), which is given by Eqs. (\ref{NcmdGauss}) and (\ref{nongaussskewness})  to detect the imprint of CSs on the CMB temperature fluctuations map.

Applying the weighted morphological measure, the normalized cmd-$n$, on the
synthetic data sets for G (Black-solid) and GS (Blue-dashed for
$G\mu= 8\times 10^{-7}$) maps are presented in
Fig.~\ref{fig_New_stat_Gauss_BN}. The lines (solid and dashed)
represent the theoretical predictions for
$\overline{\mathcal{N}}_{{\rm cmd},n=2}$ versus threshold, which are
followed by the calculation of the spectral moments from the simulated maps and taking them into account for the Gaussian term given by Eq.~(\ref{NcmdGauss}) for $n=2$. We then numerically simulated the synthetic data for the weighted morphology and
presented them as symbols with error bars for $100$ realizations. The inset maps indicate patches bounded by zero threshold value for the same seed number and window size. We expect that the G map superimposed by the strings component gets more wiggles on the circumferences and areas of iso-height contours and consequently, the normalized cmd-$n$ can capture such deformations depending on the value of cosmic string tension ($G\mu$). Taking into account the weighted version of morphology governed by the normalized cmd-$n$ for different orders, demonstrates the profound sensitivity with respect to the impact of the CS network.

To assess the impact of various values of $n$ for the normalized number density of our weighted morphological measure, $\overline{\mathcal{N}}_{{\rm cmd},n}$, as a function of threshold, we compute ensemble averaged of the normalized cmd-$n$ over 100 realizations for various values of $G\mu$ from synthetic maps numerically and compare
with theoretical prediction. Fig.~\ref{fig_Ncmd_Stats_Sim} depicts the normalized $\overline{\mathcal{N}}_{{\rm cmd},n}$ as a function of $\vartheta$ for $n=1,2,3,4,5,6$ from top left panel to bottom right panel, respectively. The filled circle symbol is for the G map
while the filled diamond, filled square and filled triangle symbols depict the GS map for $G\mu=4\times 10^{-7},\; 6\times 10^{-7},\;
8\times 10^{-7}$, respectively. The theoretical predictions
indicated by the lines for each symbol are based on Eq.
(\ref{nongaussskewness}) by neglecting the non-Gaussian contribution due to the CSs network (Eq. (\ref{NcmdGauss})). The spectral moments
related to each mock data set is derived using Eq.
(\ref{eq:spectra1}). The degree of consistency between numerical analysis of $\overline{\mathcal{N}}_{{\rm cmd},n}$ with $\overline{\mathcal{N}}^{\mathbb{G}}_{{\rm cmd},n}$ decreases by increasing the $G\mu$. This deviation becomes significant for the higher value of $n$. This behavior can be elucidated as follows: The contribution of non-Gaussianity becomes significant for higher values of $G\mu$ and even the deviation from $\overline{\mathcal{N}}^{\mathbb{G}}_{{\rm cmd},n}$ is magnified as $n$ increases. The correction terms, particularly those up to $\mathcal{O}(\sigma_0^3)$ for a given tension of CSs, exhibit an increasing trend with respect to $n$. Subsequently, the higher-order terms existing in Eq. (\ref{nongaussskewness}) must be taken into account as part of the theoretical predictions.

To quantify the level of CSs network detectability according to the normalized cmd-$n$ measures, we compute the significance of achieved deviations according to:
\begin{align}
    t^{\times}_{\diamond}(\vartheta;G\mu)=\dfrac{\langle\overline{\mathcal{N}}_{\diamond}^{\times}(\vartheta;G\mu)\rangle-\langle\overline{\mathcal{N}}_{\diamond}^{\times}(\vartheta;G\mu=0)\rangle}{\sqrt{\left[\Delta\overline{\mathcal{N}}_\diamond^\times (\vartheta;G\mu)\right]^2+\left[\Delta\overline{\mathcal{N}}_\diamond^\times(\vartheta;G\mu=0)\right]^2}}
\end{align}
where $\times$ is replaced by ideal, CMB-S4, ACT and Planck-like observations. Moreover, $\diamond\in\{({\rm cmd},n),V_1,V_2\}$.  Here, $V_1$ and $V_2$ (Eq. (\ref{eq:Minkowski1}))  are associated with the first and second types of MFs that we consider for comparison purposes. The $\langle.\rangle$ shows the ensemble averaging over 100 realizations. The $\Delta\overline{\mathcal{N}}$ in the denominator is the mean standard deviation of each corresponding term in the numerator.

According to the $t$-distribution function with $2\,N_{\text{sim}}-2$ degree of freedom, where $N_{\text{sim}}$ is the number of simulated maps, the associated p-value, $p_{\diamond}^{\times}(\vartheta;G\mu)$, is computed. Subsequently, we determine the corresponding $(\chi^2)_{\diamond}^{\times}(G\mu)$ by marginalizing over all available thresholds as $(\chi^2)_{\diamond}^{\times}(G\mu)\equiv-2\,\sum_{\vartheta}\ln p_{\diamond}^{\times}(\vartheta;G\mu) $, with $2(\vartheta_{\text{max}} -\vartheta_{\text{min}})/\Delta\vartheta - 2$ degrees of freedom, the chi-square distribution function is used to calculate the final p-value associated with $\chi^2$. In Fig.~\ref{fig_pvalues}, we depict the p-value as a function of $G\mu$ for different observational strategies, incorporating noise and beam effects to mimic a realistic scenario. This approach allows us to determine the minimum detectable value of $G\mu$.  It is noteworthy that the analysis based on the first ($V_1$) and second ($V_2$) MF exhibits weak sensitivity in distinguishing the CSs, while the normalized cmd-$n$ measure gives promising performance. Two horizontal lines in Fig. \ref{fig_pvalues} points to $2\sigma$ (95.45\%) and $3\sigma$ (99.73\%) significance level.

Notably, as the power $n$ increases, the normalized cmd-$n$ statistic becomes increasingly sensitive, enabling the probing of smaller $G\mu$ values. Therefore, the normalized cmd-$n$ statistics enhance detection sensitivity. This behavior is almost similar for all cases taken in this study, except for the Planck-like observation. For the later case, there is a non-monotonic behavior of sensitivity for  $n=1$ and $n=2$. This behavior is justified as follows: the noise level of the Planck-like observation compared to other observations is much more considerable (Table \ref{beamparameters}). On the other hand, a portion of the CS network accumulations on the Gaussian fluctuations of CMB temperature may be recognized as a part of systematic noise, particularly when we are adopting one-point statistics instead of two-point statistics. Hence, for a small value of $n$ and when the non-Gaussianity is neglected,  the approach becomes ambiguous in distinguishing the footprint of CSs from Gaussian noise. But for higher enough value of $n$ in the normalized cmd-$n$ measures, the contribution of systematic noise is no longer dominant and the expected trend is achieved. This situation is also consistent with prior research presented in \citep{Movahed:2012zt}, which demonstrated that the Gaussian CSs network was not detectable through critical sets from Gaussian noise. Subsequently, clustering of local maxima through the two-point correlation function has been employed to break the unweighted degeneracy mentioned. Here, we also expect that the unweighted TPCF of the cmd-$n$ as an extension of one-point statistics may alleviate this discrepancy. A way to diminish this degeneracy is to implement a smoothing function on the map generated by a Planck-like pipeline. We have smoothed the Planck-like by convolving with an additional proper Gaussian kernel and our prediction was confirmed.

\begin{figure}
    \centering
    \includegraphics[width=1\columnwidth]{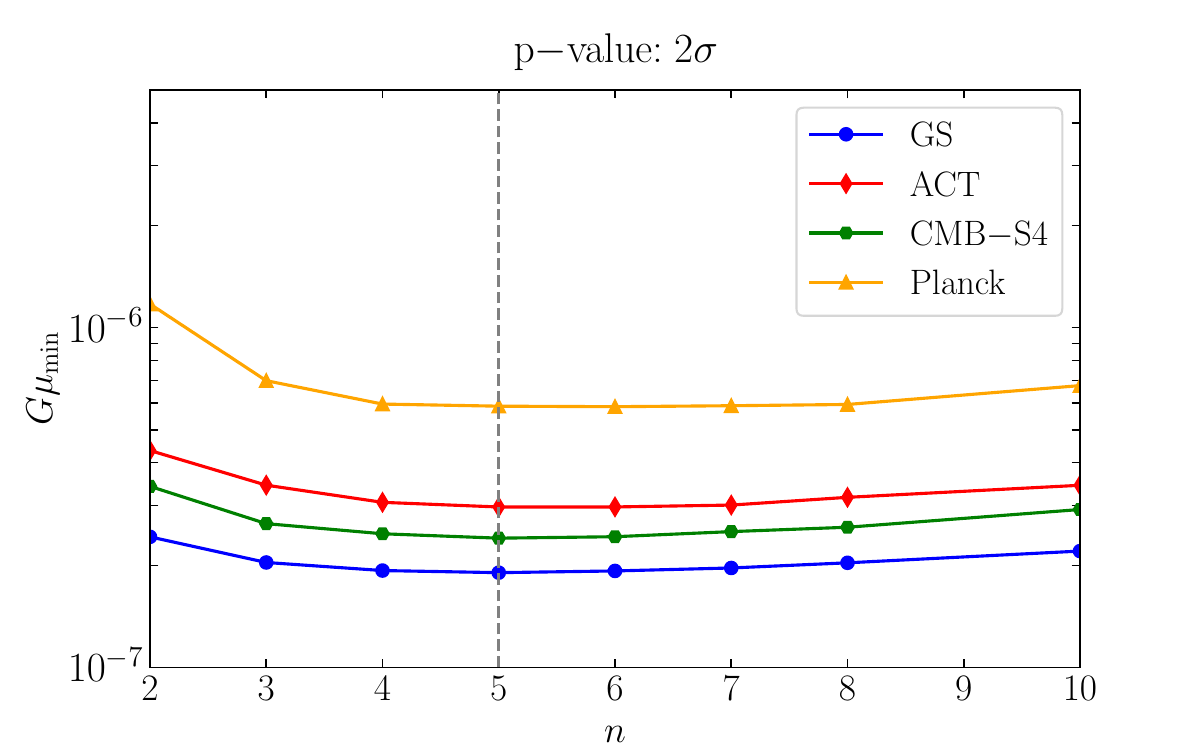}
    \caption{The $G\mu_{\rm min}$ vs. $n$ for p-value at $2\sigma$ level for GS , ACT CMB-S4 and Planck-like experiments. The minimum is found at $n=5$ presented by a vertical gray dashed line, which puts the lower bound on $G\mu$. The corresponding bounds have been summarized in Table~\ref{Table_p-value}.}
    \label{fig_Gmuvsn}%
\end{figure}

Investigating the significance of power $n$ in the normalized cmd-$n$ measures shows that higher values of $n$ lead to increased statistical uncertainty. As a result, we necessarily require a larger number of samples to achieve robust results. The quantification of error propagation, as indicated by the p-value, reveals the appropriate value of $n$ for our available simulation, taking into account the specified sample size and resolution, as illustrated in Fig.~\ref{fig_Gmuvsn}. We compute the minimum detectable value of $G\mu$ versus $n$ at $2\sigma$ confidence level for different synthetic data sets. The minimum occurs at $n=5$ (the vertical grey dashed line), which denotes the highest value of $n$ that is suitable for our synthetic data. Table~\ref{Table_p-value} summarizes the minimum detectable value of $G\mu$ at the $2\sigma$ confidence interval for the normalized cmd-$n$ statistics when $n=5$.

\begin{table}
    \centering
    \begin{tabular}{|l|c|c|}
        \hline
        Map & $G\mu_{\rm min}$ at $2\sigma$ \\
        \hline
        GS & $1.9\times 10^{-7}$\\
        \hline
        ACT &$2.9\times 10^{-7}$\\
        \hline
        CMB-S4 & $2.4\times 10^{-7}$ \\
        \hline
        Planck & $5.8\times 10^{-7}$ \\
        \hline
    \end{tabular}
    \caption{The minimum detectable value of $G\mu$ at $2\sigma$ significance, coming from the analysis of
        the normalized cmd-$(n=5)$ measure for different observational strategies.}
    \label{Table_p-value}
\end{table}
\begin{figure*}
    \centering
    \includegraphics[width=1\columnwidth]{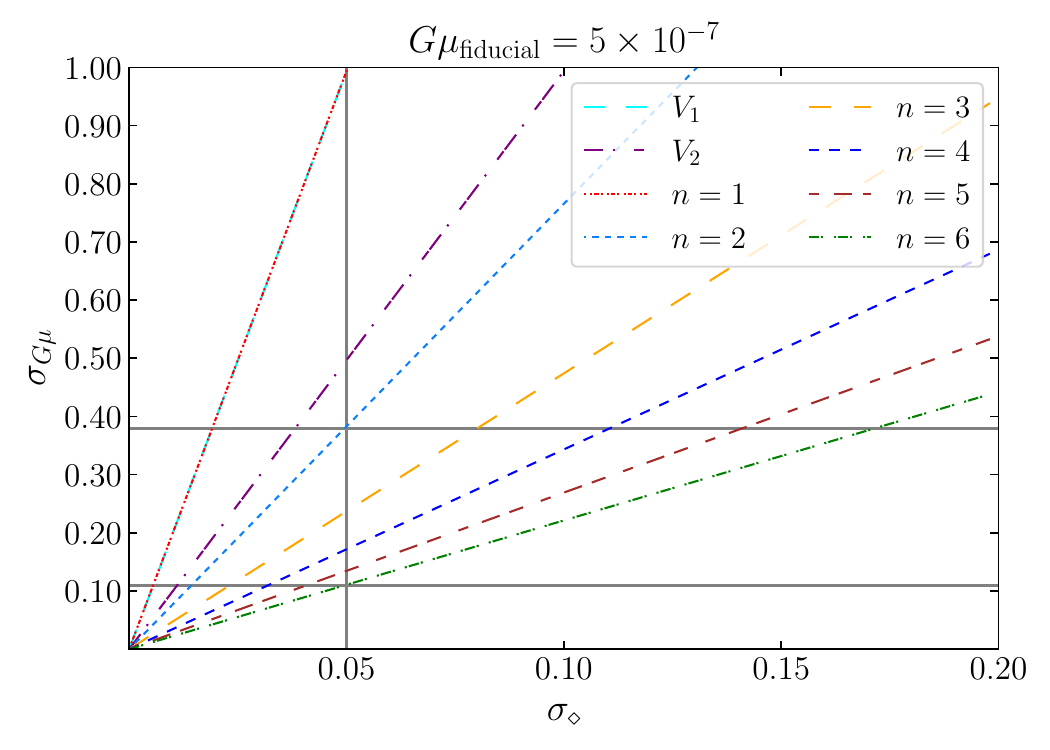}
    \includegraphics[width=1\columnwidth]{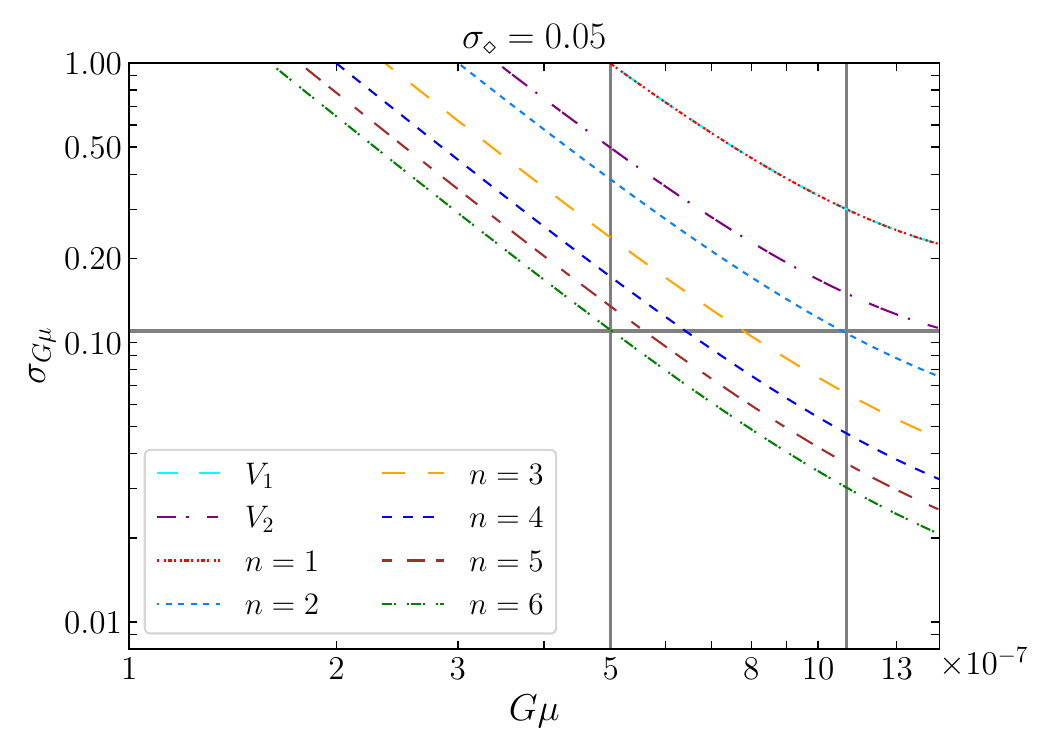}
    \includegraphics[width=1\columnwidth]{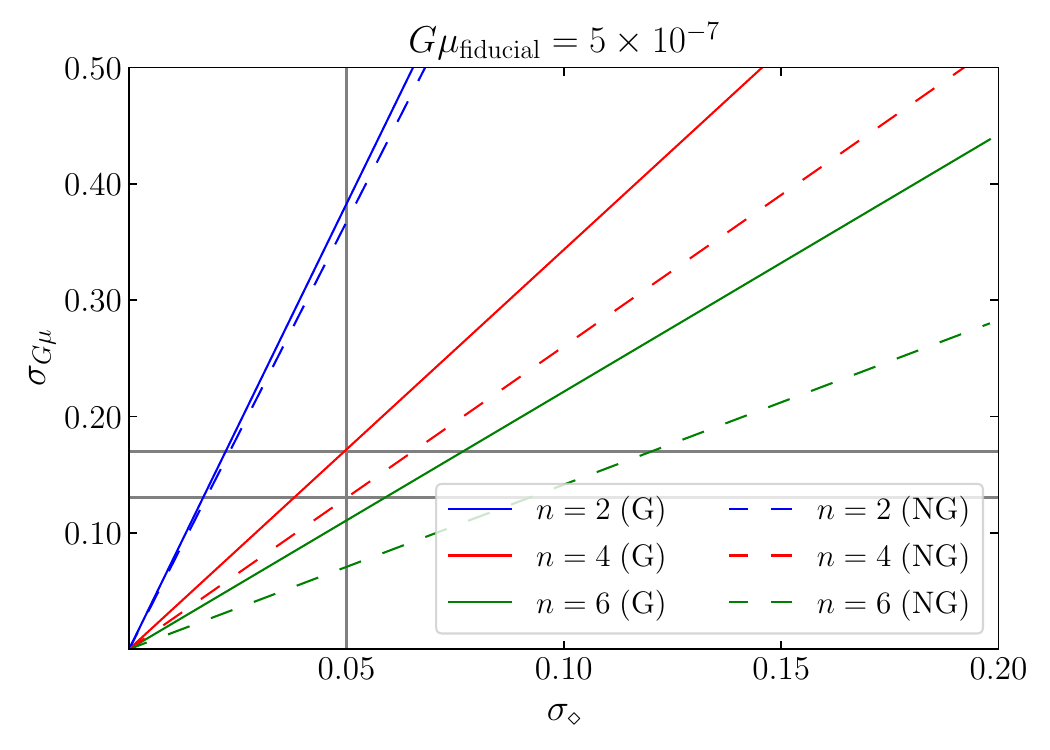}
    \includegraphics[width=1\columnwidth]{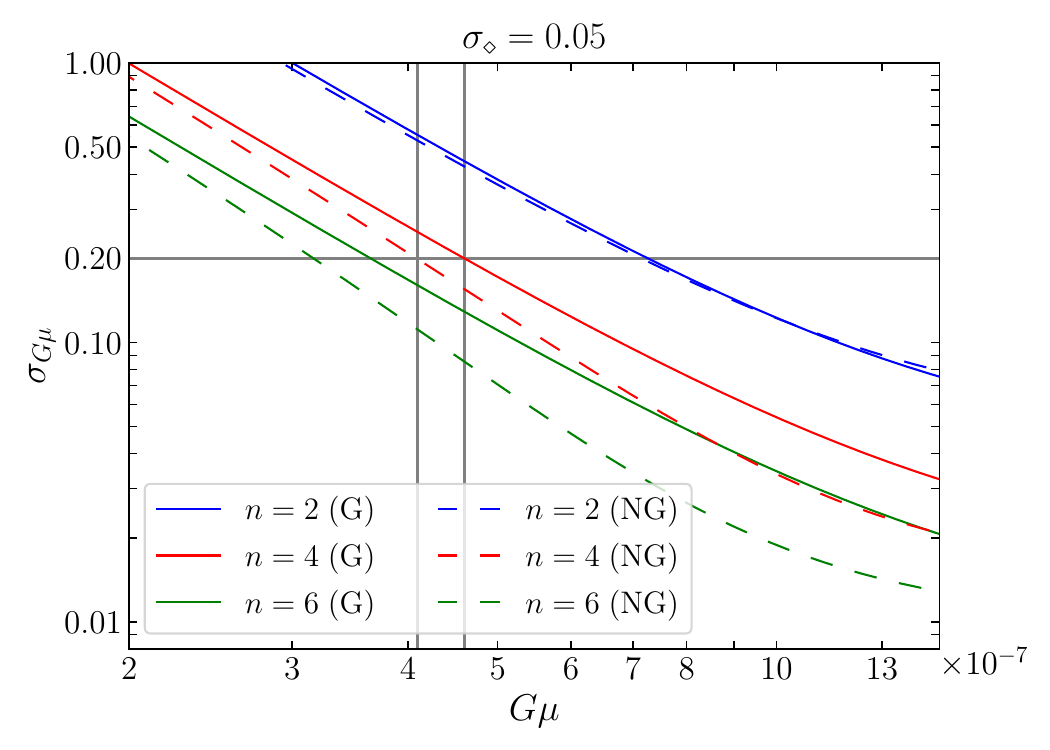}
    \caption{The Gaussian statistical error associated with $G\mu$ vs. the error associated with the measures ($\diamond\equiv$ cmd-$n,\;V_1,\;V_2$) (upper left panel) for the fiducial value of $G\mu_{\rm fiducial}=5\times 10^{-7}$. The vertical line in this panel shows 5\% relative error given for $\sigma_{\diamond}$, which leads to different levels of relative errors on $G\mu$ depending on the measurement method. The upper right panel depicts the relative statistical error, $\sigma_{G\mu}$, as a function of $G\mu$, with a fixed $\sigma_{\diamond}=0.05$. The horizontal gray line in the indicated panel corresponds to a relative statistical error of $11\%$. Two vertical gray lines illustrate the minimum measurable $G\mu$ values for the normalized cmd-$(n=6)$ and normalized cmd-$(n=2)$. The lower panels present a comparative analysis of two distinct cases:
        the first case disregards the non-Gaussian correction (illustrated by the solid line), and another where correction terms are included up to $\mathcal{O}(\sigma_0^3)$ (depicted by the dash-dot line).}
    \label{fig_sigmaGmu}%
\end{figure*}
To give an estimation for the relative error on the $G\mu$ by the normalized cmd-$n$, $V_1$ and $V_2$ statistics, we define following dimensionless quantity:
\begin{eqnarray}
    \mathcal{B}^{\mathbb{NG}}_{\diamond}(\vartheta,G\mu)&\equiv&\dfrac{\overline{\mathcal{N}}^{\mathbb{NG}}_{\diamond}(\vartheta,G\mu)}{\overline{\mathcal{N}}^{\mathbb{G}}_{\diamond}(\vartheta,G\mu=0)}
\end{eqnarray}
where $\diamond$ can be replaced by cmd-$n$, $V_1$ and $V_2$. By marginalizing over all available thresholds, we obtain:
\begin{eqnarray}
    \overline{\mathcal{B}}^{\mathbb{NG}}_{\diamond}(G\mu)&\equiv&\int d\vartheta\;\mathcal{B}^{\mathbb{NG}}_{\diamond}(\vartheta,G\mu)
\end{eqnarray}
By measuring the $\overline{\mathcal{B}}_{\diamond}$ from observed map, and according to the appropriate emulator such as Gaussian process emulator \citep{2014bda..book.....G}, we can use this new observable quantity to infer the cosmological parameters and their uncertainties. Instead, here, we utilize the error propagation framework to determine the degree of uncertainty in measuring $\overline{\mathcal{B}}^{\mathbb{NG}}_{{\rm cmd},n}$, which relies on the statistical error associated with the normalized cmd-$n$ statistics obtained from the observations. Consequently, the relative error associated with $G\mu$, resulting from the statistical error derived from the normalized cmd-$n$ statistics to the first order, is expressed as follows \citep{appleby2019ensemble,2024ApJ...963...31K}:
\begin{table*}
    \centering
    \begin{tabular}{|l|c|c|r|}
        \hline
        Measure& {\rm Data} & {\rm Limit} & {\rm References}\\
        \hline
        Wavelets and Curvelets&{\rm SPT-3G}&$G\mu \gtrsim 1.4\times 10^{-7}$ \;($3\sigma$)&\cite{Hergt:2016xup}\\
        \hline
        Wavelet-Bayesian inference&{\rm CMB} &$G\mu \gtrsim 5\times 10^{-7}$&\cite{2017MNRAS.472.4081M}\\
        \hline
        Wavelet domain Bayesian&{\rm CMB} &$G\mu \gtrsim 5\times 10^{-7}$& \cite{Hammond:2008fg}\\
        \hline
        &&& \cite{Amsel:2007ki}\\
        CANNY algorithm&{\rm SPT} &$G\mu \gtrsim 5.5\times 10^{-8}$ \;($3\sigma$)&\cite{Stewart:2008zq}\\
        &&&\cite{2010IJMPD..19..183D,2010JCAP...02..033D,2010IJMPD..19..183D}\\
        \hline
        PDF&{\rm Noise\;free\;map} & $G\mu\gtrsim 4.3\times 10^{-10}$ \; ($3\sigma$) &\\
        Peak-Peak&{\rm Planck}&  $G\mu\gtrsim8.9\times 10^{-7}$\;($3\sigma$)&\\
        Crossing-Peak&{\rm CMB-S4}&$G\mu\gtrsim 2.4\times 10^{-7}$ \; ($3\sigma$)&\cite{vafaei2017multiscale}\\
        Crossing-Crossing&{\rm Planck}&$G\mu\gtrsim 8.4\times 10^{-7}$ \; ($3\sigma$)&\\
        \hline
        &{\rm Noise\;free\; map}  & $G\mu\gtrsim 1.6\times 10^{-8}$\; ($3\sigma$)& \\
        Peak-Peak&{\rm G+S+N\; map} & $G\mu\gtrsim 1.2\times 10^{-7}$\;($3\sigma$)&\cite{Movahed:2012zt} \\
        &{\rm G+S+B} & $G\mu\gtrsim 1.5\times 10^{-7}$\;($3\sigma$)& \\
        &{\rm G+S+B+N} & $G\mu\gtrsim 2.2 \times 10^{-7}$\; ($3\sigma$)& \\
        \hline
        CNN- Machine learning&&$G\mu\gtrsim 7.7 \times 10^{-7}$\; ($2\sigma$)&\\
        LightGBM Machine learning&&$G\mu\gtrsim 3.8 \times 10^{-7}$\; ($2\sigma$)&\cite{2022MNRAS.509.2169T}\\
        LightGBM and CNN&{\rm CMB-S4}&$G\mu\gtrsim 1.9 \times 10^{-7}$\; ($3\sigma$)&\\
        \hline
        GB Machine learning&{\rm Planck-like}&$G\mu\gtrsim 7.0 \times 10^{-7}$\; ($3\sigma$)&\\
        RF Machine learning&{\rm Planck-like}&$G\mu\gtrsim 5.0 \times 10^{-7}$\; ($3\sigma$)&\\
        GB Machine learning&{\rm CMB-S4}&$G\mu\gtrsim 1.2 \times 10^{-7}$\; ($3\sigma$)&\cite{sadr2018cosmic}\\
        RF Machine learning&{\rm CMB-S4}&$G\mu\gtrsim 3.0 \times 10^{-8}$\; ($3\sigma$)&\\
        \hline
        &{\rm Noise\; free\; map}&$G\mu\gtrsim 1.9 \times 10^{-7}$\; ($2\sigma$)&\\
        cmd-$(n=5)$     &{\rm CMB-S4}&$G\mu\gtrsim 2.4 \times 10^{-7}$\; ($2\sigma$)&this work\\
        &{\rm ACT}&$G\mu\gtrsim 2.9 \times 10^{-7}$\; ($2\sigma$)&\\
        &{\rm Planck}&$G\mu\gtrsim 5.8 \times 10^{-7}$\; ($2\sigma$)&\\
        \hline
    \end{tabular}
    \caption{The minimum detectable value of $G\mu$ derived by various methods and different data sets when CMB data is considered. For other bounds on the CSs tension through different surveys, see the main text.}
    \label{tab:tab1}
\end{table*}

\begin{align}
    \sigma_{G\mu}^2(G\mu)=\left(\dfrac{\partial\ln\overline{\mathcal{B}}^{\mathbb{NG}}_{{\rm cmd},n}(G\mu)}{\partial\ln G\mu}\right)^{-2}\sigma_{{\rm cmd},n}^2(G\mu)
\end{align}
Neglecting the non-Gaussian parts of  $\overline{\mathcal{N}}_{{\rm cmd},n}$ and for $G\mu_{\rm fiducial}=5\times 10^{-7}$, the upper left panel of Fig.~\ref{fig_sigmaGmu} illustrates $\sigma_{G\mu}$ versus $\sigma_{\diamond}$ in the vicinity of ideal case, where $\diamond$ is replaced by the  cmd-$n$, $V_1$ and $V_2$. Supposing five percent relative error (vertical line in the upper left panel of Fig.~\ref{fig_sigmaGmu}) for $\sigma_{{\rm cmd},n}$ in observation for fiducial value $G\mu=5\times 10^{-7}$ gives rise a $\sim 38\%$  and $\sim 11\%$ statistical errors on $G\mu$ for the normalized cmd-$(n=2)$ and  cmd-$(n=6)$, respectively. While for the second MFs, the relative statistical error associated with the measurement of $G\mu_{\rm fiducial}$ is $\sim 50\%$ at the same level of relative error. The first MFs results in a relative error of approximately $\sim 100\%$ when measuring the cosmic string tension for specified $G\mu_{\rm fiducial}$.  The upper right panel of Fig.~\ref{fig_sigmaGmu} displays the statistical error on determining cosmic string tension as a function of $G\mu$ supposing a 5\% accurate measurement of the normalized cmd-$n$, $V_1$ and $V_2$. The horizontal gray lines show the $\sim 11\%$ statistical errors on $G\mu$. At the mentioned relative error, the minimum measurable values of $G\mu$ are $G\mu\gtrsim5\times 10^{-7}$ and   $G\mu\gtrsim11\times 10^{-7}$, when we utilize the normalized cmd-$(n=6)$ and  cmd-$(n=2)$, respectively. Including the non-Gaussian parts of  $\overline{\mathcal{N}}_{{\rm cmd},n}$, the estimation of statistical error becomes more accurate. To compare the signature of correction terms in Eq. (\ref{nongaussskewness}), we repeat the relative statistical error evaluation and the comparative analysis is depicted in the lower panel of Fig.~\ref{fig_sigmaGmu}. The results indicate two distinct cases: the first case ignores the non-Gaussian correction (solid line) while the second scenario includes the correction terms are included up to $\mathcal{O}(\sigma_0^3)$ (dash-dot line). According to the lower left panel, the estimated relative statistical error, $\sigma_{G\mu}$, decreases from approximately $\sim 17\%$ to $\sim 13\%$ for cmd-$(n=4)$. This finding implies disregarding the non-Gaussianity due to $G\mu$  leads to an overestimation for $\sigma_{G\mu}$. The lower right panel shows that the lowest measurable  $G\mu$ with 20\% relative error reduces from $G\mu\gtrsim4.6\times 10^{-7}$ to $G\mu\gtrsim4.1\times 10^{-7}$, for the normalized cmd-$(n=4)$. As the value of $G\mu$ increases, the contribution of non-Gaussian terms is amplified, and this effect is further enhanced by a higher value of $n$. Therefore, in the vicinity of a large sample size,  it is anticipated that as the value of $n$ increases, while maintaining a constant ratio of $\sigma_{G\mu}/\sigma_{\diamond}$, the minimum measurable $G\mu$  will reduce. In addition, the results demonstrate that the relative statistical error for measuring $G\mu$ increases by decreasing $G\mu$. Furthermore, increasing the power $n$ while neglecting sample size uncertainties results in a lower $\sigma_{G\mu}$, thus enhancing the sensitivity for CSs detection. It is worth mentioning that in a realistic case, and for a fixed sample size, increasing the value of power $n$ in the normalized cmd-$n$ produces an additional statistical error which should be taken into account.

\begin{figure*}
    \centering
    \includegraphics[width=2\columnwidth]{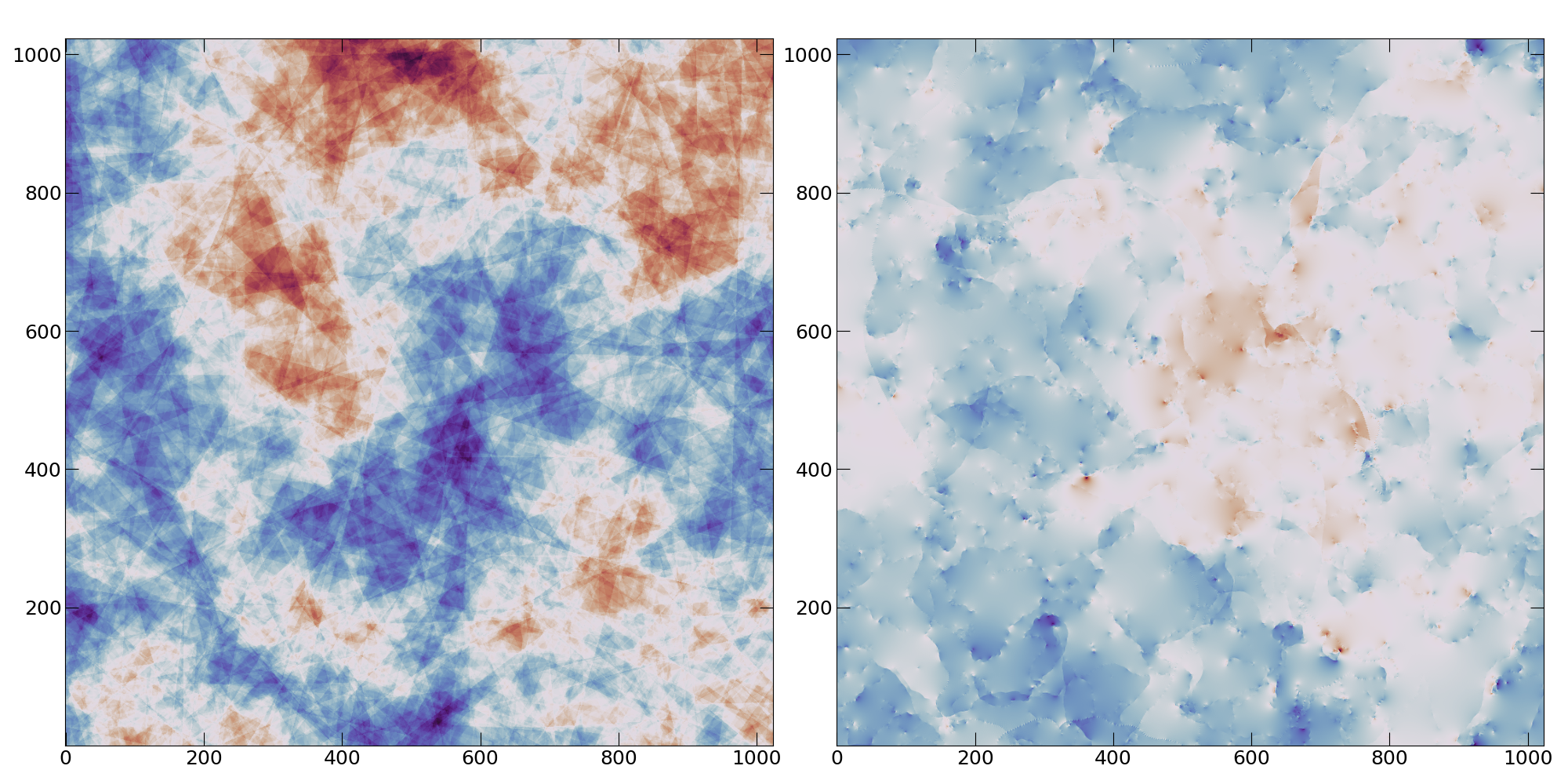}
    \caption{Comparison of two synthetic CSs maps.  Left panel shows the cosmic string component simulated by random kicks algorithm introduced by \citep{Stewart:2008zq}. Right panel represents a realization generated by Bennett-Bouchet-Ringeval code~\citep{Bennett:1990,Ringeval:2005kr}. The size of two maps is similar and equates to $7.2^{\circ}\times 7.2^{\circ}$ ($1024\times 1024$ pixels) with resolution corresponds to $0.42$ arcminutes.  }
    \label{fig.comparison}%
\end{figure*}

Finally, we have compiled a selection of related studies that pursue the same general goal. Table \ref{tab:tab1} presents a summary of works that investigate the impact of CS networks on the CMB field and assess the effectiveness of various methods in detecting CSs by determining the minimum detectable value of $G\mu$. It is important to note that the variation in the reported bounds on the string tension is not solely due to differences in detection techniques. These discrepancies also arise from the diversity of models and algorithms used to simulate CS networks as components of CMB fluctuation maps\footnote{A comprehensive overview of different types of cosmic strings is provided in \cite{Ade:2013xla}}. For instance, a toy model based on the Gott-Kaiser-Stebbins effect, proposed by \cite{Stewart:2008zq}, has been utilized in subsequent studies \citep{movahed2011level,Movahed:2012zt}. In contrast, more realistic simulations have been carried out using Nambu-Goto string networks, implemented through the Bennett-Bouchet-Ringeval code~\citep{Bennett:1990, Ringeval:2005kr}, which includes string loops and accounts for the ISW effect generated along the line of sight. By stacking maps over a broad range of redshifts, these simulations yield reliable representations at small angular scales. To illustrate the differences between the two approaches, Fig.~\ref{fig.comparison} shows a single realization from each. While both exhibit identical power spectra, they reveal notable differences in morphological characteristics.

\section{Summary and Conclusions}
\label{Sec_Conclusions}
A prominent remnant of phase transitions in the early universe is represented by line-like topological defects, which are known as cosmic strings. In this paper, we relied on a new version of the weighed morphological measure introduced by \cite{2024ApJ...963...31K} and tried to revisit the imprint of line-like topological defects, cosmic strings, on the temperature fluctuation of the CMB map. We proposed an extended version of the weighted morphological measure, cmd-$n$. These statistics allocate the specific weight through the amount of the field associated with each excursion set point. The CSs network can induce discontinuities in the gravitational potential, resulting in a series of discontinuities in the CMB temperature fluctuations. The implications of these discontinuities can alter the morphology of the CMB map. Consequently, we embraced the advantage of incorporating the derivative term into the general functional representation of $\mathcal{G}$ as presented in Eq. (\ref{eq:cmd1}) to assess the weighted morphology of the string-induced CMB map through Eq. (\ref{eq:MVs1}).

Additionally, we demonstrated that the specific form of $\mathcal{G}$ for the cmd-$n$ has an analytic formula for the Gaussian field. According to the probabilistic framework, we derived a perturbative expression up to $\mathcal{O}(\sigma_0^3)$ for the weakly non-Gaussian regime in terms of spectral moments derived either from the power spectrum of field (Eq. (\ref{eq:spectra1})) or numerically computed from field-level itself, as well as higher-order cumulants which manifest the non-Gaussianity (Eqs. (\ref{nongaussskewness}) and (\ref{eq:skewness1})).

Utilizing the cmd-$n$, we concentrated on assessing the detectability of the CSs network through weighted morphology. We evaluated our proposed methodology on the simulated CMB maps that were altered by the CSs network, employing high-resolution mock data sets of CSs. To include the beam and systematic noise, the CMB-S4, ACT and Planck experiments have been utilized. This was subsequently compared with the first and second types of MFs.  Our result demonstrated that the normalized cmd-$n$ attains higher values at all thresholds when the CMB intensity field is influenced by the CSs network, as demonstrated in Figs. \ref{fig_New_stat_Gauss_BN} and \ref{fig_Ncmd_Stats_Sim}.

The higher value of power $n$ leads to improved distinguishability of CSs networks on the CMB map, when we ignore the uncertainty due to the finite sample size (Fig. \ref{fig_Ncmd_Stats_Sim}). Accounting for statistical uncertainties, our analysis indicates that the optimal performance of the normalized cmd-$n$ method occurs at $n=5$ (Fig. \ref{fig_Gmuvsn}). Taking the normalized cmd-$(n=5)$, we argued that the CSs are detectable if $G\mu\gtrsim 1.9 \times 10^{-7}$ up to $2\sigma$ significance for an ideal CMB map. Mimicking a more realistic scenario and taking into account the experimental beam and noise effects, this limit increases to $G\mu\gtrsim 2.9 \times 10^{-7}$ for the ACT, $G\mu\gtrsim 2.4 \times 10^{-7}$ for the CMB-S4 and $G\mu\gtrsim 5.8\times 10^{-7}$  for the Planck-like experiments up to $2\sigma$ confidence level (Fig. \ref{fig_pvalues}). Considering the $V_1$ and $V_2$, confirmed that $G\mu\gtrsim 3.9\times 10^{-7}$ and $G\mu\gtrsim 4.0\times 10^{-7}$ for ideal observation, respectively.  For the Planck-like experiment, the normalized cmd-$(n=2)$ has lower capability to detect the CSs network compared to the normalized cmd-$(n=1)$, which is in contradiction to CMB-S4 and ACT. A reason is related to the fact that the noise level of Planck is much more considerable than CMB-S4 and ACT (Table \ref{beamparameters}). For this case, the imprint of CSs from the lens of one-point statistics of weighted morphology looks like the systematic noise and this approach becomes ambiguous in distinguishing the footprint of CSs from Gaussian noise. To mitigate such a discrepancy, the additional smoothing procedure should be applied to the data to reduce the noise. Going beyond one-point statistics and use the unweighed Two-Point Correlation Function of the cmd-$n$ may diminish this challenge and it has been left for future research.

By applying the error propagation approach and considering a relative error of approximately 5\% for $\overline{\mathcal{B}}^{\mathbb{NG}}_{{\rm cmd},n}$, while omitting the non-Gaussian correction terms (as specified in Eq. 16), demonstrated that the normalized cmd-$(n=2)$ and cmd-$(n=6)$ provides measurement on the $G\mu=5\times 10^{-7}$ with a $\sim$38\%  and  $\sim$11\% relative statistical error, respectively (upper left panels of Fig. \ref{fig_sigmaGmu}). Moreover, the relative statistical error associated with the measurement of $G\mu$ through the normalized cmd-$n$ decreased as the cosmic string tension increased (upper right panel of Fig. \ref{fig_sigmaGmu}). Including the non-Gaussian correction term to estimate the statistical relative error revealed a decrease in the lowest measurable $G\mu$ for a fixed relative error compared to the situation where the correction terms are neglected in Eq. \ref{nongaussskewness}) (lower panels of Fig. \ref{fig_sigmaGmu}).

The comparison of different studies with the same goal revealed different lower bounds on the detectability of methods. One may note that, in addition to the different proposed methodologies for detecting CSs on the CMB map, several parameters associated with cosmic string including the fraction of the CMB power spectrum by CSs, $f_{\ell=10}$, the intercommuting probability, the number of distinct strings at each Hubble volume as well as underlying theory for topological defects production,  result in different values of $G\mu_{\rm min}$ (Table \ref{tab:tab1} and Fig. \ref{fig.comparison}).

We can suggest performing the following tasks as complementary subjects to look for the CS network through a diverse range of observations, which can be tracers and can accommodate the smoking gun of topological phase transitions with weighted morphological measures: the 21cm map intensity
\citep{Brandenberger:2010hn, Hernandez:2011ym, Hernandez:2012qs,Pagano:2012cx,2013JCAP...02..045M,
    Hernandez:2014jda,2019JCAP...09..009B,2013JCAP...02..045M,2021JCAP...10..046T,2021MNRAS.508..408H}, large scale structures \citep{1991PhRvL..67.1057V,Ade:2013xla,2023PhRvD.108d3510J}, CMB polarization map \citep{2010PhRvD..82b3513D,2022JCAP...06..033Y}, examining the stochastic gravitational wave background produced by CSs network detected by pulsar timing array \citep{2023ApJ...951L..11A}. Computing the clustering of cmd-$n$ via the excess probability of finding pairs of features in the banner of weighted morphology, which is similar to what has been done in \citep{2021MNRAS.503..815V} is another interesting suggestion. A useful criterion via clustering of features known as ``exclusion zone'' introduced in \citep{2024MNRAS.528.1604S} is another idea for employing TPCF of the cmd-$n$.

Incorporating algebraic topology and computational geometry to
extract topological invariants such as $k$-holes and examining their
persistence through Persistent Homology for the CMB map distorted by
CSs is another intriguing direction for future research
\citep{2024JCAP...09..034Y,2024MNRAS.535..657J}. Mapping the CMB
fluctuation to a cosmic network with graph theory is also another
approach for further investigation of the CSs on the constructed network
\citep{barabasi2013network,latora2017complex,zou2019complex,hong2016discriminating,hong2020constraining}.
Combining the cmd-$n$ with the notion of the density split and
making multi-teasers to achieve sample variance cancelation could
result in a tighter bound on the $G\mu$ \citep{2025JCAP...01..026M}.
Dealing with observational data to put pristine constraints on the
CSs parameters can be accomplished through a novel method known as
likelihood-free analysis in the context of simulation-based
inference (SBI), incorporating summary statistics with the
cmd-$n$~\citep{tejero2020sbi,2016arXiv160506376P,2019MNRAS.488.4440A,2020PNAS..11730055C}.
Inspired by Field-level Bayesian inference and Implicit-likelihood
or simulation-based inference, an implicit likelihood learns for a
given $G\mu$ from simulation. Thus, the posterior distribution of the CS
tension is derived from the corresponding observational data
\citep{2024PhRvL.133v1006N}.

\section*{Acknowledgements}

The authors are grateful to C. Ringeval, F. R. Bouchet, and the Planck Collaboration, as they generously provided the simulations of the cosmic string maps used in this work. AA thanks Shahid Beheshti University for its hospitality.

\section*{Data Availability}

The new data generated and the computational program underlying this article will be shared on reasonable request to the corresponding author.

\appendix
\section{The cmd-$n$ for 3-dimensional field}
For a generic $(1+3)$-dimensional field denoted by $\delta(\mathbf{r})$, which possesses the characteristics $\langle \delta(\mathbf{r})\rangle=0$ and $\langle \delta^2(\mathbf{r})\rangle=\sigma_0^2$, we construct the associated
smoothed field in the Fourier space as:
\begin{equation}\label{eq:smoothed3}
    {\tilde{\delta}}_{\rm smoothed}(\mathbf{k};R)= \mathcal{W}(kR){\tilde{\delta}}(\mathbf{k})
\end{equation}
where $\mathcal{W}(kR)$ is smoothing window function and $R$ is smoothing scale. For the Gaussian and the Top-Hat window functions, we have: $\mathcal{W}_{\mathbb{G}}(kR)=\exp\left(-\frac{(kR)^2}{2}\right)$ and $\mathcal{W}_{\mathbb{T}}(kR)=3\frac{\sin(kR)-kR\cos(kR)}{(kR)^3}$, respectively. Now we turn to derive the cmd-$n$ for the  aforementioned field in weakly non-Gaussian and isotropic regimes, while taking into account the estimator as $\frac{1}{3}\sum_{i=1}^3 2\delta_D\left(\delta(\mathbf{r})-\vartheta\sigma_0\right)\Theta\left(\eta_{i}\right)\left(\eta_{i}\right)^n$. Therefore, we obtain:
\begin{align}
    &\langle\mathcal{N}_{\text{cmd},n}\rangle^{\mathbb{NG}}=\nonumber\\
    &\langle\mathcal{N}_{\text{cmd},n}\rangle^{\mathbb{G}}\times
    \Big[1 + \left( \frac{1}{6} S^{(0)} H_3 (\vartheta) +  \frac{n}{3} S^{(1)} H_1 (\vartheta) \right) \sigma_0 + \nonumber\\
    &\Big( \frac{1}{72} (S^{(0)})^2 H_6 (\vartheta) + \left( \frac{1}{24} K^{(0)} + \frac{n}{18} S^{(0)} S^{(1)} \right)  H_4 (\vartheta)\nonumber\\
    & + \left( \frac{n}{8} K^{(1)} + \frac{n(n-2)}{18} (S^{(1)})^2 \right) H_2 (\vartheta) - \nonumber\\
    &\frac{n(n-2)}{16} K^{(2)} H_0 (\vartheta) \Big) \sigma _0^2 + \mathcal{O} (\sigma_0^3)
    \Big]
\end{align}
The expression for the $\langle\mathcal{N}_{\text{cmd},n}\rangle^{\mathbb{G}}$ is clarified by Eq. (\ref{NcmdGauss}), where $\sigma_1(R)$ and $\sigma_0(R)$ are averaged over a smoothing scale, denoted by $R$, in 3-dimensional space. The corresponding smoothed $m$th spectral moment for the $(1+3)$-dimensional field is represented as follows:
\begin{align}\label{eq:spectra3}
    \sigma_m^2(R)=\dfrac{1}{(2\pi)^3}\int d^3\mathbf{k}\, k^{2m}\mathcal{W}^2(kR)P(k)
\end{align}
where $\langle {\tilde{\delta}}(\mathbf{k}){\tilde{\delta}}(\mathbf{k}') \rangle= (2\pi)^3\delta_d(\mathbf{k}+\mathbf{k}') P(k)$ is power spectrum. Additional cumulants are analogous to the definitions outlined in Eq. (\ref{eq:skewness1}) which must be computed for $(1+3)$-dimensional field.


%
%

    \bibliographystyle{mnras}

\begin{thebibliography}{}
\makeatletter
\relax
\def\mn@urlcharsother{\let\do\@makeother \do\$\do\&\do\#\do\^\do\_\do\%\do\~}
\def\mn@doi{\begingroup\mn@urlcharsother \@ifnextchar [ {\mn@doi@}
  {\mn@doi@[]}}
\def\mn@doi@[#1]#2{\def\@tempa{#1}\ifx\@tempa\@empty \href
  {http://dx.doi.org/#2} {doi:#2}\else \href {http://dx.doi.org/#2} {#1}\fi
  \endgroup}
\def\mn@eprint#1#2{\mn@eprint@#1:#2::\@nil}
\def\mn@eprint@arXiv#1{\href {http://arxiv.org/abs/#1} {{\tt arXiv:#1}}}
\def\mn@eprint@dblp#1{\href {http://dblp.uni-trier.de/rec/bibtex/#1.xml}
  {dblp:#1}}
\def\mn@eprint@#1:#2:#3:#4\@nil{\def\@tempa {#1}\def\@tempb {#2}\def\@tempc
  {#3}\ifx \@tempc \@empty \let \@tempc \@tempb \let \@tempb \@tempa \fi \ifx
  \@tempb \@empty \def\@tempb {arXiv}\fi \@ifundefined
  {mn@eprint@\@tempb}{\@tempb:\@tempc}{\expandafter \expandafter \csname
  mn@eprint@\@tempb\endcsname \expandafter{\@tempc}}}

\bibitem[\protect\citeauthoryear{{Abazajian} et~al.,}{{Abazajian}
  et~al.}{2016}]{2016arXiv161002743A}
{Abazajian} K.~N.,  et~al., 2016, \mn@doi [arXiv e-prints]
  {10.48550/arXiv.1610.02743}, \href
  {https://ui.adsabs.harvard.edu/abs/2016arXiv161002743A} {p. arXiv:1610.02743}

\bibitem[\protect\citeauthoryear{{Abazajian} et~al.,}{{Abazajian}
  et~al.}{2019}]{2019arXiv190704473A}
{Abazajian} K.,  et~al., 2019, \mn@doi [arXiv e-prints]
  {10.48550/arXiv.1907.04473}, \href
  {https://ui.adsabs.harvard.edu/abs/2019arXiv190704473A} {p. arXiv:1907.04473}

\bibitem[\protect\citeauthoryear{{Abbott} et~al.,}{{Abbott}
  et~al.}{2021a}]{2021PhRvD.104j2001A}
{Abbott} R.,  et~al., 2021a, \mn@doi [\prd] {10.1103/PhysRevD.104.102001},
  \href {https://ui.adsabs.harvard.edu/abs/2021PhRvD.104j2001A} {104, 102001}

\bibitem[\protect\citeauthoryear{{Abbott} et~al.,}{{Abbott}
  et~al.}{2021b}]{LIGOScientific:2021nrg}
{Abbott} R.,  et~al., 2021b, \mn@doi [\prl] {10.1103/PhysRevLett.126.241102},
  \href {https://ui.adsabs.harvard.edu/abs/2021PhRvL.126x1102A} {126, 241102}

\bibitem[\protect\citeauthoryear{{Abitbol} et~al.,}{{Abitbol}
  et~al.}{2017}]{2017arXiv170602464A}
{Abitbol} M.~H.,  et~al., 2017, \mn@doi [arXiv e-prints]
  {10.48550/arXiv.1706.02464}, \href
  {https://ui.adsabs.harvard.edu/abs/2017arXiv170602464A} {p. arXiv:1706.02464}

\bibitem[\protect\citeauthoryear{{Ade} et~al.,}{{Ade}
  et~al.}{2019}]{2019JCAP...02..056A}
{Ade} P.,  et~al., 2019, \mn@doi [\jcap] {10.1088/1475-7516/2019/02/056}, \href
  {https://ui.adsabs.harvard.edu/abs/2019JCAP...02..056A} {2019, 056}

\bibitem[\protect\citeauthoryear{{Ade} et~al.,}{{Ade}
  et~al.}{2022}]{2022ApJ...927..174A}
{Ade} P.~A.~R.,  et~al., 2022, \mn@doi [\apj] {10.3847/1538-4357/ac20df}, \href
  {https://ui.adsabs.harvard.edu/abs/2022ApJ...927..174A} {927, 174}

\bibitem[\protect\citeauthoryear{Adler}{Adler}{2010}]{adler81}
Adler R.~J.,  2010, The Geometry of Random Fields.
Society for Industrial and Applied Mathematics,
  \mn@doi{10.1137/1.9780898718980}, \url
  {https://epubs.siam.org/doi/abs/10.1137/1.9780898718980}

\bibitem[\protect\citeauthoryear{Adler \& Taylor}{Adler \&
  Taylor}{2011}]{adler2011topological}
Adler R.~J.,  Taylor J.~E.,  2011, Topological complexity of smooth random
  functions: {\'E}cole d'{\'E}t{\'e} de Probabilit{\'e}s de Saint-Flour
  XXXIX-2009.
Lecture notes in mathematics, Springer, Berlin, Heidelberg

\bibitem[\protect\citeauthoryear{Adler, Bobrowski, Borman, Subag  \&
  Weinberger}{Adler et~al.}{2010}]{adler2010persistent}
Adler R.~J.,  Bobrowski O.,  Borman M.~S.,  Subag E.,   Weinberger S.,  2010,
  in , Institute of {Mathematical} {Statistics} {Collections}.
Institute of Mathematical Statistics, Beachwood, Ohio, USA, pp 124--143,
  \mn@doi{10.1214/10-IMSCOLL609}, \url
  {http://projecteuclid.org/euclid.imsc/1288099016}

\bibitem[\protect\citeauthoryear{{Afzal} et~al.,}{{Afzal}
  et~al.}{2023}]{2023ApJ...951L..11A}
{Afzal} A.,  et~al., 2023, \mn@doi [\apjl] {10.3847/2041-8213/acdc91}, \href
  {https://ui.adsabs.harvard.edu/abs/2023ApJ...951L..11A} {951, L11}

\bibitem[\protect\citeauthoryear{{Afzal}, {Shafi}  \& {Tiwari}}{{Afzal}
  et~al.}{2024}]{2024PhLB..85038516A}
{Afzal} A.,  {Shafi} Q.,   {Tiwari} A.,  2024, \mn@doi [Physics Letters B]
  {10.1016/j.physletb.2024.138516}, \href
  {https://ui.adsabs.harvard.edu/abs/2024PhLB..85038516A} {850, 138516}

\bibitem[\protect\citeauthoryear{{Agazie} et~al.,}{{Agazie}
  et~al.}{2023}]{2023ApJ...951L...8A}
{Agazie} G.,  et~al., 2023, \mn@doi [\apjl] {10.3847/2041-8213/acdac6}, \href
  {https://ui.adsabs.harvard.edu/abs/2023ApJ...951L...8A} {951, L8}

\bibitem[\protect\citeauthoryear{{Ahmed}, {Kavic}, {Liebling}, {Lippert},
  {Mian}  \& {Simonetti}}{{Ahmed} et~al.}{2024}]{2024arXiv240704743A}
{Ahmed} S.,  {Kavic} M.~J.,  {Liebling} S.~L.,  {Lippert} M.,  {Mian} M.,
  {Simonetti} J.,  2024, \mn@doi [arXiv e-prints] {10.48550/arXiv.2407.04743},
  \href {https://ui.adsabs.harvard.edu/abs/2024arXiv240704743A} {p.
  arXiv:2407.04743}

\bibitem[\protect\citeauthoryear{{Aiola} et~al.,}{{Aiola}
  et~al.}{2020}]{2020JCAP...12..047A}
{Aiola} S.,  et~al., 2020, \mn@doi [\jcap] {10.1088/1475-7516/2020/12/047},
  \href {https://ui.adsabs.harvard.edu/abs/2020JCAP...12..047A} {2020, 047}

\bibitem[\protect\citeauthoryear{Alesker}{Alesker}{1999}]{Alesker1999}
Alesker S.,  1999, Geometriae Dedicata, 74, 241

\bibitem[\protect\citeauthoryear{{Allen}, {Caldwell}, {Dodelson}, {Knox},
  {Shellard}  \& {Stebbins}}{{Allen} et~al.}{1997}]{Allen:1997ag}
{Allen} B.,  {Caldwell} R.~R.,  {Dodelson} S.,  {Knox} L.,  {Shellard}
  E.~P.~S.,   {Stebbins} A.,  1997, \mn@doi [\prl]
  {10.1103/PhysRevLett.79.2624}, \href
  {https://ui.adsabs.harvard.edu/abs/1997PhRvL..79.2624A} {79, 2624}

\bibitem[\protect\citeauthoryear{{Alsing}, {Charnock}, {Feeney}  \&
  {Wandelt}}{{Alsing} et~al.}{2019}]{2019MNRAS.488.4440A}
{Alsing} J.,  {Charnock} T.,  {Feeney} S.,   {Wandelt} B.,  2019, \mn@doi
  [\mnras] {10.1093/mnras/stz1960}, \href
  {https://ui.adsabs.harvard.edu/abs/2019MNRAS.488.4440A} {488, 4440}

\bibitem[\protect\citeauthoryear{{Aluri} et~al.,}{{Aluri}
  et~al.}{2023}]{2023CQGra..40i4001A}
{Aluri} P.~k.,  et~al., 2023, \mn@doi [Classical and Quantum Gravity]
  {10.1088/1361-6382/acbefc}, \href
  {https://ui.adsabs.harvard.edu/abs/2023CQGra..40i4001A} {40, 094001}

\bibitem[\protect\citeauthoryear{{Amaro-Seoane} et~al.,}{{Amaro-Seoane}
  et~al.}{2017}]{2017arXiv170200786A}
{Amaro-Seoane} P.,  et~al., 2017, \mn@doi [arXiv e-prints]
  {10.48550/arXiv.1702.00786}, \href
  {https://ui.adsabs.harvard.edu/abs/2017arXiv170200786A} {p. arXiv:1702.00786}

\bibitem[\protect\citeauthoryear{{Amsel}, {Berger}  \& {Brandenberger}}{{Amsel}
  et~al.}{2008}]{Amsel:2007ki}
{Amsel} S.,  {Berger} J.,   {Brandenberger} R.~H.,  2008, \mn@doi [\jcap]
  {10.1088/1475-7516/2008/04/015}, \href
  {https://ui.adsabs.harvard.edu/abs/2008JCAP...04..015A} {2008, 015}

\bibitem[\protect\citeauthoryear{{Appleby}, {Chingangbam}, {Park}, {Hong},
  {Kim}  \& {Ganesan}}{{Appleby} et~al.}{2018}]{appleby2018minkowski}
{Appleby} S.,  {Chingangbam} P.,  {Park} C.,  {Hong} S.~E.,  {Kim} J.,
  {Ganesan} V.,  2018, \mn@doi [\apj] {10.3847/1538-4357/aabb53}, \href
  {https://ui.adsabs.harvard.edu/abs/2018ApJ...858...87A} {858, 87}

\bibitem[\protect\citeauthoryear{{Appleby}, {Kochappan}, {Chingangbam}  \&
  {Park}}{{Appleby} et~al.}{2019}]{appleby2019ensemble}
{Appleby} S.,  {Kochappan} J.~P.,  {Chingangbam} P.,   {Park} C.,  2019,
  \mn@doi [\apj] {10.3847/1538-4357/ab5057}, \href
  {https://ui.adsabs.harvard.edu/abs/2019ApJ...887..128A} {887, 128}

\bibitem[\protect\citeauthoryear{{Appleby}, {Kochappan}, {Chingangbam}  \&
  {Park}}{{Appleby} et~al.}{2023}]{Appleby2023}
{Appleby} S.,  {Kochappan} J.~P.,  {Chingangbam} P.,   {Park} C.,  2023,
  \mn@doi [\apj] {10.3847/1538-4357/aca530}, \href
  {https://ui.adsabs.harvard.edu/abs/2023ApJ...942..110A} {942, 110}

\bibitem[\protect\citeauthoryear{{Arzoumanian} et~al.,}{{Arzoumanian}
  et~al.}{2020}]{2020ApJ...905L..34A}
{Arzoumanian} Z.,  et~al., 2020, \mn@doi [\apjl] {10.3847/2041-8213/abd401},
  \href {https://ui.adsabs.harvard.edu/abs/2020ApJ...905L..34A} {905, L34}

\bibitem[\protect\citeauthoryear{{Barabasi}}{{Barabasi}}{2013}]{barabasi2013network}
{Barabasi} A.~L.,  2013, \mn@doi [Philosophical Transactions of the Royal
  Society of London Series A] {10.1098/rsta.2012.0375}, \href
  {https://ui.adsabs.harvard.edu/abs/2013RSPTA.37120375B} {371, 20120375}

\bibitem[\protect\citeauthoryear{{Bardeen}, {Bond}, {Kaiser}  \&
  {Szalay}}{{Bardeen} et~al.}{1986}]{Bardeen:1985tr}
{Bardeen} J.~M.,  {Bond} J.~R.,  {Kaiser} N.,   {Szalay} A.~S.,  1986, \mn@doi
  [\apj] {10.1086/164143}, \href
  {https://ui.adsabs.harvard.edu/abs/1986ApJ...304...15B} {304, 15}

\bibitem[\protect\citeauthoryear{{Battye} \& {Moss}}{{Battye} \&
  {Moss}}{2010}]{Battye:2010xz}
{Battye} R.,  {Moss} A.,  2010, \mn@doi [\prd] {10.1103/PhysRevD.82.023521},
  \href {https://ui.adsabs.harvard.edu/abs/2010PhRvD..82b3521B} {82, 023521}

\bibitem[\protect\citeauthoryear{{Beisbart}, {Buchert}  \& {Wagner}}{{Beisbart}
  et~al.}{2001}]{2001PhyA..293..592B}
{Beisbart} C.,  {Buchert} T.,   {Wagner} H.,  2001, \mn@doi [Physica A
  Statistical Mechanics and its Applications] {10.1016/S0378-4371(00)00612-9},
  \href {https://ui.adsabs.harvard.edu/abs/2001PhyA..293..592B} {293, 592}

\bibitem[\protect\citeauthoryear{{Beisbart}, {Dahlke}, {Mecke}  \&
  {Wagner}}{{Beisbart} et~al.}{2002}]{Beisbart2002}
{Beisbart} C.,  {Dahlke} R.,  {Mecke} K.,   {Wagner} H.,  2002, in {Mecke} K.,
  {Stoyan} D.,  eds, , Vol.~600, Morphology of Condensed Matter.
pp 238--260, \mn@doi{10.48550/arXiv.physics/0203072}

\bibitem[\protect\citeauthoryear{{Bennett} \& {Bouchet}}{{Bennett} \&
  {Bouchet}}{1990}]{Bennett:1990}
{Bennett} D.~P.,  {Bouchet} F.~R.,  1990, \mn@doi [\prd]
  {10.1103/PhysRevD.41.2408}, \href
  {https://ui.adsabs.harvard.edu/abs/1990PhRvD..41.2408B} {41, 2408}

\bibitem[\protect\citeauthoryear{{Bernardeau}, {Colombi}, {Gazta{\~n}aga}  \&
  {Scoccimarro}}{{Bernardeau} et~al.}{2002}]{Bernardeau:2001qr}
{Bernardeau} F.,  {Colombi} S.,  {Gazta{\~n}aga} E.,   {Scoccimarro} R.,  2002,
  \mn@doi [\physrep] {10.1016/S0370-1573(02)00135-7}, \href
  {https://ui.adsabs.harvard.edu/abs/2002PhR...367....1B} {367, 1}

\bibitem[\protect\citeauthoryear{{Bevis}, {Hindmarsh}, {Kunz}  \&
  {Urrestilla}}{{Bevis} et~al.}{2007a}]{Bevis:2006m}
{Bevis} N.,  {Hindmarsh} M.,  {Kunz} M.,   {Urrestilla} J.,  2007a, \mn@doi
  [\prd] {10.1103/PhysRevD.75.065015}, \href
  {https://ui.adsabs.harvard.edu/abs/2007PhRvD..75f5015B} {75, 065015}

\bibitem[\protect\citeauthoryear{{Bevis}, {Hindmarsh}, {Kunz}  \&
  {Urrestilla}}{{Bevis} et~al.}{2007b}]{2007PhRvD..76d3005B}
{Bevis} N.,  {Hindmarsh} M.,  {Kunz} M.,   {Urrestilla} J.,  2007b, \mn@doi
  [\prd] {10.1103/PhysRevD.76.043005}, \href
  {https://ui.adsabs.harvard.edu/abs/2007PhRvD..76d3005B} {76, 043005}

\bibitem[\protect\citeauthoryear{{Bevis}, {Hindmarsh}, {Kunz}  \&
  {Urrestilla}}{{Bevis} et~al.}{2008}]{Bevis:2007gh}
{Bevis} N.,  {Hindmarsh} M.,  {Kunz} M.,   {Urrestilla} J.,  2008, \mn@doi
  [\prl] {10.1103/PhysRevLett.100.021301}, \href
  {https://ui.adsabs.harvard.edu/abs/2008PhRvL.100b1301B} {100, 021301}

\bibitem[\protect\citeauthoryear{{Bevis}, {Hindmarsh}, {Kunz}  \&
  {Urrestilla}}{{Bevis} et~al.}{2010}]{Bevis:2010gj}
{Bevis} N.,  {Hindmarsh} M.,  {Kunz} M.,   {Urrestilla} J.,  2010, \mn@doi
  [\prd] {10.1103/PhysRevD.82.065004}, \href
  {https://ui.adsabs.harvard.edu/abs/2010PhRvD..82f5004B} {82, 065004}

\bibitem[\protect\citeauthoryear{{Blamart}, {Fronenberg}  \&
  {Brandenberger}}{{Blamart} et~al.}{2022}]{2022JCAP...11..012B}
{Blamart} M.,  {Fronenberg} H.,   {Brandenberger} R.,  2022, \mn@doi [\jcap]
  {10.1088/1475-7516/2022/11/012}, \href
  {https://ui.adsabs.harvard.edu/abs/2022JCAP...11..012B} {2022, 012}

\bibitem[\protect\citeauthoryear{{Blanco-Pillado} \& {Olum}}{{Blanco-Pillado}
  \& {Olum}}{2017}]{Blanco-Pillado:2017oxo}
{Blanco-Pillado} J.~J.,  {Olum} K.~D.,  2017, \mn@doi [\prd]
  {10.1103/PhysRevD.96.104046}, \href
  {https://ui.adsabs.harvard.edu/abs/2017PhRvD..96j4046B} {96, 104046}

\bibitem[\protect\citeauthoryear{{Blanco-Pillado}, {Olum}  \&
  {Shlaer}}{{Blanco-Pillado} et~al.}{2011}]{2011PhRvD..83h3514B}
{Blanco-Pillado} J.~J.,  {Olum} K.~D.,   {Shlaer} B.,  2011, \mn@doi [\prd]
  {10.1103/PhysRevD.83.083514}, \href
  {https://ui.adsabs.harvard.edu/abs/2011PhRvD..83h3514B} {83, 083514}

\bibitem[\protect\citeauthoryear{{Blanco-Pillado}, {Olum}  \&
  {Shlaer}}{{Blanco-Pillado} et~al.}{2015}]{2015PhRvD..92f3528B}
{Blanco-Pillado} J.~J.,  {Olum} K.~D.,   {Shlaer} B.,  2015, \mn@doi [\prd]
  {10.1103/PhysRevD.92.063528}, \href
  {https://ui.adsabs.harvard.edu/abs/2015PhRvD..92f3528B} {92, 063528}

\bibitem[\protect\citeauthoryear{{Blanco-Pillado}, {Olum}  \&
  {Siemens}}{{Blanco-Pillado} et~al.}{2018}]{Blanco-Pillado:2017rnf}
{Blanco-Pillado} J.~J.,  {Olum} K.~D.,   {Siemens} X.,  2018, \mn@doi [Physics
  Letters B] {10.1016/j.physletb.2018.01.050}, \href
  {https://ui.adsabs.harvard.edu/abs/2018PhLB..778..392B} {778, 392}

\bibitem[\protect\citeauthoryear{{Blanco-Pillado}, {Olum}  \&
  {Wachter}}{{Blanco-Pillado} et~al.}{2021}]{2021PhRvD.103j3512B}
{Blanco-Pillado} J.~J.,  {Olum} K.~D.,   {Wachter} J.~M.,  2021, \mn@doi [\prd]
  {10.1103/PhysRevD.103.103512}, \href
  {https://ui.adsabs.harvard.edu/abs/2021PhRvD.103j3512B} {103, 103512}

\bibitem[\protect\citeauthoryear{{Bond} \& {Efstathiou}}{{Bond} \&
  {Efstathiou}}{1987}]{bond1987statistics}
{Bond} J.~R.,  {Efstathiou} G.,  1987, \mn@doi [\mnras]
  {10.1093/mnras/226.3.655}, \href
  {https://ui.adsabs.harvard.edu/abs/1987MNRAS.226..655B} {226, 655}

\bibitem[\protect\citeauthoryear{{Bouchet}, {Bennett}  \& {Stebbins}}{{Bouchet}
  et~al.}{1988}]{Bouchet:1988hh}
{Bouchet} F.~R.,  {Bennett} D.~P.,   {Stebbins} A.,  1988, \mn@doi [\nat]
  {10.1038/335410a0}, \href
  {https://ui.adsabs.harvard.edu/abs/1988Natur.335..410B} {335, 410}

\bibitem[\protect\citeauthoryear{{Brandenberger}}{{Brandenberger}}{2014}]{2014NuPhS.246...45B}
{Brandenberger} R.~H.,  2014, \mn@doi [Nuclear Physics B Proceedings
  Supplements] {10.1016/j.nuclphysbps.2013.10.064}, \href
  {https://ui.adsabs.harvard.edu/abs/2014NuPhS.246...45B} {246, 45}

\bibitem[\protect\citeauthoryear{{Brandenberger}, {Danos}, {Hern{\'a}ndez}  \&
  {Holder}}{{Brandenberger} et~al.}{2010}]{Brandenberger:2010hn}
{Brandenberger} R.~H.,  {Danos} R.~J.,  {Hern{\'a}ndez} O.~F.,   {Holder}
  G.~P.,  2010, \mn@doi [\jcap] {10.1088/1475-7516/2010/12/028}, \href
  {https://ui.adsabs.harvard.edu/abs/2010JCAP...12..028B} {2010, 028}

\bibitem[\protect\citeauthoryear{{Brandenberger}, {Cyr}  \&
  {Shi}}{{Brandenberger} et~al.}{2019}]{2019JCAP...09..009B}
{Brandenberger} R.,  {Cyr} B.,   {Shi} R.,  2019, \mn@doi [\jcap]
  {10.1088/1475-7516/2019/09/009}, \href
  {https://ui.adsabs.harvard.edu/abs/2019JCAP...09..009B} {2019, 009}

\bibitem[\protect\citeauthoryear{{Caprini}, {Figueroa}, {Flauger}, {Nardini},
  {Peloso}, {Pieroni}, {Ricciardone}  \& {Tasinato}}{{Caprini}
  et~al.}{2019}]{2019JCAP...11..017C}
{Caprini} C.,  {Figueroa} D.~G.,  {Flauger} R.,  {Nardini} G.,  {Peloso} M.,
  {Pieroni} M.,  {Ricciardone} A.,   {Tasinato} G.,  2019, \mn@doi [\jcap]
  {10.1088/1475-7516/2019/11/017}, \href
  {https://ui.adsabs.harvard.edu/abs/2019JCAP...11..017C} {2019, 017}

\bibitem[\protect\citeauthoryear{{Carones}, {Carr{\'o}nDuque}, {Marinucci},
  {Migliaccio}  \& {Vittorio}}{{Carones} et~al.}{2024}]{2024MNRAS.527..756C}
{Carones} A.,  {Carr{\'o}nDuque} J.,  {Marinucci} D.,  {Migliaccio} M.,
  {Vittorio} N.,  2024, \mn@doi [\mnras] {10.1093/mnras/stad3002}, \href
  {https://ui.adsabs.harvard.edu/abs/2024MNRAS.527..756C} {527, 756}

\bibitem[\protect\citeauthoryear{{Carr{\'o}n Duque}, {Carones}, {Marinucci},
  {Migliaccio}  \& {Vittorio}}{{Carr{\'o}n Duque}
  et~al.}{2024}]{2024JCAP...01..039C}
{Carr{\'o}n Duque} J.,  {Carones} A.,  {Marinucci} D.,  {Migliaccio} M.,
  {Vittorio} N.,  2024, \mn@doi [\jcap] {10.1088/1475-7516/2024/01/039}, \href
  {https://ui.adsabs.harvard.edu/abs/2024JCAP...01..039C} {2024, 039}

\bibitem[\protect\citeauthoryear{{Charnock}, {Avgoustidis}, {Copeland}  \&
  {Moss}}{{Charnock} et~al.}{2016}]{Charnock:2016nzm}
{Charnock} T.,  {Avgoustidis} A.,  {Copeland} E.~J.,   {Moss} A.,  2016,
  \mn@doi [\prd] {10.1103/PhysRevD.93.123503}, \href
  {https://ui.adsabs.harvard.edu/abs/2016PhRvD..93l3503C} {93, 123503}

\bibitem[\protect\citeauthoryear{{Christiansen}, {Albin}, {Fletcher},
  {Goldman}, {Teng}, {Foley}  \& {Smoot}}{{Christiansen}
  et~al.}{2011}]{Christiansen:2010zi}
{Christiansen} J.~L.,  {Albin} E.,  {Fletcher} T.,  {Goldman} J.,  {Teng}
  I.~P.~W.,  {Foley} M.,   {Smoot} G.~F.,  2011, \mn@doi [\prd]
  {10.1103/PhysRevD.83.122004}, \href
  {https://ui.adsabs.harvard.edu/abs/2011PhRvD..83l2004C} {83, 122004}

\bibitem[\protect\citeauthoryear{{Ciuca} \& {Hern{\'a}ndez}}{{Ciuca} \&
  {Hern{\'a}ndez}}{2017}]{Ciuca:2017wrk}
{Ciuca} R.,  {Hern{\'a}ndez} O.~F.,  2017, \mn@doi [\jcap]
  {10.1088/1475-7516/2017/08/028}, \href
  {https://ui.adsabs.harvard.edu/abs/2017JCAP...08..028C} {2017, 028}

\bibitem[\protect\citeauthoryear{{Ciuca}, {Hern{\'a}ndez}  \& {Wolman}}{{Ciuca}
  et~al.}{2019}]{Ciuca:2017gca}
{Ciuca} R.,  {Hern{\'a}ndez} O.~F.,   {Wolman} M.,  2019, \mn@doi [\mnras]
  {10.1093/mnras/stz491}, \href
  {https://ui.adsabs.harvard.edu/abs/2019MNRAS.485.1377C} {485, 1377}

\bibitem[\protect\citeauthoryear{{Codis}, {Pichon}, {Pogosyan}, {Bernardeau}
  \& {Matsubara}}{{Codis} et~al.}{2013}]{codis2013non}
{Codis} S.,  {Pichon} C.,  {Pogosyan} D.,  {Bernardeau} F.,   {Matsubara} T.,
  2013, \mn@doi [\mnras] {10.1093/mnras/stt1316}, \href
  {https://ui.adsabs.harvard.edu/abs/2013MNRAS.435..531C} {435, 531}

\bibitem[\protect\citeauthoryear{{Cooray} \& {Sheth}}{{Cooray} \&
  {Sheth}}{2002}]{Cooray:2002dia}
{Cooray} A.,  {Sheth} R.,  2002, \mn@doi [\physrep]
  {10.1016/S0370-1573(02)00276-4}, \href
  {https://ui.adsabs.harvard.edu/abs/2002PhR...372....1C} {372, 1}

\bibitem[\protect\citeauthoryear{{Copeland} \& {Kibble}}{{Copeland} \&
  {Kibble}}{2010}]{Copeland:2009ga}
{Copeland} E.~J.,  {Kibble} T.~W.~B.,  2010, \mn@doi [Proceedings of the Royal
  Society of London Series A] {10.1098/rspa.2009.0591}, \href
  {https://ui.adsabs.harvard.edu/abs/2010RSPSA.466..623C} {466, 623}

\bibitem[\protect\citeauthoryear{{Copeland}, {Liddle}, {Lyth}, {Stewart}  \&
  {Wands}}{{Copeland} et~al.}{1994}]{Copeland:1994vg}
{Copeland} E.~J.,  {Liddle} A.~R.,  {Lyth} D.~H.,  {Stewart} E.~D.,   {Wands}
  D.,  1994, \mn@doi [\prd] {10.1103/PhysRevD.49.6410}, \href
  {https://ui.adsabs.harvard.edu/abs/1994PhRvD..49.6410C} {49, 6410}

\bibitem[\protect\citeauthoryear{{Copeland}, {Myers}  \&
  {Polchinski}}{{Copeland} et~al.}{2004}]{Copeland:2003bj}
{Copeland} E.~J.,  {Myers} R.~C.,   {Polchinski} J.,  2004, \mn@doi [Journal of
  High Energy Physics] {10.1088/1126-6708/2004/06/013}, \href
  {https://ui.adsabs.harvard.edu/abs/2004JHEP...06..013C} {2004, 013}

\bibitem[\protect\citeauthoryear{{Cranmer}, {Brehmer}  \& {Louppe}}{{Cranmer}
  et~al.}{2020}]{2020PNAS..11730055C}
{Cranmer} K.,  {Brehmer} J.,   {Louppe} G.,  2020, \mn@doi [Proceedings of the
  National Academy of Science] {10.1073/pnas.1912789117}, \href
  {https://ui.adsabs.harvard.edu/abs/2020PNAS..11730055C} {117, 30055}

\bibitem[\protect\citeauthoryear{{Cyr}, {Jiao}  \& {Brandenberger}}{{Cyr}
  et~al.}{2022}]{2022MNRAS.517.2221C}
{Cyr} B.,  {Jiao} H.,   {Brandenberger} R.,  2022, \mn@doi [\mnras]
  {10.1093/mnras/stac1939}, \href
  {https://ui.adsabs.harvard.edu/abs/2022MNRAS.517.2221C} {517, 2221}

\bibitem[\protect\citeauthoryear{{Damour} \& {Vilenkin}}{{Damour} \&
  {Vilenkin}}{2001}]{2001PhRvD..64f4008D}
{Damour} T.,  {Vilenkin} A.,  2001, \mn@doi [\prd]
  {10.1103/PhysRevD.64.064008}, \href
  {https://ui.adsabs.harvard.edu/abs/2001PhRvD..64f4008D} {64, 064008}

\bibitem[\protect\citeauthoryear{{Damour} \& {Vilenkin}}{{Damour} \&
  {Vilenkin}}{2005}]{Damour:2004kw}
{Damour} T.,  {Vilenkin} A.,  2005, \mn@doi [\prd]
  {10.1103/PhysRevD.71.063510}, \href
  {https://ui.adsabs.harvard.edu/abs/2005PhRvD..71f3510D} {71, 063510}

\bibitem[\protect\citeauthoryear{{Danos} \& {Brandenberger}}{{Danos} \&
  {Brandenberger}}{2010a}]{2010IJMPD..19..183D}
{Danos} R.~J.,  {Brandenberger} R.~H.,  2010a, \mn@doi [International Journal
  of Modern Physics D] {10.1142/S0218271810016324}, \href
  {https://ui.adsabs.harvard.edu/abs/2010IJMPD..19..183D} {19, 183}

\bibitem[\protect\citeauthoryear{{Danos} \& {Brandenberger}}{{Danos} \&
  {Brandenberger}}{2010b}]{2010JCAP...02..033D}
{Danos} R.~J.,  {Brandenberger} R.~H.,  2010b, \mn@doi [\jcap]
  {10.1088/1475-7516/2010/02/033}, \href
  {https://ui.adsabs.harvard.edu/abs/2010JCAP...02..033D} {2010, 033}

\bibitem[\protect\citeauthoryear{{Danos}, {Brandenberger}  \& {Holder}}{{Danos}
  et~al.}{2010}]{2010PhRvD..82b3513D}
{Danos} R.~J.,  {Brandenberger} R.~H.,   {Holder} G.,  2010, \mn@doi [\prd]
  {10.1103/PhysRevD.82.023513}, \href
  {https://ui.adsabs.harvard.edu/abs/2010PhRvD..82b3513D} {82, 023513}

\bibitem[\protect\citeauthoryear{{Depies} \& {Hogan}}{{Depies} \&
  {Hogan}}{2007}]{Depies:2009im}
{Depies} M.~R.,  {Hogan} C.~J.,  2007, \mn@doi [\prd]
  {10.1103/PhysRevD.75.125006}, \href
  {https://ui.adsabs.harvard.edu/abs/2007PhRvD..75l5006D} {75, 125006}

\bibitem[\protect\citeauthoryear{{Desjacques}, {Jeong}  \&
  {Schmidt}}{{Desjacques} et~al.}{2018}]{desjacques2018large}
{Desjacques} V.,  {Jeong} D.,   {Schmidt} F.,  2018, \mn@doi [\physrep]
  {10.1016/j.physrep.2017.12.002}, \href
  {https://ui.adsabs.harvard.edu/abs/2018PhR...733....1D} {733, 1}

\bibitem[\protect\citeauthoryear{{Dewdney}, {Hall}, {Schilizzi}  \&
  {Lazio}}{{Dewdney} et~al.}{2009}]{5136190}
{Dewdney} P.~E.,  {Hall} P.~J.,  {Schilizzi} R.~T.,   {Lazio} T.~J.~L.~W.,
  2009, \mn@doi [IEEE Proceedings] {10.1109/JPROC.2009.2021005}, \href
  {https://ui.adsabs.harvard.edu/abs/2009IEEEP..97.1482D} {97, 1482}

\bibitem[\protect\citeauthoryear{{Dey} et~al.,}{{Dey} et~al.}{2019}]{Dey_2019}
{Dey} A.,  et~al., 2019, \mn@doi [\aj] {10.3847/1538-3881/ab089d}, \href
  {https://ui.adsabs.harvard.edu/abs/2019AJ....157..168D} {157, 168}

\bibitem[\protect\citeauthoryear{Dodelson \& Schmidt}{Dodelson \&
  Schmidt}{2020}]{dodelson2020modern}
Dodelson S.,  Schmidt F.,  2020, Modern Cosmology.
Elsevier Science, \url {https://books.google.com/books?id=GGjfywEACAAJ}

\bibitem[\protect\citeauthoryear{{Ducout}, {Bouchet}, {Colombi}, {Pogosyan}  \&
  {Prunet}}{{Ducout} et~al.}{2013}]{Ducout:2012it}
{Ducout} A.,  {Bouchet} F.~R.,  {Colombi} S.,  {Pogosyan} D.,   {Prunet} S.,
  2013, \mn@doi [\mnras] {10.1093/mnras/sts483}, \href
  {https://ui.adsabs.harvard.edu/abs/2013MNRAS.429.2104D} {429, 2104}

\bibitem[\protect\citeauthoryear{Durrer}{Durrer}{2020}]{durrer2020cosmic}
Durrer R.,  2020, The Cosmic Microwave Background.
Cambridge University Press

\bibitem[\protect\citeauthoryear{{Dvali} \& {Vilenkin}}{{Dvali} \&
  {Vilenkin}}{2004}]{Dvali:2003zj}
{Dvali} G.,  {Vilenkin} A.,  2004, \mn@doi [\jcap]
  {10.1088/1475-7516/2004/03/010}, \href
  {https://ui.adsabs.harvard.edu/abs/2004JCAP...03..010D} {2004, 010}

\bibitem[\protect\citeauthoryear{{Eifler} et~al.,}{{Eifler}
  et~al.}{2021}]{10.1093/mnras/stab1762}
{Eifler} T.,  et~al., 2021, \mn@doi [\mnras] {10.1093/mnras/stab1762}, \href
  {https://ui.adsabs.harvard.edu/abs/2021MNRAS.507.1746E} {507, 1746}

\bibitem[\protect\citeauthoryear{{Ellis} \& {Lewicki}}{{Ellis} \&
  {Lewicki}}{2021}]{2021PhRvL.126d1304E}
{Ellis} J.,  {Lewicki} M.,  2021, \mn@doi [\prl]
  {10.1103/PhysRevLett.126.041304}, \href
  {https://ui.adsabs.harvard.edu/abs/2021PhRvL.126d1304E} {126, 041304}

\bibitem[\protect\citeauthoryear{{Essinger-Hileman} et~al.,}{{Essinger-Hileman}
  et~al.}{2014}]{10.1117/12.2056701}
{Essinger-Hileman} T.,  et~al., 2014, in {Holland} W.~S.,  {Zmuidzinas} J.,
  eds,  Society of Photo-Optical Instrumentation Engineers (SPIE) Conference
  Series Vol. 9153, Millimeter, Submillimeter, and Far-Infrared Detectors and
  Instrumentation for Astronomy VII. p. 91531I (\mn@eprint {arXiv}
  {1408.4788}), \mn@doi{10.1117/12.2056701}

\bibitem[\protect\citeauthoryear{{Euclid Collaboration} et~al.,}{{Euclid
  Collaboration} et~al.}{2020}]{2020A&A...642A.191E}
{Euclid Collaboration} et~al., 2020, \mn@doi [\aap]
  {10.1051/0004-6361/202038071}, \href
  {https://ui.adsabs.harvard.edu/abs/2020A&A...642A.191E} {642, A191}

\bibitem[\protect\citeauthoryear{{Fraisse}, {Ringeval}, {Spergel}  \&
  {Bouchet}}{{Fraisse} et~al.}{2008}]{Fraisse:2007nu}
{Fraisse} A.~A.,  {Ringeval} C.,  {Spergel} D.~N.,   {Bouchet} F.~R.,  2008,
  \mn@doi [\prd] {10.1103/PhysRevD.78.043535}, \href
  {https://ui.adsabs.harvard.edu/abs/2008PhRvD..78d3535F} {78, 043535}

\bibitem[\protect\citeauthoryear{{Ganesan} \& {Chingangbam}}{{Ganesan} \&
  {Chingangbam}}{2017}]{2017JCAP...06..023G}
{Ganesan} V.,  {Chingangbam} P.,  2017, \mn@doi [\jcap]
  {10.1088/1475-7516/2017/06/023}, \href
  {https://ui.adsabs.harvard.edu/abs/2017JCAP...06..023G} {2017, 023}

\bibitem[\protect\citeauthoryear{{Gay}, {Pichon}  \& {Pogosyan}}{{Gay}
  et~al.}{2012}]{gay2012non}
{Gay} C.,  {Pichon} C.,   {Pogosyan} D.,  2012, \mn@doi [\prd]
  {10.1103/PhysRevD.85.023011}, \href
  {https://ui.adsabs.harvard.edu/abs/2012PhRvD..85b3011G} {85, 023011}

\bibitem[\protect\citeauthoryear{{Gelman}, {Carlin}, {Stern}, {Dunson},
  {Vehtari}  \& {Rubin}}{{Gelman} et~al.}{2014}]{2014bda..book.....G}
{Gelman} A.,  {Carlin} J.~B.,  {Stern} H.~S.,  {Dunson} D.~B.,  {Vehtari} A.,
  {Rubin} D.~B.,  2014, {Bayesian Data Analysis}

\bibitem[\protect\citeauthoryear{{Goldenfeld}}{{Goldenfeld}}{2018}]{2018lopt.book.....G}
{Goldenfeld} N.,  2018, {Lectures On Phase Transitions And The Renormalization
  Group}

\bibitem[\protect\citeauthoryear{{Gong} et~al.,}{{Gong}
  et~al.}{2019}]{2019ApJ883}
{Gong} Y.,  et~al., 2019, \mn@doi [Astrophysical Journal]
  {10.3847/1538-4357/ab391e}, \href
  {https://ui.adsabs.harvard.edu/abs/2019ApJ...883..203G} {883, 203}

\bibitem[\protect\citeauthoryear{{Gott}~III}{{Gott}~III}{1985}]{Gott:1985}
{Gott}~III J.~R.,  1985, \mn@doi [\apj] {10.1086/162808}, \href
  {http://adsabs.harvard.edu/abs/1985ApJ...288..422G} {288, 422}

\bibitem[\protect\citeauthoryear{{Griffin}, {Lilienblum}, {Delaney}, {Kumagai},
  {Fiebig}  \& {Spaldin}}{{Griffin} et~al.}{2012}]{2012PhRvX...2d1022G}
{Griffin} S.~M.,  {Lilienblum} M.,  {Delaney} K.~T.,  {Kumagai} Y.,  {Fiebig}
  M.,   {Spaldin} N.~A.,  2012, \mn@doi [Physical Review X]
  {10.1103/PhysRevX.2.041022}, \href
  {https://ui.adsabs.harvard.edu/abs/2012PhRvX...2d1022G} {2, 041022}

\bibitem[\protect\citeauthoryear{{Guth}}{{Guth}}{1981}]{Guth:1980zm}
{Guth} A.~H.,  1981, \mn@doi [\prd] {10.1103/PhysRevD.23.347}, \href
  {https://ui.adsabs.harvard.edu/abs/1981PhRvD..23..347G} {23, 347}

\bibitem[\protect\citeauthoryear{{Hahn}, {Abidi}, {Eickenberg}, {Ho}, {Lemos},
  {Massara}, {Moradinezhad Dizgah}  \& {R{\'e}galdo-Saint Blancard}}{{Hahn}
  et~al.}{2022}]{2022mla..confE..24H}
{Hahn} C.,  {Abidi} M.,  {Eickenberg} M.,  {Ho} S.,  {Lemos} P.,  {Massara} E.,
   {Moradinezhad Dizgah} A.,   {R{\'e}galdo-Saint Blancard} B.,  2022, in
  Machine Learning for Astrophysics. p.~24

\bibitem[\protect\citeauthoryear{{Hammond}, {Wiaux}  \&
  {Vandergheynst}}{{Hammond} et~al.}{2009}]{Hammond:2008fg}
{Hammond} D.~K.,  {Wiaux} Y.,   {Vandergheynst} P.,  2009, \mn@doi [\mnras]
  {10.1111/j.1365-2966.2009.14978.x}, \href
  {https://ui.adsabs.harvard.edu/abs/2009MNRAS.398.1317H} {398, 1317}

\bibitem[\protect\citeauthoryear{{Hanany} et~al.,}{{Hanany}
  et~al.}{2019}]{2019arXiv190210541H}
{Hanany} S.,  et~al., 2019, \mn@doi [arXiv e-prints]
  {10.48550/arXiv.1902.10541}, \href
  {https://ui.adsabs.harvard.edu/abs/2019arXiv190210541H} {p. arXiv:1902.10541}

\bibitem[\protect\citeauthoryear{{Hergt}, {Amara}, {Brandenberger}, {Kacprzak}
  \& {R{\'e}fr{\'e}gier}}{{Hergt} et~al.}{2017}]{Hergt:2016xup}
{Hergt} L.,  {Amara} A.,  {Brandenberger} R.,  {Kacprzak} T.,
  {R{\'e}fr{\'e}gier} A.,  2017, \mn@doi [\jcap]
  {10.1088/1475-7516/2017/06/004}, \href
  {https://ui.adsabs.harvard.edu/abs/2017JCAP...06..004H} {2017, 004}

\bibitem[\protect\citeauthoryear{{Hern{\'a}ndez}}{{Hern{\'a}ndez}}{2014}]{Hernandez:2014jda}
{Hern{\'a}ndez} O.~F.,  2014, \mn@doi [\prd] {10.1103/PhysRevD.90.123504},
  \href {https://ui.adsabs.harvard.edu/abs/2014PhRvD..90l3504H} {90, 123504}

\bibitem[\protect\citeauthoryear{{Hern{\'a}ndez}}{{Hern{\'a}ndez}}{2021}]{2021MNRAS.508..408H}
{Hern{\'a}ndez} O.~F.,  2021, \mn@doi [\mnras] {10.1093/mnras/stab2634}, \href
  {https://ui.adsabs.harvard.edu/abs/2021MNRAS.508..408H} {508, 408}

\bibitem[\protect\citeauthoryear{{Hern{\'a}ndez} \&
  {Brandenberger}}{{Hern{\'a}ndez} \& {Brandenberger}}{2012}]{Hernandez:2012qs}
{Hern{\'a}ndez} O.~F.,  {Brandenberger} R.~H.,  2012, \mn@doi [\jcap]
  {10.1088/1475-7516/2012/07/032}, \href
  {https://ui.adsabs.harvard.edu/abs/2012JCAP...07..032H} {2012, 032}

\bibitem[\protect\citeauthoryear{{Hern{\'a}ndez}, {Wang}, {Brandenberger}  \&
  {Fong}}{{Hern{\'a}ndez} et~al.}{2011}]{Hernandez:2011ym}
{Hern{\'a}ndez} O.~F.,  {Wang} Y.,  {Brandenberger} R.,   {Fong} J.,  2011,
  \mn@doi [\jcap] {10.1088/1475-7516/2011/08/014}, \href
  {https://ui.adsabs.harvard.edu/abs/2011JCAP...08..014H} {2011, 014}

\bibitem[\protect\citeauthoryear{{Hikage}, {Komatsu}  \& {Matsubara}}{{Hikage}
  et~al.}{2006}]{hikage2006primordial}
{Hikage} C.,  {Komatsu} E.,   {Matsubara} T.,  2006, \mn@doi [\apj]
  {10.1086/508653}, \href
  {https://ui.adsabs.harvard.edu/abs/2006ApJ...653...11H} {653, 11}

\bibitem[\protect\citeauthoryear{{Hindmarsh}}{{Hindmarsh}}{1994}]{Hindmarsh:1993pu}
{Hindmarsh} M.,  1994, \mn@doi [\apj] {10.1086/174505}, \href
  {https://ui.adsabs.harvard.edu/abs/1994ApJ...431..534H} {431, 534}

\bibitem[\protect\citeauthoryear{{Hindmarsh} \& {Kibble}}{{Hindmarsh} \&
  {Kibble}}{1995}]{Hindmarsh:1994re}
{Hindmarsh} M.~B.,  {Kibble} T.~W.~B.,  1995, \mn@doi [Reports on Progress in
  Physics] {10.1088/0034-4885/58/5/001}, \href
  {https://ui.adsabs.harvard.edu/abs/1995RPPh...58..477H} {58, 477}

\bibitem[\protect\citeauthoryear{{Hindmarsh} \& {Kume}}{{Hindmarsh} \&
  {Kume}}{2023}]{2023JCAP...04..045H}
{Hindmarsh} M.,  {Kume} J.,  2023, \mn@doi [\jcap]
  {10.1088/1475-7516/2023/04/045}, \href
  {https://ui.adsabs.harvard.edu/abs/2023JCAP...04..045H} {2023, 045}

\bibitem[\protect\citeauthoryear{{Hindmarsh}, {Ringeval}  \&
  {Suyama}}{{Hindmarsh} et~al.}{2009}]{Hindmarsh:2009qk}
{Hindmarsh} M.,  {Ringeval} C.,   {Suyama} T.,  2009, \mn@doi [\prd]
  {10.1103/PhysRevD.80.083501}, \href
  {https://ui.adsabs.harvard.edu/abs/2009PhRvD..80h3501H} {80, 083501}

\bibitem[\protect\citeauthoryear{{Hindmarsh}, {Ringeval}  \&
  {Suyama}}{{Hindmarsh} et~al.}{2010}]{Hindmarsh:2009es}
{Hindmarsh} M.,  {Ringeval} C.,   {Suyama} T.,  2010, \mn@doi [\prd]
  {10.1103/PhysRevD.81.063505}, \href
  {https://ui.adsabs.harvard.edu/abs/2010PhRvD..81f3505H} {81, 063505}

\bibitem[\protect\citeauthoryear{{Hobson}, {Jones}  \& {Lasenby}}{{Hobson}
  et~al.}{1999}]{Hobson:1998av}
{Hobson} M.~P.,  {Jones} A.~W.,   {Lasenby} A.~N.,  1999, \mn@doi [\mnras]
  {10.1046/j.1365-8711.1999.02824.x}, \href
  {https://ui.adsabs.harvard.edu/abs/1999MNRAS.309..125H} {309, 125}

\bibitem[\protect\citeauthoryear{{Hong}, {Coutinho}, {Dey}, {Barab{\'a}si},
  {Vogelsberger}, {Hernquist}  \& {Gebhardt}}{{Hong}
  et~al.}{2016}]{hong2016discriminating}
{Hong} S.,  {Coutinho} B.~C.,  {Dey} A.,  {Barab{\'a}si} A.-L.,  {Vogelsberger}
  M.,  {Hernquist} L.,   {Gebhardt} K.,  2016, \mn@doi [\mnras]
  {10.1093/mnras/stw803}, \href
  {https://ui.adsabs.harvard.edu/abs/2016MNRAS.459.2690H} {459, 2690}

\bibitem[\protect\citeauthoryear{{Hong} et~al.,}{{Hong}
  et~al.}{2020}]{hong2020constraining}
{Hong} S.,  et~al., 2020, \mn@doi [\mnras] {10.1093/mnras/staa566}, \href
  {https://ui.adsabs.harvard.edu/abs/2020MNRAS.493.5972H} {493, 5972}

\bibitem[\protect\citeauthoryear{{Hu} \& {Dodelson}}{{Hu} \&
  {Dodelson}}{2002}]{hu2002cosmic}
{Hu} W.,  {Dodelson} S.,  2002, \mn@doi [\araa]
  {10.1146/annurev.astro.40.060401.093926}, \href
  {https://ui.adsabs.harvard.edu/abs/2002ARA&A..40..171H} {40, 171}

\bibitem[\protect\citeauthoryear{{Huber} et~al.,}{{Huber}
  et~al.}{2024}]{2024SPIE13102E..22H}
{Huber} Z.~B.,  et~al., 2024, in {Zmuidzinas} J.,  {Gao} J.-R.,  eds,  Society
  of Photo-Optical Instrumentation Engineers (SPIE) Conference Series Vol.
  13102, Millimeter, Submillimeter, and Far-Infrared Detectors and
  Instrumentation for Astronomy XII. p. 1310222, \mn@doi{10.1117/12.3020373}

\bibitem[\protect\citeauthoryear{Hug, Schneider  \& Schuster}{Hug
  et~al.}{2007}]{Hug2007TheSO}
Hug D.,  Schneider R.,   Schuster R.,  2007, \mn@doi [St Petersburg
  Mathematical Journal] {10.1090/S1061-0022-07-00990-9}, 19

\bibitem[\protect\citeauthoryear{{Jalali Kanafi} \& {Movahed}}{{Jalali Kanafi}
  \& {Movahed}}{2024}]{2024ApJ...963...31K}
{Jalali Kanafi} M.~H.,  {Movahed} S.~M.~S.,  2024, \mn@doi [The Astrophysical
  Journal] {10.3847/1538-4357/ad1880}, \href
  {https://ui.adsabs.harvard.edu/abs/2024ApJ...963...31K} {963, 31}

\bibitem[\protect\citeauthoryear{{Jalali Kanafi}, {Ansarifard}  \&
  {Movahed}}{{Jalali Kanafi} et~al.}{2024}]{2024MNRAS.535..657J}
{Jalali Kanafi} M.~H.,  {Ansarifard} S.,   {Movahed} S.~M.~S.,  2024, \mn@doi
  [\mnras] {10.1093/mnras/stae2044}, \href
  {https://ui.adsabs.harvard.edu/abs/2024MNRAS.535..657J} {535, 657}

\bibitem[\protect\citeauthoryear{{Jamieson}, {Li}, {de Oliveira},
  {Villaescusa-Navarro}, {Ho}  \& {Spergel}}{{Jamieson}
  et~al.}{2023}]{2023ApJ...952..145J}
{Jamieson} D.,  {Li} Y.,  {de Oliveira} R.~A.,  {Villaescusa-Navarro} F.,  {Ho}
  S.,   {Spergel} D.~N.,  2023, \mn@doi [\apj] {10.3847/1538-4357/acdb6c},
  \href {https://ui.adsabs.harvard.edu/abs/2023ApJ...952..145J} {952, 145}

\bibitem[\protect\citeauthoryear{{Jasche} \& {Wandelt}}{{Jasche} \&
  {Wandelt}}{2013}]{fieldlevelLSS}
{Jasche} J.,  {Wandelt} B.~D.,  2013, \mn@doi [\mnras] {10.1093/mnras/stt449},
  \href {https://ui.adsabs.harvard.edu/abs/2013MNRAS.432..894J} {432, 894}

\bibitem[\protect\citeauthoryear{{Jenet} et~al.,}{{Jenet}
  et~al.}{2006}]{Jenet:2006sv}
{Jenet} F.~A.,  et~al., 2006, \mn@doi [\apj] {10.1086/508702}, \href
  {https://ui.adsabs.harvard.edu/abs/2006ApJ...653.1571J} {653, 1571}

\bibitem[\protect\citeauthoryear{{Jiao}, {Brandenberger}  \&
  {Refregier}}{{Jiao} et~al.}{2023}]{2023PhRvD.108d3510J}
{Jiao} H.,  {Brandenberger} R.,   {Refregier} A.,  2023, \mn@doi [\prd]
  {10.1103/PhysRevD.108.043510}, \href
  {https://ui.adsabs.harvard.edu/abs/2023PhRvD.108d3510J} {108, 043510}

\bibitem[\protect\citeauthoryear{{Jiao}, {Brandenberger}  \&
  {Refregier}}{{Jiao} et~al.}{2024}]{2024PhRvD.109l3524J}
{Jiao} H.,  {Brandenberger} R.,   {Refregier} A.,  2024, \mn@doi [\prd]
  {10.1103/PhysRevD.109.123524}, \href
  {https://ui.adsabs.harvard.edu/abs/2024PhRvD.109l3524J} {109, 123524}

\bibitem[\protect\citeauthoryear{{Kaiser}}{{Kaiser}}{1984}]{kaiser1984spatial}
{Kaiser} N.,  1984, \mn@doi [\apjl] {10.1086/184341}, \href
  {https://ui.adsabs.harvard.edu/abs/1984ApJ...284L...9K} {284, L9}

\bibitem[\protect\citeauthoryear{{Kaiser} \& {Stebbins}}{{Kaiser} \&
  {Stebbins}}{1984}]{1984Natur.310..391K}
{Kaiser} N.,  {Stebbins} A.,  1984, \mn@doi [\nat] {10.1038/310391a0}, \href
  {https://ui.adsabs.harvard.edu/abs/1984Natur.310..391K} {310, 391}

\bibitem[\protect\citeauthoryear{{Kardar}}{{Kardar}}{2007}]{2007spf..book.....K}
{Kardar} M.,  2007, {Statistical Physics of Fields}

\bibitem[\protect\citeauthoryear{{Kibble}}{{Kibble}}{1976}]{Kibble:1976sj}
{Kibble} T.~W.~B.,  1976, \mn@doi [Journal of Physics A Mathematical General]
  {10.1088/0305-4470/9/8/029}, \href
  {https://ui.adsabs.harvard.edu/abs/1976JPhA....9.1387K} {9, 1387}

\bibitem[\protect\citeauthoryear{{Kibble}}{{Kibble}}{1980}]{Kibble:1980mv22}
{Kibble} T.~W.~B.,  1980, \mn@doi [\physrep] {10.1016/0370-1573(80)90091-5},
  \href {https://ui.adsabs.harvard.edu/abs/1980PhR....67..183K} {67, 183}

\bibitem[\protect\citeauthoryear{{Kibble}}{{Kibble}}{2004}]{Kibble:2004hq}
{Kibble} T.~W.~B.,  2004, \mn@doi [arXiv e-prints]
  {10.48550/arXiv.astro-ph/0410073}, \href
  {https://ui.adsabs.harvard.edu/abs/2004astro.ph.10073K} {pp
  astro--ph/0410073}

\bibitem[\protect\citeauthoryear{{Kosowsky}}{{Kosowsky}}{2006}]{2006NewAR..50..969K}
{Kosowsky} A.,  2006, \mn@doi [\nar] {10.1016/j.newar.2006.09.021}, \href
  {https://ui.adsabs.harvard.edu/abs/2006NewAR..50..969K} {50, 969}

\bibitem[\protect\citeauthoryear{{Kume} \& {Hindmarsh}}{{Kume} \&
  {Hindmarsh}}{2024}]{2024JCAP...12..001K}
{Kume} J.,  {Hindmarsh} M.,  2024, \mn@doi [\jcap]
  {10.1088/1475-7516/2024/12/001}, \href
  {https://ui.adsabs.harvard.edu/abs/2024JCAP...12..001K} {2024, 001}

\bibitem[\protect\citeauthoryear{{Kuroyanagi}, {Miyamoto}, {Sekiguchi},
  {Takahashi}  \& {Silk}}{{Kuroyanagi} et~al.}{2013}]{Kuroyanagi:2012jf}
{Kuroyanagi} S.,  {Miyamoto} K.,  {Sekiguchi} T.,  {Takahashi} K.,   {Silk} J.,
   2013, \mn@doi [\prd] {10.1103/PhysRevD.87.023522}, \href
  {https://ui.adsabs.harvard.edu/abs/2013PhRvD..87b3522K} {87, 023522}

\bibitem[\protect\citeauthoryear{{Kuroyanagi}, {Chiba}  \&
  {Takahashi}}{{Kuroyanagi} et~al.}{2018}]{2018JCAP...11..038K}
{Kuroyanagi} S.,  {Chiba} T.,   {Takahashi} T.,  2018, \mn@doi [\jcap]
  {10.1088/1475-7516/2018/11/038}, \href
  {https://ui.adsabs.harvard.edu/abs/2018JCAP...11..038K} {2018, 038}

\bibitem[\protect\citeauthoryear{{Latora}, {Nicosia}  \& {Russo}}{{Latora}
  et~al.}{2017}]{latora2017complex}
{Latora} V.,  {Nicosia} V.,   {Russo} G.,  2017, {Complex Networks}

\bibitem[\protect\citeauthoryear{{Lazanu} \& {Shellard}}{{Lazanu} \&
  {Shellard}}{2015}]{Lazanu:2014eya}
{Lazanu} A.,  {Shellard} P.,  2015, \mn@doi [\jcap]
  {10.1088/1475-7516/2015/02/024}, \href
  {https://ui.adsabs.harvard.edu/abs/2015JCAP...02..024L} {2015, 024}

\bibitem[\protect\citeauthoryear{{Lesgourges}}{{Lesgourges}}{2013}]{Lesgourgues:2013qba}
{Lesgourges} J.,  2013, in {Matchev} K.,  {et al.} eds, Searching for New
  Physics at Small and Large Scales (TASI 2012) - Proceedings of the 2012
  Theoretical Advanced Study Institute in Elementary Particle Physics. Edited
  by Schmaltz Martin \& Pierpaoli Elena. Published by World Scientific
  Publishing Co. Pte. Ltd. pp 29--97, \mn@doi{10.1142/9789814525220_0002}

\bibitem[\protect\citeauthoryear{{Lewis}, {Challinor}  \& {Lasenby}}{{Lewis}
  et~al.}{2000}]{Lewis:1999bs}
{Lewis} A.,  {Challinor} A.,   {Lasenby} A.,  2000, \mn@doi [\apj]
  {10.1086/309179}, \href
  {https://ui.adsabs.harvard.edu/abs/2000ApJ...538..473L} {538, 473}

\bibitem[\protect\citeauthoryear{{Li}, {Liu}, {Li}, {Li}  \& {Zhang}}{{Li}
  et~al.}{2017a}]{2017arXiv170909053L}
{Li} Y.-P.,  {Liu} Y.,  {Li} S.-Y.,  {Li} H.,   {Zhang} X.,  2017a, \mn@doi
  [arXiv e-prints] {10.48550/arXiv.1709.09053}, \href
  {https://ui.adsabs.harvard.edu/abs/2017arXiv170909053L} {p. arXiv:1709.09053}

\bibitem[\protect\citeauthoryear{{Li} et~al.,}{{Li}
  et~al.}{2017b}]{2017arXiv171003047L}
{Li} H.,  et~al., 2017b, \mn@doi [arXiv e-prints] {10.48550/arXiv.1710.03047},
  \href {https://ui.adsabs.harvard.edu/abs/2017arXiv171003047L} {p.
  arXiv:1710.03047}

\bibitem[\protect\citeauthoryear{Li et~al.,}{Li
  et~al.}{2018}]{10.1093/nsr/nwy019}
Li H.,  et~al., 2018, \mn@doi [National Science Review] {10.1093/nsr/nwy019},
  6, 145

\bibitem[\protect\citeauthoryear{{Liddle}}{{Liddle}}{1999}]{Liddle:1999mq}
{Liddle} A.~R.,  1999, in {Masiero} A.,  {Senjanovic} G.,   {Smirnov} A.,  eds,
  High Energy Physics and Cosmology, 1998 Summer School. p.~260 (\mn@eprint
  {arXiv} {astro-ph/9901124}), \mn@doi{10.48550/arXiv.astro-ph/9901124}

\bibitem[\protect\citeauthoryear{{Liddle} \& {Lyth}}{{Liddle} \&
  {Lyth}}{1993}]{Liddle:1993fq}
{Liddle} A.~R.,  {Lyth} D.~H.,  1993, \mn@doi [\physrep]
  {10.1016/0370-1573(93)90114-S}, \href
  {https://ui.adsabs.harvard.edu/abs/1993PhR...231....1L} {231, 1}

\bibitem[\protect\citeauthoryear{{Lizarraga}, {Urrestilla}, {Daverio},
  {Hindmarsh}  \& {Kunz}}{{Lizarraga} et~al.}{2016}]{Lizarraga:2016onn}
{Lizarraga} J.,  {Urrestilla} J.,  {Daverio} D.,  {Hindmarsh} M.,   {Kunz} M.,
  2016, \mn@doi [\jcap] {10.1088/1475-7516/2016/10/042}, \href
  {https://ui.adsabs.harvard.edu/abs/2016JCAP...10..042L} {2016, 042}

\bibitem[\protect\citeauthoryear{{Lorenz}, {Ringeval}  \&
  {Sakellariadou}}{{Lorenz} et~al.}{2010}]{2010JCAP...10..003L}
{Lorenz} L.,  {Ringeval} C.,   {Sakellariadou} M.,  2010, \mn@doi [\jcap]
  {10.1088/1475-7516/2010/10/003}, \href
  {https://ui.adsabs.harvard.edu/abs/2010JCAP...10..003L} {2010, 003}

\bibitem[\protect\citeauthoryear{{LSST Science Collaboration} et~al.,}{{LSST
  Science Collaboration} et~al.}{2017}]{2017arXiv170804058L}
{LSST Science Collaboration} et~al., 2017, \mn@doi [arXiv e-prints]
  {10.48550/arXiv.1708.04058}, \href
  {https://ui.adsabs.harvard.edu/abs/2017arXiv170804058L} {p. arXiv:1708.04058}

\bibitem[\protect\citeauthoryear{{Majumdar} \& {Davis}}{{Majumdar} \&
  {Davis}}{2002}]{Majumdar:2002hy}
{Majumdar} M.,  {Davis} A.-C.,  2002, \mn@doi [Journal of High Energy Physics]
  {10.1088/1126-6708/2002/03/056}, \href
  {https://ui.adsabs.harvard.edu/abs/2002JHEP...03..056M} {2002, 056}

\bibitem[\protect\citeauthoryear{{Martins} \& {Shellard}}{{Martins} \&
  {Shellard}}{2006}]{2006PhRvD..73d3515M}
{Martins} C.~J.,  {Shellard} E.~P.,  2006, \mn@doi [\prd]
  {10.1103/PhysRevD.73.043515}, \href
  {https://ui.adsabs.harvard.edu/abs/2006PhRvD..73d3515M} {73, 043515}

\bibitem[\protect\citeauthoryear{{Matsubara}}{{Matsubara}}{2003}]{matsubara03}
{Matsubara} T.,  2003, \mn@doi [\apj] {10.1086/345521}, \href
  {https://ui.adsabs.harvard.edu/abs/2003ApJ...584....1M} {584, 1}

\bibitem[\protect\citeauthoryear{{Matsubara}}{{Matsubara}}{2010}]{matsubara2010analytic}
{Matsubara} T.,  2010, \mn@doi [\prd] {10.1103/PhysRevD.81.083505}, \href
  {https://ui.adsabs.harvard.edu/abs/2010PhRvD..81h3505M} {81, 083505}

\bibitem[\protect\citeauthoryear{{Matsubara}}{{Matsubara}}{2020}]{matsubara2020statistics}
{Matsubara} T.,  2020, \mn@doi [\prd] {10.1103/PhysRevD.101.043532}, \href
  {https://ui.adsabs.harvard.edu/abs/2020PhRvD.101d3532M} {101, 043532}

\bibitem[\protect\citeauthoryear{{Matsubara}}{{Matsubara}}{2024a}]{2024PhRvD.110f3543M}
{Matsubara} T.,  2024a, \mn@doi [\prd] {10.1103/PhysRevD.110.063543}, \href
  {https://ui.adsabs.harvard.edu/abs/2024PhRvD.110f3543M} {110, 063543}

\bibitem[\protect\citeauthoryear{{Matsubara}}{{Matsubara}}{2024b}]{2024PhRvD.110f3544M}
{Matsubara} T.,  2024b, \mn@doi [\prd] {10.1103/PhysRevD.110.063544}, \href
  {https://ui.adsabs.harvard.edu/abs/2024PhRvD.110f3544M} {110, 063544}

\bibitem[\protect\citeauthoryear{{Matsubara}}{{Matsubara}}{2024c}]{2024PhRvD.110f3545M}
{Matsubara} T.,  2024c, \mn@doi [\prd] {10.1103/PhysRevD.110.063545}, \href
  {https://ui.adsabs.harvard.edu/abs/2024PhRvD.110f3545M} {110, 063545}

\bibitem[\protect\citeauthoryear{{Matsubara}}{{Matsubara}}{2024d}]{2024PhRvD.110f3546M}
{Matsubara} T.,  2024d, \mn@doi [\prd] {10.1103/PhysRevD.110.063546}, \href
  {https://ui.adsabs.harvard.edu/abs/2024PhRvD.110f3546M} {110, 063546}

\bibitem[\protect\citeauthoryear{{Matsubara} \& {Kuriki}}{{Matsubara} \&
  {Kuriki}}{2021}]{matsubara2021weakly}
{Matsubara} T.,  {Kuriki} S.,  2021, \mn@doi [\prd]
  {10.1103/PhysRevD.104.103522}, \href
  {https://ui.adsabs.harvard.edu/abs/2021PhRvD.104j3522M} {104, 103522}

\bibitem[\protect\citeauthoryear{{Matsubara} \& {Yokoyama}}{{Matsubara} \&
  {Yokoyama}}{1996}]{matsubara1996genus}
{Matsubara} T.,  {Yokoyama} J.,  1996, \mn@doi [\apj] {10.1086/177257}, \href
  {https://ui.adsabs.harvard.edu/abs/1996ApJ...463..409M} {463, 409}

\bibitem[\protect\citeauthoryear{{Matsubara}, {Hikage}  \&
  {Kuriki}}{{Matsubara} et~al.}{2022}]{matsubara2022minkowski}
{Matsubara} T.,  {Hikage} C.,   {Kuriki} S.,  2022, \mn@doi [\prd]
  {10.1103/PhysRevD.105.023527}, \href
  {https://ui.adsabs.harvard.edu/abs/2022PhRvD.105b3527M} {105, 023527}

\bibitem[\protect\citeauthoryear{{McDonough} \& {Brandenberger}}{{McDonough} \&
  {Brandenberger}}{2013}]{2013JCAP...02..045M}
{McDonough} E.,  {Brandenberger} R.~H.,  2013, \mn@doi [\jcap]
  {10.1088/1475-7516/2013/02/045}, \href
  {https://ui.adsabs.harvard.edu/abs/2013JCAP...02..045M} {2013, 045}

\bibitem[\protect\citeauthoryear{{McEwen}, {Feeney}, {Peiris}, {Wiaux},
  {Ringeval}  \& {Bouchet}}{{McEwen} et~al.}{2017}]{2017MNRAS.472.4081M}
{McEwen} J.~D.,  {Feeney} S.~M.,  {Peiris} H.~V.,  {Wiaux} Y.,  {Ringeval} C.,
   {Bouchet} F.~R.,  2017, \mn@doi [\mnras] {10.1093/mnras/stx2268}, \href
  {https://ui.adsabs.harvard.edu/abs/2017MNRAS.472.4081M} {472, 4081}

\bibitem[\protect\citeauthoryear{McMullen}{McMullen}{1997}]{McMullen1997}
McMullen P.,  1997, Rend. Circ. Mat. Palermo (2) Suppl, 50, 259

\bibitem[\protect\citeauthoryear{{Mecke}, {Buchert}  \& {Wagner}}{{Mecke}
  et~al.}{1994}]{mecke1993robust}
{Mecke} K.~R.,  {Buchert} T.,   {Wagner} H.,  1994, \mn@doi [\aap]
  {10.48550/arXiv.astro-ph/9312028}, \href
  {https://ui.adsabs.harvard.edu/abs/1994A&A...288..697M} {288, 697}

\bibitem[\protect\citeauthoryear{{Morawetz}, {Paillas}  \&
  {Percival}}{{Morawetz} et~al.}{2025}]{2025JCAP...01..026M}
{Morawetz} J.,  {Paillas} E.,   {Percival} W.~J.,  2025, \mn@doi [\jcap]
  {10.1088/1475-7516/2025/01/026}, \href
  {https://ui.adsabs.harvard.edu/abs/2025JCAP...01..026M} {2025, 026}

\bibitem[\protect\citeauthoryear{{Movahed} \& {Khosravi}}{{Movahed} \&
  {Khosravi}}{2011}]{movahed2011level}
{Movahed} S.~M.~S.,  {Khosravi} S.,  2011, \mn@doi [\jcap]
  {10.1088/1475-7516/2011/03/012}, \href
  {https://ui.adsabs.harvard.edu/abs/2011JCAP...03..012S} {2011, 012}

\bibitem[\protect\citeauthoryear{{Movahed}, {Javanmardi}  \& {Sheth}}{{Movahed}
  et~al.}{2013}]{Movahed:2012zt}
{Movahed} M.~S.,  {Javanmardi} B.,   {Sheth} R.~K.,  2013, \mn@doi [\mnras]
  {10.1093/mnras/stt1284}, \href
  {https://ui.adsabs.harvard.edu/abs/2013MNRAS.434.3597M} {434, 3597}

\bibitem[\protect\citeauthoryear{{Nguyen}, {Schmidt}, {Tucci}, {Reinecke}  \&
  {Kosti{\'c}}}{{Nguyen} et~al.}{2024}]{2024PhRvL.133v1006N}
{Nguyen} N.-M.,  {Schmidt} F.,  {Tucci} B.,  {Reinecke} M.,   {Kosti{\'c}} A.,
  2024, \mn@doi [\prl] {10.1103/PhysRevLett.133.221006}, \href
  {https://ui.adsabs.harvard.edu/abs/2024PhRvL.133v1006N} {133, 221006}

\bibitem[\protect\citeauthoryear{{Oknyanskij}}{{Oknyanskij}}{2002}]{Oknyanskij:2005pd}
{Oknyanskij} V.~L.,  2002, in {Rosada} M.,  {Binette} L.,   {Arias} L.,  eds,
  Astronomical Society of the Pacific Conference Series Vol. 282, Galaxies: the
  Third Dimension. p.~352

\bibitem[\protect\citeauthoryear{{Pagano} \& {Brandenberger}}{{Pagano} \&
  {Brandenberger}}{2012}]{Pagano:2012cx}
{Pagano} M.,  {Brandenberger} R.,  2012, \mn@doi [\jcap]
  {10.1088/1475-7516/2012/05/014}, \href
  {https://ui.adsabs.harvard.edu/abs/2012JCAP...05..014P} {2012, 014}

\bibitem[\protect\citeauthoryear{{Papamakarios} \& {Murray}}{{Papamakarios} \&
  {Murray}}{2016}]{2016arXiv160506376P}
{Papamakarios} G.,  {Murray} I.,  2016, \mn@doi [arXiv e-prints]
  {10.48550/arXiv.1605.06376}, \href
  {https://ui.adsabs.harvard.edu/abs/2016arXiv160506376P} {p. arXiv:1605.06376}

\bibitem[\protect\citeauthoryear{Peebles}{Peebles}{1980}]{peeb80}
Peebles P.,  1980, The Large-scale Structure of the Universe.
Princeton Series in Physics, Princeton University Press, \url
  {https://books.google.com/books?id=O_BPaHFtX1YC}

\bibitem[\protect\citeauthoryear{Peebles}{Peebles}{2020}]{peebles2020large}
Peebles P.,  2020, The Large-Scale Structure of the Universe.
Princeton Series in Physics, Princeton University Press, \url
  {https://books.google.com/books?id=L5HUDwAAQBAJ}

\bibitem[\protect\citeauthoryear{{Pen}, {Seljak}  \& {Turok}}{{Pen}
  et~al.}{1997}]{Pen:1997ae}
{Pen} U.-L.,  {Seljak} U.,   {Turok} N.,  1997, \mn@doi [\prl]
  {10.1103/PhysRevLett.79.1611}, \href
  {https://ui.adsabs.harvard.edu/abs/1997PhRvL..79.1611P} {79, 1611}

\bibitem[\protect\citeauthoryear{{Perivolaropoulos} \&
  {Skara}}{{Perivolaropoulos} \& {Skara}}{2022}]{2022NewAR..9501659P}
{Perivolaropoulos} L.,  {Skara} F.,  2022, \mn@doi [\nar]
  {10.1016/j.newar.2022.101659}, \href
  {https://ui.adsabs.harvard.edu/abs/2022NewAR..9501659P} {95, 101659}

\bibitem[\protect\citeauthoryear{{Planck Collaboration} et~al.,}{{Planck
  Collaboration} et~al.}{2014a}]{planck2013results}
{Planck Collaboration} et~al., 2014a, \mn@doi [\aap]
  {10.1051/0004-6361/201321534}, \href
  {https://ui.adsabs.harvard.edu/abs/2014A&A...571A..23P} {571, A23}

\bibitem[\protect\citeauthoryear{{Planck Collaboration} et~al.,}{{Planck
  Collaboration} et~al.}{2014b}]{Ade:2013xla}
{Planck Collaboration} et~al., 2014b, \mn@doi [\aap]
  {10.1051/0004-6361/201321621}, \href
  {https://ui.adsabs.harvard.edu/abs/2014A&A...571A..25P} {571, A25}

\bibitem[\protect\citeauthoryear{{Planck Collaboration} et~al.,}{{Planck
  Collaboration} et~al.}{2020a}]{aghanim2020planck}
{Planck Collaboration} et~al., 2020a, \mn@doi [\aap]
  {10.1051/0004-6361/201833880}, \href
  {https://ui.adsabs.harvard.edu/abs/2020A&A...641A...1P} {641, A1}

\bibitem[\protect\citeauthoryear{{Planck Collaboration} et~al.,}{{Planck
  Collaboration} et~al.}{2020b}]{aghanim2020planckvi}
{Planck Collaboration} et~al., 2020b, \mn@doi [\aap]
  {10.1051/0004-6361/201833910}, \href
  {https://ui.adsabs.harvard.edu/abs/2020A&A...641A...6P} {641, A6}

\bibitem[\protect\citeauthoryear{{Planck Collaboration} et~al.,}{{Planck
  Collaboration} et~al.}{2020c}]{akrami2020planck}
{Planck Collaboration} et~al., 2020c, \mn@doi [\aap]
  {10.1051/0004-6361/201935201}, \href
  {https://ui.adsabs.harvard.edu/abs/2020A&A...641A...7P} {641, A7}

\bibitem[\protect\citeauthoryear{{Planck Collaboration} et~al.,}{{Planck
  Collaboration} et~al.}{2020d}]{2020A&A...641A..10P}
{Planck Collaboration} et~al., 2020d, \mn@doi [\aap]
  {10.1051/0004-6361/201833887}, \href
  {https://ui.adsabs.harvard.edu/abs/2020A&A...641A..10P} {641, A10}

\bibitem[\protect\citeauthoryear{{Pogosian}, {Tye}, {Wasserman}  \&
  {Wyman}}{{Pogosian} et~al.}{2003}]{Pogosian:2003mz}
{Pogosian} L.,  {Tye} S. H.~H.,  {Wasserman} I.,   {Wyman} M.,  2003, \mn@doi
  [\prd] {10.1103/PhysRevD.68.023506}, \href
  {https://ui.adsabs.harvard.edu/abs/2003PhRvD..68b3506P} {68, 023506}

\bibitem[\protect\citeauthoryear{{Pogosyan}, {Pichon}, {Gay}, {Prunet},
  {Cardoso}, {Sousbie}  \& {Colombi}}{{Pogosyan}
  et~al.}{2009}]{Pogosyan:2008jb}
{Pogosyan} D.,  {Pichon} C.,  {Gay} C.,  {Prunet} S.,  {Cardoso} J.~F.,
  {Sousbie} T.,   {Colombi} S.,  2009, \mn@doi [\mnras]
  {10.1111/j.1365-2966.2009.14753.x}, \href
  {https://ui.adsabs.harvard.edu/abs/2009MNRAS.396..635P} {396, 635}

\bibitem[\protect\citeauthoryear{{Polchinski}}{{Polchinski}}{2005}]{Polchinski:2004hb}
{Polchinski} J.,  2005, \mn@doi [International Journal of Modern Physics A]
  {10.1142/S0217751X05026686}, \href
  {https://ui.adsabs.harvard.edu/abs/2005IJMPA..20.3413P} {20, 3413}

\bibitem[\protect\citeauthoryear{{Pranav} et~al.,}{{Pranav}
  et~al.}{2019}]{pranav2019topology}
{Pranav} P.,  et~al., 2019, \mn@doi [\mnras] {10.1093/mnras/stz541}, \href
  {https://ui.adsabs.harvard.edu/abs/2019MNRAS.485.4167P} {485, 4167}

\bibitem[\protect\citeauthoryear{{Pshirkov} \& {Tuntsov}}{{Pshirkov} \&
  {Tuntsov}}{2010}]{Pshirkov:2009vb}
{Pshirkov} M.~S.,  {Tuntsov} A.~V.,  2010, \mn@doi [\prd]
  {10.1103/PhysRevD.81.083519}, \href
  {https://ui.adsabs.harvard.edu/abs/2010PhRvD..81h3519P} {81, 083519}

\bibitem[\protect\citeauthoryear{{Punturo} et~al.,}{{Punturo}
  et~al.}{2010}]{Punturo_2010}
{Punturo} M.,  et~al., 2010, \mn@doi [Classical and Quantum Gravity]
  {10.1088/0264-9381/27/19/194002}, \href
  {https://ui.adsabs.harvard.edu/abs/2010CQGra..27s4002P} {27, 194002}

\bibitem[\protect\citeauthoryear{{Regan} \& {Hindmarsh}}{{Regan} \&
  {Hindmarsh}}{2015}]{Regan:2015cfa}
{Regan} D.,  {Hindmarsh} M.,  2015, \mn@doi [\jcap]
  {10.1088/1475-7516/2015/10/030}, \href
  {https://ui.adsabs.harvard.edu/abs/2015JCAP...10..030R} {2015, 030}

\bibitem[\protect\citeauthoryear{{Regan} \& {Shellard}}{{Regan} \&
  {Shellard}}{2010}]{2010PhRvD..82f3527R}
{Regan} D.~M.,  {Shellard} E.~P.~S.,  2010, \mn@doi [\prd]
  {10.1103/PhysRevD.82.063527}, \href
  {https://ui.adsabs.harvard.edu/abs/2010PhRvD..82f3527R} {82, 063527}

\bibitem[\protect\citeauthoryear{{Rice}}{{Rice}}{1944}]{rice44a}
{Rice} S.~O.,  1944, \mn@doi [Bell System Technical Journal]
  {10.1002/j.1538-7305.1944.tb00874.x}, \href
  {https://ui.adsabs.harvard.edu/abs/1944BSTJ...23..282R} {23, 282}

\bibitem[\protect\citeauthoryear{{Rice}}{{Rice}}{1945}]{rice44b}
{Rice} S.~O.,  1945, \mn@doi [Bell System Technical Journal]
  {10.1002/j.1538-7305.1945.tb00453.x}, \href
  {https://ui.adsabs.harvard.edu/abs/1945BSTJ...24...46R} {24, 46}

\bibitem[\protect\citeauthoryear{Rice}{Rice}{1954}]{rice1954selected}
Rice S.,  1954, in Wax N.,  ed., Dover Books on Engineering, Selected Papers on
  Noise and Stochastic Processes.
Dover Publications, \url {https://books.google.com/books?id=i5mkDgAAQBAJ}

\bibitem[\protect\citeauthoryear{{Ringeval}}{{Ringeval}}{2010}]{Ringeval:2010ca}
{Ringeval} C.,  2010, \mn@doi [Advances in Astronomy] {10.1155/2010/380507},
  \href {https://ui.adsabs.harvard.edu/abs/2010AdAst2010E..66R} {2010, 380507}

\bibitem[\protect\citeauthoryear{{Ringeval} \& {Bouchet}}{{Ringeval} \&
  {Bouchet}}{2012}]{Ringeval:2012tk}
{Ringeval} C.,  {Bouchet} F.~R.,  2012, \mn@doi [\prd]
  {10.1103/PhysRevD.86.023513}, \href
  {https://ui.adsabs.harvard.edu/abs/2012PhRvD..86b3513R} {86, 023513}

\bibitem[\protect\citeauthoryear{{Ringeval} \& {Suyama}}{{Ringeval} \&
  {Suyama}}{2017}]{Ringeval:2017eww}
{Ringeval} C.,  {Suyama} T.,  2017, \mn@doi [\jcap]
  {10.1088/1475-7516/2017/12/027}, \href
  {https://ui.adsabs.harvard.edu/abs/2017JCAP...12..027R} {2017, 027}

\bibitem[\protect\citeauthoryear{{Ringeval}, {Sakellariadou}  \&
  {Bouchet}}{{Ringeval} et~al.}{2007}]{Ringeval:2005kr}
{Ringeval} C.,  {Sakellariadou} M.,   {Bouchet} F.~R.,  2007, \mn@doi [\jcap]
  {10.1088/1475-7516/2007/02/023}, \href
  {https://ui.adsabs.harvard.edu/abs/2007JCAP...02..023R} {2007, 023}

\bibitem[\protect\citeauthoryear{{Rybak}, {Martins}, {Peter}  \&
  {Shellard}}{{Rybak} et~al.}{2024}]{2024PhRvD.110b3534R}
{Rybak} I.~Y.,  {Martins} C.~J.~A.~P.,  {Peter} P.,   {Shellard} E.~P.~S.,
  2024, \mn@doi [\prd] {10.1103/PhysRevD.110.023534}, \href
  {https://ui.adsabs.harvard.edu/abs/2024PhRvD.110b3534R} {110, 023534}

\bibitem[\protect\citeauthoryear{{Ryden}}{{Ryden}}{1988}]{ryden1988area}
{Ryden} B.~S.,  1988, \mn@doi [\apjl] {10.1086/185284}, \href
  {https://ui.adsabs.harvard.edu/abs/1988ApJ...333L..41R} {333, L41}

\bibitem[\protect\citeauthoryear{{Ryden}, {Melott}, {Craig}, {Gott},
  {Weinberg}, {Scherrer}, {Bhavsar}  \& {Miller}}{{Ryden} et~al.}{1989}]{ryd89}
{Ryden} B.~S.,  {Melott} A.~L.,  {Craig} D.~A.,  {Gott} III J.~R.,  {Weinberg}
  D.~H.,  {Scherrer} R.~J.,  {Bhavsar} S.~P.,   {Miller} J.~M.,  1989, \mn@doi
  [\apj] {10.1086/167426}, \href
  {https://ui.adsabs.harvard.edu/abs/1989ApJ...340..647R} {340, 647}

\bibitem[\protect\citeauthoryear{{Sakellariadou}}{{Sakellariadou}}{1997}]{Sakellariadou:1997zt}
{Sakellariadou} M.,  1997, \mn@doi [International Journal of Theoretical
  Physics] {10.1007/BF02768939}, \href
  {https://ui.adsabs.harvard.edu/abs/1997IJTP...36.2503S} {36, 2503}

\bibitem[\protect\citeauthoryear{{Sakellariadou}}{{Sakellariadou}}{2007}]{Sakellariadou:2006qs}
{Sakellariadou} M.,  2007, in {Unruh} W.~G.,  {Sch{\"u}tzhold} R.,  eds, ,
  Vol.~718, Lecture Notes in Physics, Berlin Springer Verlag.
p.~247, \mn@doi{10.1007/3-540-70859-6_10}

\bibitem[\protect\citeauthoryear{{Sarangi} \& {Tye}}{{Sarangi} \&
  {Tye}}{2002}]{Sarangi:2002yt}
{Sarangi} S.,  {Tye} S. H.~H.,  2002, \mn@doi [Physics Letters B]
  {10.1016/S0370-2693(02)01824-5}, \href
  {https://ui.adsabs.harvard.edu/abs/2002PhLB..536..185S} {536, 185}

\bibitem[\protect\citeauthoryear{{Schmalzing} \& {Buchert}}{{Schmalzing} \&
  {Buchert}}{1997}]{schmalzing1997beyond}
{Schmalzing} J.,  {Buchert} T.,  1997, \mn@doi [\apjl] {10.1086/310680}, \href
  {https://ui.adsabs.harvard.edu/abs/1997ApJ...482L...1S} {482, L1}

\bibitem[\protect\citeauthoryear{{Schmalzing} \& {Gorski}}{{Schmalzing} \&
  {Gorski}}{1998}]{1998MNRAS.297..355S}
{Schmalzing} J.,  {Gorski} K.~M.,  1998, \mn@doi [\mnras]
  {10.1046/j.1365-8711.1998.01467.x}, \href
  {https://ui.adsabs.harvard.edu/abs/1998MNRAS.297..355S} {297, 355}

\bibitem[\protect\citeauthoryear{{Schmalzing}, {Kerscher}  \&
  {Buchert}}{{Schmalzing} et~al.}{1996}]{schmalzing1995minkowski}
{Schmalzing} J.,  {Kerscher} M.,   {Buchert} T.,  1996, in {Bonometto} S.,
  {Primack} J.~R.,   {Provenzale} A.,  eds, Dark Matter in the Universe. p.~281
  (\mn@eprint {arXiv} {astro-ph/9508154}),
  \mn@doi{10.48550/arXiv.astro-ph/9508154}

\bibitem[\protect\citeauthoryear{{Shellard}}{{Shellard}}{1987}]{Shellard:1987bv}
{Shellard} E.~P.~S.,  1987, \mn@doi [Nuclear Physics B]
  {10.1016/0550-3213(87)90290-2}, \href
  {https://ui.adsabs.harvard.edu/abs/1987NuPhB.283..624S} {283, 624}

\bibitem[\protect\citeauthoryear{{Shim}, {Pichon}, {Pogosyan}, {Appleby},
  {Cadiou}, {Kim}, {Kraljic}  \& {Park}}{{Shim}
  et~al.}{2024}]{2024MNRAS.528.1604S}
{Shim} J.,  {Pichon} C.,  {Pogosyan} D.,  {Appleby} S.,  {Cadiou} C.,  {Kim}
  J.,  {Kraljic} K.,   {Park} C.,  2024, \mn@doi [\mnras]
  {10.1093/mnras/stae151}, \href
  {https://ui.adsabs.harvard.edu/abs/2024MNRAS.528.1604S} {528, 1604}

\bibitem[\protect\citeauthoryear{{Shlaer}, {Vilenkin}  \& {Loeb}}{{Shlaer}
  et~al.}{2012}]{Shlaer:2012rj}
{Shlaer} B.,  {Vilenkin} A.,   {Loeb} A.,  2012, \mn@doi [\jcap]
  {10.1088/1475-7516/2012/05/026}, \href
  {https://ui.adsabs.harvard.edu/abs/2012JCAP...05..026S} {2012, 026}

\bibitem[\protect\citeauthoryear{{Silk} \& {Vilenkin}}{{Silk} \&
  {Vilenkin}}{1984}]{1984PhRvL..53.1700S}
{Silk} J.,  {Vilenkin} A.,  1984, \mn@doi [\prl] {10.1103/PhysRevLett.53.1700},
  \href {https://ui.adsabs.harvard.edu/abs/1984PhRvL..53.1700S} {53, 1700}

\bibitem[\protect\citeauthoryear{{Stebbins}}{{Stebbins}}{1988}]{Stebbins:1988}
{Stebbins} A.,  1988, \mn@doi [\apj] {10.1086/166218}, \href
  {https://ui.adsabs.harvard.edu/abs/1988ApJ...327..584S} {327, 584}

\bibitem[\protect\citeauthoryear{{Stebbins} \& {Veeraraghavan}}{{Stebbins} \&
  {Veeraraghavan}}{1995}]{Stebbins:1995}
{Stebbins} A.,  {Veeraraghavan} S.,  1995, \mn@doi [\prd]
  {10.1103/PhysRevD.51.1465}, \href
  {https://ui.adsabs.harvard.edu/abs/1995PhRvD..51.1465S} {51, 1465}

\bibitem[\protect\citeauthoryear{{Steinhardt}}{{Steinhardt}}{1995}]{Steinhardt:1995uf}
{Steinhardt} P.~J.,  1995, in {Kolb} E.~W.,  {Peccei} R.~D.,  eds, Particle and
  Nuclear Astrophysics and Cosmology in the Next Millenium. p.~51 (\mn@eprint
  {arXiv} {astro-ph/9502024}), \mn@doi{10.48550/arXiv.astro-ph/9502024}

\bibitem[\protect\citeauthoryear{{Stewart} \& {Brandenberger}}{{Stewart} \&
  {Brandenberger}}{2009}]{Stewart:2008zq}
{Stewart} A.,  {Brandenberger} R.,  2009, \mn@doi [\jcap]
  {10.1088/1475-7516/2009/02/009}, \href
  {https://ui.adsabs.harvard.edu/abs/2009JCAP...02..009S} {2009, 009}

\bibitem[\protect\citeauthoryear{{Sugai} et~al.,}{{Sugai}
  et~al.}{2020}]{2020JLTP..199.1107S}
{Sugai} H.,  et~al., 2020, \mn@doi [Journal of Low Temperature Physics]
  {10.1007/s10909-019-02329-w}, \href
  {https://ui.adsabs.harvard.edu/abs/2020JLTP..199.1107S} {199, 1107}

\bibitem[\protect\citeauthoryear{{Szalay}}{{Szalay}}{1988}]{szalay88}
{Szalay} A.~S.,  1988, \mn@doi [\apj] {10.1086/166721}, \href
  {https://ui.adsabs.harvard.edu/abs/1988ApJ...333...21S} {333, 21}

\bibitem[\protect\citeauthoryear{{Tamura} et~al.,}{{Tamura}
  et~al.}{2016}]{2016SPIE.9908E..1MT}
{Tamura} N.,  et~al., 2016, in {Evans} C.~J.,  {Simard} L.,   {Takami} H.,
  eds,  Society of Photo-Optical Instrumentation Engineers (SPIE) Conference
  Series Vol. 9908, Ground-based and Airborne Instrumentation for Astronomy VI.
  p. 99081M (\mn@eprint {arXiv} {1608.01075}), \mn@doi{10.1117/12.2232103}

\bibitem[\protect\citeauthoryear{{Tejero-Cantero}, {Boelts}, {Deistler},
  {Lueckmann}, {Durkan}, {Gon{\c{c}}alves}, {Greenberg}  \&
  {Macke}}{{Tejero-Cantero} et~al.}{2020}]{tejero2020sbi}
{Tejero-Cantero} A.,  {Boelts} J.,  {Deistler} M.,  {Lueckmann} J.-M.,
  {Durkan} C.,  {Gon{\c{c}}alves} P.,  {Greenberg} D.,   {Macke} J.,  2020,
  \mn@doi [The Journal of Open Source Software] {10.21105/joss.02505}, \href
  {https://ui.adsabs.harvard.edu/abs/2020JOSS....5.2505T} {5, 2505}

\bibitem[\protect\citeauthoryear{{The CMB-HD Collaboration} et~al.,}{{The
  CMB-HD Collaboration} et~al.}{2022}]{2022arXiv220305728T}
{The CMB-HD Collaboration} et~al., 2022, \mn@doi [arXiv e-prints]
  {10.48550/arXiv.2203.05728}, \href
  {https://ui.adsabs.harvard.edu/abs/2022arXiv220305728T} {p. arXiv:2203.05728}

\bibitem[\protect\citeauthoryear{{Th{\'e}riault}, {Mirocha}  \&
  {Brandenberger}}{{Th{\'e}riault} et~al.}{2021}]{2021JCAP...10..046T}
{Th{\'e}riault} R.,  {Mirocha} J.~T.,   {Brandenberger} R.,  2021, \mn@doi
  [\jcap] {10.1088/1475-7516/2021/10/046}, \href
  {https://ui.adsabs.harvard.edu/abs/2021JCAP...10..046T} {2021, 046}

\bibitem[\protect\citeauthoryear{{Torki}, {Hajizadeh}, {Farhang}, {Vafaei Sadr}
   \& {Movahed}}{{Torki} et~al.}{2022}]{2022MNRAS.509.2169T}
{Torki} M.,  {Hajizadeh} H.,  {Farhang} M.,  {Vafaei Sadr} A.,   {Movahed}
  S.~M.~S.,  2022, \mn@doi [\mnras] {10.1093/mnras/stab3030}, \href
  {https://ui.adsabs.harvard.edu/abs/2022MNRAS.509.2169T} {509, 2169}

\bibitem[\protect\citeauthoryear{{Tuntsov} \& {Pshirkov}}{{Tuntsov} \&
  {Pshirkov}}{2010}]{Tuntsov:2010fu}
{Tuntsov} A.~V.,  {Pshirkov} M.~S.,  2010, \mn@doi [\prd]
  {10.1103/PhysRevD.81.063523}, \href
  {https://ui.adsabs.harvard.edu/abs/2010PhRvD..81f3523T} {81, 063523}

\bibitem[\protect\citeauthoryear{{Tye}}{{Tye}}{2008}]{HenryTye:2006uv}
{Tye} S. H.~H.,  2008, in {Gasperini} M.,  {Maharana} J.,  eds, , Vol.~737,
  String Theory and Fundamental Interactions.
p.~949, \mn@doi{10.48550/arXiv.hep-th/0610221}

\bibitem[\protect\citeauthoryear{{Vachaspati} \& {Vilenkin}}{{Vachaspati} \&
  {Vilenkin}}{1984}]{Vachaspati:1984dz}
{Vachaspati} T.,  {Vilenkin} A.,  1984, \mn@doi [\prd]
  {10.1103/PhysRevD.30.2036}, \href
  {https://ui.adsabs.harvard.edu/abs/1984PhRvD..30.2036V} {30, 2036}

\bibitem[\protect\citeauthoryear{{Vachaspati} \& {Vilenkin}}{{Vachaspati} \&
  {Vilenkin}}{1985}]{1985PhRvD..31.3052V}
{Vachaspati} T.,  {Vilenkin} A.,  1985, \mn@doi [\prd]
  {10.1103/PhysRevD.31.3052}, \href
  {https://ui.adsabs.harvard.edu/abs/1985PhRvD..31.3052V} {31, 3052}

\bibitem[\protect\citeauthoryear{{Vachaspati} \& {Vilenkin}}{{Vachaspati} \&
  {Vilenkin}}{1991}]{1991PhRvL..67.1057V}
{Vachaspati} T.,  {Vilenkin} A.,  1991, \mn@doi [\prl]
  {10.1103/PhysRevLett.67.1057}, \href
  {https://ui.adsabs.harvard.edu/abs/1991PhRvL..67.1057V} {67, 1057}

\bibitem[\protect\citeauthoryear{{Vafaei Sadr} \& {Movahed}}{{Vafaei Sadr} \&
  {Movahed}}{2021}]{2021MNRAS.503..815V}
{Vafaei Sadr} A.,  {Movahed} S.~M.~S.,  2021, \mn@doi [\mnras]
  {10.1093/mnras/stab368}, \href
  {https://ui.adsabs.harvard.edu/abs/2021MNRAS.503..815V} {503, 815}

\bibitem[\protect\citeauthoryear{{Vafaei Sadr}, {Movahed}, {Farhang},
  {Ringeval}  \& {Bouchet}}{{Vafaei Sadr} et~al.}{2018a}]{vafaei2017multiscale}
{Vafaei Sadr} A.,  {Movahed} S.~M.~S.,  {Farhang} M.,  {Ringeval} C.,
  {Bouchet} F.~R.,  2018a, \mn@doi [\mnras] {10.1093/mnras/stx3126}, \href
  {https://ui.adsabs.harvard.edu/abs/2018MNRAS.475.1010V} {475, 1010}

\bibitem[\protect\citeauthoryear{{Vafaei Sadr}, {Farhang}, {Movahed}, {Bassett}
   \& {Kunz}}{{Vafaei Sadr} et~al.}{2018b}]{sadr2018cosmic}
{Vafaei Sadr} A.,  {Farhang} M.,  {Movahed} S.~M.~S.,  {Bassett} B.,   {Kunz}
  M.,  2018b, \mn@doi [\mnras] {10.1093/mnras/sty1055}, \href
  {https://ui.adsabs.harvard.edu/abs/2018MNRAS.478.1132V} {478, 1132}

\bibitem[\protect\citeauthoryear{{Vilenkin}}{{Vilenkin}}{1981}]{Vilenkin:1981iu}
{Vilenkin} A.,  1981, \mn@doi [\prl] {10.1103/PhysRevLett.46.1169}, \href
  {https://ui.adsabs.harvard.edu/abs/1981PhRvL..46.1169V} {46, 1169}

\bibitem[\protect\citeauthoryear{{Vilenkin}}{{Vilenkin}}{1985}]{Vilenkin:1984ib}
{Vilenkin} A.,  1985, \mn@doi [\physrep] {10.1016/0370-1573(85)90033-X}, \href
  {https://ui.adsabs.harvard.edu/abs/1985PhR...121..263V} {121, 263}

\bibitem[\protect\citeauthoryear{{Vilenkin} \& {Shellard}}{{Vilenkin} \&
  {Shellard}}{2000}]{Vilenkin:2000jqa}
{Vilenkin} A.,  {Shellard} E.~P.~S.,  2000, {Cosmic Strings and Other
  Topological Defects}

\bibitem[\protect\citeauthoryear{White, Carlstrom  \& Dragovan}{White
  et~al.}{1999}]{White:1997wq}
White M.~J.,  Carlstrom J.~E.,   Dragovan M.,  1999, \mn@doi [Astrophys. J.]
  {10.1086/306911}, 514, 12

\bibitem[\protect\citeauthoryear{{Yamauchi}, {Takahashi}, {Sendouda}, {Yoo}  \&
  {Sasaki}}{{Yamauchi} et~al.}{2010}]{2010PhRvD..82f3518Y}
{Yamauchi} D.,  {Takahashi} K.,  {Sendouda} Y.,  {Yoo} C.-M.,   {Sasaki} M.,
  2010, \mn@doi [\prd] {10.1103/PhysRevD.82.063518}, \href
  {https://ui.adsabs.harvard.edu/abs/2010PhRvD..82f3518Y} {82, 063518}

\bibitem[\protect\citeauthoryear{{Yin}, {Dai}  \& {Ferraro}}{{Yin}
  et~al.}{2022}]{2022JCAP...06..033Y}
{Yin} W.~W.,  {Dai} L.,   {Ferraro} S.,  2022, \mn@doi [\jcap]
  {10.1088/1475-7516/2022/06/033}, \href
  {https://ui.adsabs.harvard.edu/abs/2022JCAP...06..033Y} {2022, 033}

\bibitem[\protect\citeauthoryear{{Yip}, {Biagetti}, {Cole}, {Viswanathan}  \&
  {Shiu}}{{Yip} et~al.}{2024}]{2024JCAP...09..034Y}
{Yip} J. H.~T.,  {Biagetti} M.,  {Cole} A.,  {Viswanathan} K.,   {Shiu} G.,
  2024, \mn@doi [\jcap] {10.1088/1475-7516/2024/09/034}, \href
  {https://ui.adsabs.harvard.edu/abs/2024JCAP...09..034Y} {2024, 034}

\bibitem[\protect\citeauthoryear{{Zeldovich}}{{Zeldovich}}{1980}]{Zeldovich:1980gh}
{Zeldovich} I.~B.,  1980, \mn@doi [\mnras] {10.1093/mnras/192.4.663}, \href
  {https://ui.adsabs.harvard.edu/abs/1980MNRAS.192..663Z} {192, 663}

\bibitem[\protect\citeauthoryear{{Zhan}}{{Zhan}}{2011}]{2011SSPMA}
{Zhan} H.,  2011, \mn@doi [Scientia Sinica Physica, Mechanica \& Astronomica]
  {10.1360/132011-961}, \href
  {https://ui.adsabs.harvard.edu/abs/2011SSPMA..41.1441Z} {41, 1441}

\bibitem[\protect\citeauthoryear{{Zhang} et~al.,}{{Zhang}
  et~al.}{2024}]{2024ApJS..274...26Z}
{Zhang} J.,  et~al., 2024, \mn@doi [\apjs] {10.3847/1538-4365/ad5c63}, \href
  {https://ui.adsabs.harvard.edu/abs/2024ApJS..274...26Z} {274, 26}

\bibitem[\protect\citeauthoryear{{Zou}, {Donner}, {Marwan}, {Donges}  \&
  {Kurths}}{{Zou} et~al.}{2019}]{zou2019complex}
{Zou} Y.,  {Donner} R.~V.,  {Marwan} N.,  {Donges} J.~F.,   {Kurths} J.,  2019,
  \mn@doi [\physrep] {10.1016/j.physrep.2018.10.005}, \href
  {https://ui.adsabs.harvard.edu/abs/2019PhR...787....1Z} {787, 1}

\makeatother
\end{thebibliography}

\end{document}